\documentclass[12pt]{article}
\pdfoutput=1
\usepackage[margin=1.0in]{geometry}
\usepackage[T1]{fontenc}
\usepackage[font=small, labelfont=bf]{caption}
\usepackage{graphicx}
\usepackage{amsmath}
\usepackage{heck}
\usepackage{amssymb}
\usepackage{amsthm}
\usepackage{longtable}
\usepackage{slashed}
\usepackage{verbatim}
\usepackage{cite}
\usepackage{enumerate}
\usepackage{subfig}  
\usepackage{soul}
\usepackage{url}
\usepackage{hyperref}
\usepackage[usenames,dviptsnames]{xcolor}
\usepackage{bm}

\newcommand{\Tr}{{\rm Tr}}

\newcommand{\mf}{\mathfrak}

\newcommand{\rmd}{\textrm{d}}
\newcommand{\JE}{J^{\textrm{E}}}
\newcommand{\JM}{J^{\textrm{M}}}

\newcommand{\Dp}{{\mathrm{D}p}}
\theoremstyle{definition}

\newcommand{\beq}{\begin{eqnarray}}
\newcommand{\eeq}{\end{eqnarray}}

\def\simgt{\stackrel{>}{{}_\sim}}
\usepackage{titlesec}

\numberwithin{equation}{section}

\newcommand{\assign}{:=}
\newcommand{\mathd}{\mathrm{d}}
\newcommand{\tmop}[1]{\ensuremath{\operatorname{#1}}}

\def\beqa{\begin{eqnarray}}
\def\eeqa{\end{eqnarray}}

\date{\today}
\title{
Chern-Weil Global Symmetries\\
and How Quantum Gravity Avoids Them
}
\institution{AMHERST}{\centerline{${}^{1}$Department of Physics, University of Massachusetts, Amherst, MA 01003 USA}}
\institution{HARVARD}{\centerline{${}^{2}$Department of Physics, Harvard University, Cambridge, MA 02138 USA}}
\institution{BERKELEY}{\centerline{${}^{3}$Physics Department, University of California, Berkeley CA 94720 USA}}
\institution{PRINCETON}{\centerline{${}^{4}$School of Natural Sciences, Institute for Advanced Study, Princeton, NJ 08540 USA}}
\authors{Ben Heidenreich\worksat{\AMHERST}\footnote{e-mail: {\tt bheidenreich@umass.edu}}, Jacob McNamara\worksat{\HARVARD}\footnote{e-mail: {\tt jmcnamara@g.harvard.edu}}, Miguel Montero\worksat{\HARVARD}\footnote{e-mail: {\tt mmontero@g.harvard.edu}}, Matthew Reece\worksat{\HARVARD}\footnote{e-mail: {\tt mreece@g.harvard.edu}}, Tom Rudelius\worksat{\BERKELEY, \PRINCETON}\footnote{e-mail: {\tt rudelius@berkeley.edu}}, and Irene Valenzuela\worksat{\HARVARD}\footnote{e-mail: {\tt ivalenzuela@g.harvard.edu}}}

\abstract{
We draw attention to a class of generalized global symmetries, which we call ``Chern-Weil global symmetries,'' that arise ubiquitously in gauge theories. The Noether currents of these Chern-Weil global symmetries are given by wedge products of gauge field strengths, such as $F_2 \wedge H_3$ and $\tr(F_2^2)$, and their conservation follows from Bianchi identities. As a result, they are not easy to break. However, it is widely believed that exact global symmetries are not allowed in a consistent theory of quantum gravity. As a result, any Chern-Weil global symmetry in a low-energy effective field theory must be either broken or gauged when the theory is coupled to gravity. In this paper, we explore the processes by which Chern-Weil symmetries may be broken or gauged in effective field theory and string theory. We will see that many familiar phenomena in string theory, such as axions, Chern-Simons terms, worldvolume degrees of freedom, and branes ending on or dissolving in other branes, can be interpreted as consequences of the absence of Chern-Weil symmetries in quantum gravity, suggesting that they might be general features of quantum gravity. We further discuss implications of breaking and gauging Chern-Weil symmetries for particle phenomenology and for boundary CFTs of AdS bulk theories. Chern-Weil global symmetries thus offer a unified framework for understanding many familiar aspects of quantum field theory and quantum gravity.
}

\begin{document}
\hfill ACFI-T20-16 \vspace{-1.5\baselineskip} 
\maketitle

\setcounter{tocdepth}{2}
\tableofcontents

\section{Introduction}

Symmetries have long played a major role in theoretical physics. In recent years, it has been widely understood that the concept of symmetry is broader than previously recognized \cite{Kapustin:2013uxa, Kapustin:2014gua}, and the notion of a $p$-form {\em generalized global symmetry} has been introduced \cite{Gaiotto:2014kfa}. Such a symmetry acts on charged operators supported on $p$-dimensional manifolds and is realized (in theories with $d$ spacetime dimensions) by linking the charged operators with $(d-p-1)$-dimensional symmetry operators. This new language has led to crisp reformulations of old physics: for example, the classic topological understanding of confinement due to Polyakov and 't~Hooft \cite{Polyakov:1976fu, tHooft:1977nqb} becomes the statement that the confining phase of a gauge theory with gauge group $G$ is one where a 1-form $Z(G)$ center symmetry (acting on 1-dimensional Wilson loop operators) is preserved by the theory's ground state. 

In this paper, we point out a ubiquitous class of generalized global symmetries that have received limited attention from the symmetry point of view. These objects, which we call {\em Chern-Weil} symmetries, are generated by (gauge-invariant) products of gauge field strengths. To give a few examples, the generators could take the form of ${\rm tr}(F \wedge F)$ or ${\rm tr}(F \wedge F \wedge F \wedge \cdots)$ where $F$ is the field strength of a nonabelian gauge theory, or $F$ or  $F \wedge H$ where $F$, $H$ are field strengths of $U(1)$ gauge fields, or even (generalizing the notion of ``gauge field'' slightly) ${\rm d}\sigma$ or ${\rm d}\sigma \wedge F$ where $\sigma$ is a compact (periodic) scalar field. All of these expressions are closed forms thanks to Bianchi identities obeyed by the gauge fields. This fact may be familiar to readers from the theory of characteristic classes (see, e.g., chapter 11 of \cite{Nakahara:2003nw}), but we will review the explanation in \S\ref{ssec:CW} below. Any such closed $k$-form $\omega_k$ with quantized (integer) periods gives rise to a $U(1)$ global symmetry generated by the operators $\exp(i \alpha \int_{M_k} \omega_k)$. We refer to these as Chern-Weil symmetries owing to the two mathematicians who initially appreciated the importance of such closed forms, years before the advent of Yang-Mills theory itself.

The fact that these closed forms give rise to global symmetries can have important implications for the dynamics of gauge theories. As with any global symmetry, we can gauge a Chern-Weil symmetry, in the absence of 't~Hooft anomalies. In some cases Chern-Weil symmetries are explicitly broken, but because this requires a breakdown in the Bianchi identity for a gauge field, it imposes requirements on the ultraviolet dynamics of the gauge theory. Thus, like other global symmetries, Chern-Weil symmetries provide useful probes of a theory.

These symmetries have been occasionally discussed in the literature, especially in 5d gauge theories where ${\rm tr}(F \wedge F)$ generates an ordinary 0-form global symmetry under which instantons are charged particles \cite{Lambert:2014jna, Tachikawa:2015mha, BenettiGenolini:2020doj}, and in 6d gauge theories where such a current generates a 1-form symmetry under which strings are charged \cite{Apruzzi:2020zot, Bhardwaj:2020phs, Cordova:2020tij}.\footnote{In 6d superconformal field theories, these 1-form symmetries are necessarily gauged \cite{Cordova:2020tij}. The charged strings have been extensively studied; see, e.g., \cite{Haghighat:2013tka, Kim:2018gjo}.} However, to the best of our knowledge there has been little systematic study of general Chern-Weil global symmetries. We find that Chern-Weil symmetries provide a unifying language in which a wide variety of phenomena in quantum field theory, and especially in string theory, can be usefully rephrased.

Theories of quantum gravity are believed to have no global symmetries \cite{Hawking:1974sw, Zeldovich:1976vq, Zeldovich:1977be, Banks:1988yz}. Recent work has sharpened this old folklore and extended it to the case of generalized global symmetries \cite{Banks:2010zn, Harlow:2018jwu, Harlow:2018tng}. (For even more recent arguments, see \cite{Harlow:2020bee, Chen:2020ojn, Hsin:2020mfa,Yonekura:2020ino}.) Of course, effective field theories describing the low-energy dynamics of quantum gravity often contain gauge fields, and hence potentially have Chern-Weil global symmetries. This raises the question: how does quantum gravity avoid having Chern-Weil global symmetries? In broad strokes, the answer is clear: any candidate Chern-Weil global symmetry must be either gauged or broken. In practice, however, we find that the physics of {\em how} Chern-Weil global symmetries are gauged or broken is often rich with subtleties. By investigating this question in various corners of quantum field theory and string theory, we find a new perspective on familiar physics. For example, Chern-Simons terms are ubiquitous in string theory vacua. It has previously been argued that the presence of these terms in theories of quantum gravity is sufficient to avoid certain paradoxes involving black hole remnants \cite{Montero:2017yja}. We suggest a different reason why  such terms are so common in theories of quantum gravity: by gauging Chern-Weil symmetries, they remove potential global symmetries from the theory. On the other hand, some Chern-Weil symmetries are simply broken. For example, if a gauge group $G$ is Higgsed to a product group $G_1 \times G_2 \subset G$, then the low energy effective field theory contains two different Chern-Weil symmetry generators, ${\rm tr}(F_1 \wedge F_1)$ and ${\rm tr}(F_2 \wedge F_2)$, while the UV theory has only a single such generator ${\rm tr}(F \wedge F)$. In this case, the low-energy Chern-Weil global  symmetries are {\em accidental}, and are explicitly broken (to a diagonal subgroup) in the full theory.

We find many examples in which Chern-Weil symmetries in quantum gravity are {\em both} broken and gauged. By this we mean that a linear combination of the Chern-Weil current and other currents is gauged, but the Chern-Weil charge is not conserved independently. (Notice that here, and throughout the paper, we use the term ``current'' even for an operator that is not conserved, but which is potentially or approximately conserved.) This is not so different from, for example, the electron charge current, which is gauged by the photon but broken because the charge of an electron can be converted into the charge of (e.g.) a muon. Furthermore, we find that the breaking of the Chern-Weil symmetry is often correlated with the existence of localized degrees of freedom. For example, in a $U(1)$ gauge theory, the existence of magnetic monopoles breaks the would-be Chern-Weil symmetry generated by $F \wedge F$, so gauging this current via  a Chern-Simons term of the form $C \wedge F \wedge F$ is inconsistent. However,  when these magnetic monopoles admit dyonic excitations, they have a compact  scalar $\sigma$ localized on their worldvolume, and a  current  of the form $F \wedge F + \delta^{\rm m}_3 \wedge {\rm d}_A \sigma$ is conserved and may be gauged. In some cases, the presence of a Chern-Weil global symmetry in a UV theory may be used to argue for the existence of localized degrees of freedom on the worldvolume of objects in an IR theory, leading to a new perspective on results that could be obtained with anomaly inflow \cite{Callan:1984sa}.

It has been argued that theories of quantum gravity should admit states of all possible charges; this is known as the ``Completeness Hypothesis'' \cite{polchinski:2003bq, Harlow:2018tng}. This notion is closely linked to the absence of global symmetries. As an example, in a $U(1)$ gauge theory the existence of magnetic monopoles breaks the $(d-3)$-form Chern-Weil symmetry generated by the current $F$  (this simplest example of a Chern-Weil global symmetry is the familiar magnetic 1-form global symmetry in 4d Maxwell theory \cite{Gaiotto:2014kfa}). More generally, an incomplete spectrum in a gauge theory implies the existence of certain topological operators, which generalize global symmetries \cite{Rudelius:2020orz, completenesspaper}. In string theory, however, enumerating the set of charged objects (such as fundamental strings, D-branes, and NS5-branes) only tells a small fraction of the story. There are also extensive relationships among these objects, which can  end on or dissolve into each other. We find that these phenomena are closely linked to the absence of Chern-Weil global symmetries. In particular, the interplay between Chern-Simons terms, which gauge Chern-Weil symmetries, and brane-localized degrees of freedom, which are necessary for consistency of this gauging, leads to phenomena such as the dissolving of lower-dimensional branes in higher-dimensional branes. Many aspects of the physics of string theory, traditionally presented as consequences of direct computation from the top down in specific supersymmetric theories, are seen from our perspective as necessary ingredients to provide a consistent theory of quantum gravity containing gauge fields but lacking Chern-Weil global symmetries. 

Recently, it was conjectured that quantum gravity theories cannot have nontrivial cobordism invariants \cite{McNamara:2019rup}. Although we focus our discussion on the absence of Chern-Weil global symmetries, one could rephrase much of the discussion in the language of cobordism. While \cite{McNamara:2019rup} mainly focused on cobordisms of compactification manifolds, we could also consider cobordisms of manifolds equipped with gauge bundles. As Chern-Weil currents represent topological invariants of the gauge bundle, they serve as a first approximation to the collection of symmetries arising from cobordism classes of manifolds with gauge fields.\footnote{More precisely, they enter as the bottom row in the Atiyah-Hirzebruch spectral sequence computing the bordism homology of the classifying space of the gauge group.} Moreover, Chern-Weil currents arising from Poincar\'{e} symmetry, such as $\tr(R \wedge R)$, describe charges carried by nontrivial cobordism classes. As a result, the absence of nontrivial cobordism invariants is closely related to the absence of Chern-Weil global symmetries. 

The paper is organized as follows. In Section \ref{sec:HigherForm}, we provide a review of generalized global symmetries, their currents and charged operators, and what it means to gauge or break them. We then introduce Chern-Weil symmetries. Section \ref{sec:F} discusses the simplest example of Chern-Weil currents, those involving a single field strength. We revisit the familiar BF theory and reframe the discussion of axion monodromy models in terms of the slightly exotic notion of ``$(-1)$-form symmetry,'' which we introduce and discuss in some detail. Section \ref{sec:F2} focuses on two-field Chern-Weil currents, explaining their gauging, breaking, and how the combination of the two can lead to nontrivial constraints on the worldvolume degrees of freedom of charged objects. Section \ref{sec:String} illustrates the general considerations in examples from several string compactifications, while Section \ref{sec:adscft} does the same in the context of AdS/CFT. Section \ref{sec:pheno} discusses how careful consideration of Chern-Weil symmetries could lead to interesting phenomenological constraints, and finally, Section \ref{sec:conclus} contains our conclusions.

\section{Higher-Form Global Symmetries}\label{sec:HigherForm}

We begin this section with a review of higher-form global symmetries, as discussed in the seminal paper \cite{Gaiotto:2014kfa}. A $p$-form global symmetry in a quantum field theory in $d$ dimensions is a global symmetry for which the charged operators are supported on $p$-dimensional manifolds. The case $p=0$ corresponds to an ordinary global symmetry, as the charged operators are local operators, supported on manifolds of dimension 0, i.e., points in spacetime. Global symmetry transformations form a group $G$. $G$ may be either nonabelian or abelian in the case of a $0$-form global symmetry, but $G$ must be abelian for a $p$-form symmetry with $p > 0$.

The symmetry group $G$ may be either discrete or continuous. If $G$ is continuous, then under certain (often reasonable) assumptions, it admits a conserved $(d-p-1)$-form Noether current $J_{d-p-1}$. The statement that $J_{d-p-1}$ is conserved is equivalent to the statement that it is closed as a differential form:\footnote{Some readers may be more accustomed to the statement that a $p$-form global symmetry admits a $(p+1)$-form Noether current $j_{p+1}$, which is co-closed: $\rmd {\star j_{p+1}}=0$. This is simply a matter of convention, and the two definitions of the Noether current are related via Hodge duality, $J_{d-p-1} = \star j_{p+1}$.}
\begin{equation}
\rmd J_{d-p-1} = 0.
\label{Noethercons}
\end{equation}

The hallmark of a $p$-form global symmetry is the existence of topological operators $U_g(M^{(d-p-1)})$, which are labeled by elements $g$ of the group $G$ and supported on manifolds $M^{(d-p-1)}$ of dimension $d-p-1$. Here, the statement that the ``charge operator'' is topological means that all correlations functions with the operator are invariant under continuous deformations of the manifold $M^{(d-p-1)}$, provided that the manifold does not cross any charged operators in the deformation process. Two such charge operators supported on the same manifold fuse according to the group multiplication law:
\begin{equation}
U_g(M^{(d-p-1)}) \times U_{g'}(M^{(d-p-1)}) = U_{g''}(M^{(d-p-1)}),
\end{equation}
with $g'' = g g'$. When $G=U(1)$ and a Noether current exists, one can write the charge operators more explicitly,
 \begin{equation}
U_{g=\mathrm{e}^{i \alpha}}(M^{(d-p-1)}) = \exp\left(i \alpha \oint_{M^{(d-p-1)}} J_{d-p-1}  \right).
\end{equation}
The topological nature of the charge operators then follows immediately from the conservation of $J_{d-p-1}$.

Symmetries are associated with Ward identities. In terms of these operators, such a Ward identity is given by the relation,
\begin{equation}
U_g(S^{d-p-1}) V(C^{(p)}) = g(V) V(C^{(p)}) ,
\end{equation}
where $U_g(S^{d-p-1})$ is supported on a small $(d-p-1)$-sphere that links once with the charged operator $V(C^{(p)})$, which lives on the manifold $C^{(p)}$ of dimension $p$. Here, $g(V)$ is a representation of $g$, which is simply a phase determined by the charge of $V$ in the case that $G$ is abelian.

The prototypical example of a theory with a higher-form global symmetry is Maxwell theory in four dimensions, which has action 
\begin{equation}
S = \int \left( -\frac{1}{2 g^2} F_2 \wedge \star F_2 \right).
\end{equation}
This theory has Wilson line operators, each of which is supported on a closed 1-manifold $\gamma^{(1)}$ and labeled by an electric charge $n \in \mathbb{Z}$:
\begin{equation}
W_{n}(\gamma^{(1)}) = \exp \left( i n \oint_{\gamma^{(1)}} A_1  \right) .
\end{equation}
Wilson lines are charged under an electric 1-form $U(1)$ global symmetry. Its conserved Noether current is given by
\begin{equation}
\JE_2 = \frac{1}{g^2} {\star F_2}.
\end{equation}
The generators of this global symmetry are labeled by a phase $\alpha \in [0, 2 \pi)$ and are supported on dimension-2 manifolds $M^{(2)}$. Explicitly, such operators are given by the exponentiated integral of the Noether current over $M^{(2)}$,
\begin{equation}
U_{g=\mathrm{e}^{i \alpha}} = \exp \left(  \frac{ i \alpha}{g^2} \int_{M^{(2)}} \star F_2 \right).
\end{equation}
The Ward identity for the electric 1-form symmetry says that if a Wilson line of charge $n$ supported on a curve $\gamma^{(1)}$ is surrounded by an $S^{2}$ supporting a symmetry generator $U_{g={\rm e}^{i \alpha}}$, the symmetry generator can be removed at the expense of a phase $\exp(in \alpha)$,
\begin{equation}
 \exp \left(  \frac{ i \alpha}{g^2} \int_{S^2} \star F_2 \right) \cdot \exp \left( i n \oint_{\gamma^{(1)}} A_1  \right) =   \exp  \left(  i n \alpha \right)  \exp \left( i n \oint_{\gamma^{(1)}} A_1  \right).
\end{equation}

Pure Maxwell theory in four dimensions also has a magnetic 1-form $U(1)$ global symmetry with Noether current $\JM_2 = \frac{1}{2 \pi} F_2$ whose charged operators are 't~Hooft lines. The symmetry generators and Ward identity for this symmetry are perfectly analogous to the electric case we have just discussed, and indeed the two are related by electromagnetic duality.

\subsection{Breaking and Gauging Global Symmetries}\label{sec:bg}

Standard lore holds that exact global symmetries are not allowed in a consistent theory of quantum gravity. This means that any would-be global symmetry in a quantum field theory must be removed before such a theory can be coupled to quantum gravity. Within the framework of quantum field theory, there are two ways to remove a global symmetry: (i) gauge the symmetry or (ii) break it explicitly.\footnote{A spontaneously broken symmetry is still a global symmetry of the quantum field theory from the modern perspective, as it still has topological operators $U_{g}$. For the purposes of this paper, we work under the assumption that no such topological operators may exist in a consistent theory of quantum gravity.}

To gauge a $p$-form global symmetry with Noether current $J_{d-p-1}$, one couples the current to a dynamical $(p+1)$-form gauge field $B_{p+1}$:
\begin{equation}
S \supset \int \left( -\frac{1}{2g^2} H_{p+2} \wedge \star H_{p+2} + B_{p+1} \wedge J_{d-p-1} \right),
\label{gaugingLag}
\end{equation}
with $H_{p+2}=\rmd B_{p+1}$ (locally).
The conservation of $J_{d-p-1}$ implies that the action is invariant up to boundary terms under symmetry transformations $B_{p+1} \rightarrow B_{p+1} + \rmd \Lambda_{p}$. The equation of motion for $B_{p+1}$ then takes the form
\begin{equation}
\label{GaussLaw}
\frac{1}{g^2} \rmd\, {\star H_{p+2}} = (-1)^{p+1} J_{d-p-1}.
\end{equation}
In particular, as a result of the gauging process, $J_{d-p-1}$ is not merely closed, but exact. Equation \eqref{GaussLaw} is a generalization of Gauss's Law, and implies that the integral of $J_{d-p-1}$ over a closed manifold vanishes.

More generally, the gauging of a $p$-form symmetry may be understood in terms of the operators charged under that symmetry. Such charged operators are no longer genuine operators of the theory once the symmetry has been gauged, as the gauging process projects onto the gauge-singlet sector of the theory. Instead, the charged operators will represent boundaries of higher-dimensional gauge-invariant operators. For instance, one may consider a local operator $\phi(x)$ carrying charge $n$ under some 0-form $U(1)$ global symmetry. Upon gauging this symmetry, $\phi(x)$ is no longer a genuine local operator of the theory, but instead $\phi(x)$ represents the endpoint of a Wilson line operator of charge $n$.

In contrast, one way to break a symmetry is to add terms to the Lagrangian that violate the conservation of the Noether current. As a first example, we may consider once again the theory in (\ref{gaugingLag}). In the absence of the $B \wedge J$ coupling, $J_{d-p-2}:={\star} H_{p+2}$ is a conserved current for a $(p+1)$-form symmetry. Thus, the $B \wedge J$ coupling serves to both gauge the $p$-form symmetry with current $J_{d-p-1}$ and break the $(p+1)$-form symmetry with current $J_{d-p-2}$.\footnote{For mathematically inclined readers, this should be thought of as a differential in a cochain complex or spectral sequence, which removes both cochains from cohomology.}

As a second example of global symmetry breaking, consider the electric 1-form symmetry of pure $U(1)$ gauge theory in four dimensions discussed previously. To break this 1-form global symmetry, we add a charged particle $\varphi$ (which we take to be a complex scalar) to the gauge theory,
\begin{equation}
\int \left( -\frac{1}{2 g^2} F_2 \wedge \star F_2 - (\rmd_A  \varphi)^\dagger \wedge \star {\rmd_A \varphi} - m^2 \varphi^\dagger \varphi \star 1\right),
\end{equation}
with $\rmd_A = \rmd - i A_1$. The equation of motion for the gauge field then takes the form
\begin{equation}
\frac{1}{g^2} \rmd  \, {\star F_2}=  -j_3^\mathrm{e} = \star(i \varphi^\dagger \rmd_A \varphi - i \varphi \, \rmd_A \varphi^\dagger).
\end{equation}
In particular, $\JE_2$ is no longer conserved: the $U(1)$ electric 1-form symmetry has been broken by the addition of the charged particle. More generally, a charge $n$ particle will break this $U(1)$ symmetry to a $\mathbb{Z}_n$ subgroup.

The process of breaking a higher-form global symmetry may often be understood in terms of the operators charged under the symmetry as well.  In particular, one way to break a $p$-form global symmetry is to modify the theory with the global symmetry so as to allow the charged $p$-dimensional operators to live on $p$-dimensional manifolds with nontrivial boundary (such operators are called \emph{endable} in the language of \cite{Rudelius:2020orz}). For instance, in pure $U(1)$ gauge theory, Wilson lines may be supported only on closed manifolds without boundary, so they are not endable. But in the presence of charge $n$ matter $\phi$, the Wilson line of charge $n$ is endable, as it may end on a local excitation $\phi(x)$ of the charged particle. Any global symmetry generator $U_{g}(S^{2})$ linking the Wilson line with boundary may be smoothly unlinked, which means that the Wilson line cannot carry charge under this symmetry. This in turn restricts $g=\mathrm{e}^{i \alpha}$ to be an $n$th root of unity, which shows that indeed, the presence of the charge $n$ particle breaks the $U(1)$ electric 1-form global symmetry of the pure gauge theory down to a $\mathbb{Z}_n$ subgroup.

\subsection{Chern-Weil Global Symmetries}\label{ssec:CW}

We have seen that pure $U(1)$ gauge theory in four dimensions possesses a conserved 2-form current, $F_2$. Indeed, this is true more generally for pure $U(1)$ gauge theory in $d \geq 3$ spacetime dimensions, as the equation of motion gives $\rmd F_2 = 0$ in the absence of any magnetically-charged sources.

In fact, $F_2$ is only one of a family of conserved $p$-form currents in $U(1)$ gauge theory. In particular, we may consider the current
\begin{equation}
J_{2k} = \underset{k}{\underbrace{F_2 \wedge \cdots \wedge F_2}}:=F_2^k.
\label{CW}
\end{equation}
This satisfies
\begin{equation}
\rmd J_{2k} = k \,\rmd F_2 \wedge F_2^{k-1} ,
\end{equation}
so it is indeed conserved provided $\rmd F_2 = 0$.

A nonabelian gauge theory with field strength $F_2$ similarly possesses currents of the form
\begin{equation}
J_{2k} = \tr\Big[ \underset{k}{\underbrace{F_2 \wedge \cdots \wedge F_2}}\Big] := \tr \,F_2^k.
\end{equation}
The trace is taken in some representation, which we will suppress in our notation unless it is necessary to clarify.    The conservation of $J_{2k}$ follows from rewriting (see, e.g., Theorem 11.1 of  Ref.~\cite{Nakahara:2003nw})
\begin{align}
\rmd J_{2k} =&~ \rmd \,\tr \,F_2^k. \nonumber \\
= &~k \tr \left[\rmd F_2 \wedge F_2^{k-1} \right] \nonumber \\
= &~k \tr \left[(\rmd F_2-i [A_1,F_2]) \wedge F_2^{k-1} \right] \nonumber \\
= &~k \tr \left[\rmd_A F_2 \wedge F_2^{k-1} \right] \nonumber \\
=&~0,
\end{align}
where in the last step we have used the Bianchi identity $\rmd_A F_2 := \rmd F_2 - i [A_1, F_2]$ = 0. Note that in the nonabelian, semisimple case, $J_{2k}$ vanishes for $k=1$ since $\tr \,F_2=0$, so we do not have a conserved current in this case. Similarly, if the gauge group $G$ is associated with a simple Lie algebra $\mathfrak{g}$, $J_{2k}$ vanishes for $k=3$ unless $\mathfrak{g}=A_{n-1}$ with $n \geq 3$. We could also form multi-trace conserved currents, such as $J_{2k} \wedge J_{2l}$.

By a similar analysis, a higher-form gauge symmetry with ($p-1$)-form abelian gauge field $C_{p-1}$ also possesses a collection of conserved currents,
\begin{equation}
J_{kp} =\underset{k}{\underbrace{H_p \wedge \cdots \wedge H_p}} := H_p^k.
\end{equation}
with $H_p = \rmd C_{p-1}$ (locally). These vanish for $k > 1$ if $p$ is odd, due to antisymmetry of the wedge product.

Even more generally, a theory with multiple gauge fields will have conserved currents involving products of field strengths of distinct gauge fields. In particular, a theory with a collection of $(p_i-1)$-form gauge fields $C^{(1)}_{p_1 - 1}$, $C^{(2)}_{p_2 - 1}$,...,$C^{(N)}_{p_N - 1}$ has conserved currents of the form
\begin{equation}
J^{i_1,...,i_k} =F_{p_{i_1}}^{(i_1)} \wedge ... \wedge F^{(i_k)}_{p_{i_k}}
 \label{eq:mixedChernWeil}
\end{equation}
for an arbitrary set of indices $i_1,...,i_k \in \{ 1,..., N \}$, where an index $i \in N$ may be repeated more than once (again, provided $p_i$ is even, since $F^{(i)}_{p_i} \wedge F^{(i)}_{p_i} = 0$ for $p_i$ odd). 

We will refer to all the currents we have just constructed, in both abelian and nonabelian gauge theory, as ``Chern-Weil'' currents, as the conservation of these currents is a lemma in the construction of the Chern-Weil homomorphism, which plays a key role in the study of characteristic classes of gauge bundles. Likewise, we will refer to the symmetries generated by these currents as Chern-Weil symmetries. Note that the current $J_{kp}$ generates a $(d-kp-1)$-form global symmetry in $d$ dimensions, which is well-defined provided $d > kp$.

If the gauge groups whose field strengths appear in the Chern-Weil current $J$ are compact, $J$ will have quantized periods, thereby generating a $U(1)$ global symmetry (as opposed to $\mathbb{R}$). For $k=1$, this is simply the familiar statement of Dirac quantization for magnetic charge. For $k=2$, $J = \tr (F_2 \wedge F_2)$, this is the familiar statement that instanton number is quantized. From the physics point of view, this is related to the fact that a single BPST instanton \cite{Belavin:1975fg} is accompanied by the 't~Hooft vertex, which emits an integer number of fermions of each flavor and chirality, due to the chiral anomaly \cite{tHooft:1976rip}. The mathematical counterpart of this is the Atiyah-Singer index theorem (see, e.g., \cite{Nakahara:2003nw}), which relates the integral of $J$ to the number of zero modes of the 4d Dirac operator coupled to the instanton gauge bundle, which is an integer. Similar arguments hold for Dirac operators coupled to arbitrary bundles in any dimension, and integrality properties of the more general Chern-Weil currents are related to the integrality of these indices.\footnote{The classic approach \cite{milnor1974characteristic} uses the splitting principle to show that any bundle over a manifold $M$ is the pullback of a direct sum of $U(1)$ bundles over a certain auxiliary manifold $\text{Fl}(M)$, so integrality follows from integrality in the abelian case.} 

The case $d =kp$ is somewhat degenerate: we may still define the Chern-Weil current $J_{kp} = J_d$, but the conservation $\rmd J_d = 0$ of this current is trivial since it is a top form. Nonetheless, it can be useful to think of such a current as generating a ``($-1$)-form'' global symmetry,\footnote{The notion of a $(-1)$-form symmetry has already appeared in the literature (see, e.g., \cite{Cordova:2019jnf, Tanizaki:2019rbk, McNamara:2020uza}) and is linked to the presence of  ``dynamical parameters.''} and we will even see that there is a sense in which these symmetries can be broken or gauged.
Such a ``symmetry'' may be considered a $U(1)$ symmetry if the periods of $J_d$, $\oint_{\text{spacetime}} J_d$, are quantized, whereas it is an $\mathbb{R}$ symmetry otherwise.

While the notion of a ($-1$)-form \emph{global} symmetry is somewhat tenuous, since any top form is closed, the notion of \emph{gauging} a ($-1$)-form symmetry is on more solid footing: gauging a global symmetry renders the current exact, and it is perfectly sensible to ask if a top form $J_d$ is exact. $U(1)$ ($-1$)-form gauge symmetries are associated with compact scalar fields, also known as axions, which play an important role in model building for cosmology and particle physics.
An axion $\theta$ is thus identified as a 0-form gauge field, which leads to the Chern-Weil current $\rmd \theta$ as well as potentially mixed currents, e.g., $\rmd \theta^{(1)} \wedge \rmd \theta^{(2)}$ or $\rmd \theta \wedge F_2$, as in \eqref{eq:mixedChernWeil}.
We will revisit these ($-1$)-form symmetries periodically in the remainder of this paper.

As we have just seen, the conservation of Chern-Weil currents follows from the Bianchi identity for the field strength of a (possibly nonabelian) gauge theory. As a result, Chern-Weil symmetries are not easy to break---their conservation follows from rather general statements in the theory of characteristic classes. Nonetheless, compatibility with quantum gravity lore requires these symmetries to be either broken or gauged in quantum gravity. In what follows, we will see how exactly this may be done, examining the breaking/gauging of Chern-Weil symmetries in effective field theory, string/M-theory, and AdS/CFT.

\section{$F$ Chern-Weil Currents\label{sec:F}}

We begin by discussing Chern-Weil currents of the form $J_p=F_p$. This corresponds to the simplest case in \eqref{CW}, namely $k=1$, and reduces to the well-understood $(d-p-1)$-form magnetic global symmetry. Still, it is illustrative to consider this case first, as many of the features regarding the gauging and breaking mechanisms will also be present in the cases with $k>1$. Hence, we will use this section to introduce some concepts that will be useful for future sections, including different notions of charge \cite{Marolf:2000cb}, as well as a first encounter with the apparent puzzle of introducing charged states to break a symmetry which is already gauged by a coupling to $B_{d-p}$ field.

\subsection{BF Theory}\label{sec:BFgauging}

Let us begin with the canonical example of BF theory in four dimensions (see \cite{Banks:2010zn} for a similar discussion). 
The action is given by 
\beq
\label{SBF}
S=\int \left(-\frac{1}{2 g^2} H_3\wedge \star H_3-\frac{1}{2e^2} F_2\wedge \star F_2+\frac{m}{2\pi}\, B_2 \wedge F_2\right).
\eeq
where $F_2=\rmd A_1$, $H_3=\rmd B_2$, and $m \in \mathbb{Z}$. With the BF coupling $m$ set to zero, the theory has four independent $U(1)$ global symmetries: a 0-form symmetry with current $H_3$, an electric 1-form symmetry with current $\star F_2$, a magnetic 1-form symmetry with current $F_2$, and a 2-form symmetry with current $\star H_3$. All of them are examples of Chern-Weil symmetries with $k=1$ in an appropriate duality frame. As discussed in Section \ref{sec:bg}, these symmetries may be broken by the addition of \emph{fundamental} objects (i.e., configurations which are singular from the point of view of the low-energy effective field theory), namely instantons, electric charges, monopoles, and strings, so that we have
\beq
\label{bfcharges}
\frac{1}{2\pi} \rmd H_3 = j_4^{\textrm{inst}}, \quad \frac{1}{e^2} \rmd  {\star F_2} = j_3^{\rm e}, \quad \frac{1}{2\pi} \rmd F_2 = j_3^{\rm m}, \quad \frac{1}{g^2} \rmd  {\star H_3} = j_2^{\rm string}.
\eeq
In this section, we investigate the fate of these global symmetries in the presence of a nonzero BF coupling $m \neq 0$, both without and with additional charged objects.

Let us begin by focusing on the magnetic 1-form symmetry with current $F_2$, which is conserved,
\beq
\rmd F_2=0,
\eeq
in the absence of monopoles. In fact, with $m \neq 0$, this symmetry is gauged by the coupling to $B_2$ in \eqref{SBF} since the current is exact, as can be seen from the equation of motion for $B_2$,
\beq
\frac{1}{g^2} \rmd \, {\star H_3}=\frac{m}{2\pi} F_2.
\label{d*H}
\eeq
In this case, it is also possible to see the gauging explicitly as a Stueckelberg coupling in the dual Lagrangian description,
\beq
S_\mathrm{dual}=\int \left( -\frac{1}{2g^2} H_3\wedge \star H_3-\frac{1}{2 {\tilde e}^2}  (\rmd\tilde A_1 -m\,B_2)^2\right),
\eeq
where $\tilde A_1$ is the electromagnetic dual field to $A_1$, with field strength $\tilde F_2$ and coupling $\tilde e = 2\pi/e$. The $\tilde A_1$ field gets ``eaten'' by the 2-form gauge field, which becomes massive. The theory is only invariant under the combined transformation
\beq
\tilde A_1\rightarrow \tilde A_1+m\Lambda_1\ ,\quad B_2\rightarrow B_2+ \rmd \Lambda_1
\eeq
which indicates that the 1-form magnetic global symmetry is gauged by the 2-form gauge field $B_2$.

The 0-form global symmetry with current $H_3$ is gauged in an analogous way. The current is exact,
\beq
\label{d*F}
\rmd H_3=0\ ,\quad \frac{1}{e^2} \rmd \, {\star F_2}= -\frac{m}{2\pi} H_3,
\eeq
meaning that the 0-form symmetry is gauged by the electric 1-form gauge field $A_1$. In terms of the compact scalar field $\phi$ which is dual to the 2-form gauge field $B_2$, one recovers the typical Stueckelberg action
\beq
\label{stuckelberg}
S_\mathrm{dual}'=\int \left( -\frac{1}{2e^2} F_2\wedge \star F_2-\frac{1}{2} f^2 (\rmd \phi -m\,A_1)^2\right),
\eeq
where $f = g/(2\pi)$, and so indeed this 0-form symmetry (corresponding to shifts of $\phi$) is gauged by the 1-form gauge field $A_1$.

We have just seen that two of the four potential global symmetries of \eqref{SBF} are gauged. The other two (namely, the 2-form and the electric 1-form global symmetries) are instead broken to $\mathbb{Z}_m$ by the BF coupling. The current associated to the 2-form global symmetry is ${\star H_3}$, which by \eqref{d*H} is not conserved.
Hence, \eqref{d*H} may be simultaneously read as gauging the 1-form global symmetry with $F_2$ and breaking the 2-form global symmetry with $\star H_3$, leaving no conserved global current if $m=1$.\footnote{This possibility was emphasized in \cite{Montero:2017yja}, where it was also conjectured that Chern-Simons couplings are indeed the ubiquitous mechanism by which quantum gravity guarantees the breaking/gauging of higher-form symmetries.}

As discussed in Section \ref{sec:bg}, the breaking of global symmetries may be related to the presence of charged objects. In this case, we should expect the non-conservation of $\star H_3$ to be related to the presence of strings charged under $B_2$. Indeed, by comparison with \eqref{bfcharges}, we may identify these strings as field configurations carrying nonzero magnetic flux $\int F_2$. Though these magnetic flux strings are not fundamental defects, they are still dynamical objects of codimension 2, and carry $m$ units of string charge for each unit of magnetic flux. One might object that there are no infinitely long magnetic flux strings with finite tension, since we may calculate
\beq
\int_{\mathbb{R}^2} F_2 = \frac{2\pi }{g^2 m} \int_{S^1_\infty} \star H_3 = 0,
\eeq
using \eqref{d*H} and the fact that $H_3$ must die off faster than $1/r$ in order to have finite tension. However, this is a consequence of the fact that $B_2$ is confining in four dimensions, just as an ordinary $U(1)$ gauge field is confining in three dimensions. The magnetic flux strings are thus analogous to quarks in QCD: charged dynamical excitations in the UV, but confined in the IR.

Similarly, the current associated to the electric 1-form global symmetry is $\star F_2$, which by \eqref{d*F} is not conserved. The breaking of this symmetry corresponds to the presence of dynamical objects with $m$ units of electric charge, which in this case are the codimension-3 localized fluxes of $H_3$. Unlike fluxes of $F_2$, fluxes of $H_3$ are not confined, since the gauge field $A_1$ is not confining in four dimensions.

In summary, the BF coupling gauges the 0-form symmetry with current $H_3$ and breaks the dual current $\star H_3$. Likewise, it breaks the electric 1-form global symmetry with current  $\star F_2$ and gauges the dual magnetic symmetry, with current $F_2$. This can be understood as a consequence of the mixed `t Hooft anomaly between the 0-form and 2-form symmetries, as well as between the electric and magnetic 1-form symmetries.

Now, what if we have a BF coupling $m \neq 0$ \emph{and} fundamental charged objects? For strings and electric charges, there is no issue, as the symmetries they break are already broken. On the other hand, monopoles and instantons naively present a contradiction, as we are trying to simultaneously gauge \emph{and} break the 3-form and magnetic 1-form global symmetries. In terms of currents, we have
\beq
j_3^{\rm m} = \frac{1}{2\pi} \rmd F_2 \propto \rmd ( \rmd \,{ \star H_3} ) = 0, \quad j_4^{\rm inst} = \frac{1}{2\pi}  \rmd H_3 \propto \rmd (\rmd\,{ \star F_2}) = 0,
\eeq
so indeed, adding monopoles or instantons alone would be inconsistent. For monopoles, this can be understood as a result of the Meissner effect: the action \eqref{SBF} describes a Type I superconductor, in which the presence of a nonzero magnetic flux destroys the phase. 

The resolution to this paradox lies in the fact that there may be additional conserved currents in the presence of additional sources. In particular, suppose that in addition to monopoles, we now add fundamental strings charged under $B_2$.\footnote{In fact, if $|m| \neq 1$, such strings may be necessary in the context of quantum gravity to break the $\mathbb{Z}_m$ global 2-form symmetry that would otherwise be preserved.} The equation of motion for $B_2$ then gives
\beq
\label{F+str}
\frac{1}{g^2} \rmd \, {\star H_3}=\frac{m}{2\pi} F_2+j_2^{\rm string}.
\eeq
Thus, while monopoles break the original magnetic 1-form symmetry with current $\JM_2 =  \frac{1}{2\pi} F_2$ as in \eqref{bfcharges}, the linear combination $(\JM_2)' = \frac{m}{2\pi} F_2 +j_2^{\rm string}$ of magnetic flux and fundamental string charge is gauged. Further, taking the exterior derivative gives
\beq
0 = \rmd (\rmd \,{\star H_3}) \propto \frac{m}{2\pi} \, \rmd F_2 + \rmd j_2^{\rm string} = m j_3^{\rm m} + \rmd j_2^{\rm string}.
\eeq
which tells us that monopoles must sit at the junction of $m$ fundamental strings. By placing $m$ strings on top of $-1$ units of magnetic flux, we may form a stringlike excitation that is uncharged under $B_2$, and so is deconfined. This is what happens in the intermediate phase of a Type II superconductor, where magnetic flux strings may pierce the material without destroying the phase entirely.\footnote{In such a superconductor, the role of ``fundamental string'' is played by the codimension-2 locus where the condensate field (a complex scalar) goes to zero. The fact that a magnetic monopole must be attached to $m$ such strings can be seen as follows. The $S^2$ surrounding the monopole carries a gauge bundle of Chern class one, while the complex scalar is a section of the associated line bundle of charge $m$, and so has exactly $m$ zeroes on $S^2$, corresponding to the locations of the $m$ strings.}

Similarly, the addition of both instantons and electrically-charged particles coupled to $A_1$ leads to the equation of motion \cite{BerasaluceGonzalez:2012vb},
\beq
\frac{1}{e^2} \rmd \, {\star F_2}= -\frac{m}{2\pi} H_3+j_3^{\rm e},
\eeq
and its exterior derivative,
\beq
0 = -m j_4^{\rm inst} + \rmd j_3^{\rm e}.
\eeq	
Thus, whereas the original current $J_3 = \frac{1}{2\pi}H_3$ is no longer conserved due to the instanton current in \eqref{bfcharges}, the improved current $(J_3)' = -\frac{m}{2\pi} H_3 +j_3^{\rm e}$ is gauged by its coupling to $A_1$. Further, we see that the fundamental instantons provide a tunneling process by which one unit of $H_3$ flux can decay into $m$ electric particles.

The distinction between the original currents $\JM_2$, $J_3$ and the improved currents $(\JM_2)'$, $J_3'$ was famously discussed by Marolf in \cite{Marolf:2000cb}. The conserved charge that couples to the electromagnetic field can be carried either by localized objects (brane sources) or by fluxes. In the language of that paper, $j_2^{\rm string}$ and $j_3^{\rm e}$ are currents of (non-conserved) ``brane source charge,'' whereas $(\JM_2)'$ and $J_3'$ are gauge invariant, conserved currents with quantized charge, which play the roles of both ``Page charge'' and ``Maxwell charge'' \cite{page:1984qv}. The distinction between Page charge and Maxwell charge collapses in cases like this one, with two-field Chern-Simons terms (BF terms), but not in cases with three-field Chern-Simons terms. We will encounter the more general case in Sections \ref{sec:F2} and \ref{sec:String}.

In this paper, we will consider both ways of breaking the symmetries, either by the presence of localized fundamental objects or non-localized solitonic field configurations, as both occur in string theory. Ultimately, the distinction between these two is not sharp in a UV-complete theory, since if one were able to take the zero-size limit of the flux tube carrying charge $\int F_2$, one would get a localized fundamental string. As the flux tube starts to shrink, at some point the energy density at its core will surpass the EFT cutoff, so the shrinking cannot be described purely within the low-energy EFT. This situation is common in string theory, where higher-form global symmetries are typically broken both by the presence of localized charged fundamental objects as well as Chern-Simons/BF couplings. We will elaborate on this point in Section \ref{sec:String}.

\subsection{Completeness Hypothesis and Abelian Higgs Model\label{sec:Higgsing}}

Standard lore holds that the Completeness Hypothesis follows from the absence of higher-form global symmetries, at least for abelian gauge groups.\footnote{For nonabelian gauge groups, this lore is false as stated; see \cite{Harlow:2018jwu,Rudelius:2020orz, completenesspaper} .} For the case of the BF theory considered in the previous section at $m = 0$, this lore is borne out by equation \eqref{bfcharges}, and indeed the breaking of all four symmetries of the theory implies the presence of dynamical instantons, electric particles, monopoles, and strings. However, for $m \neq 0$, we have seen that the story becomes more complicated. In this section, we discuss the implications of a nonzero BF coupling $m$ for these two Swampland conditions.

To begin, let us consider the electric symmetries of the BF theory in \eqref{SBF}. When $m=0$, there is an electric $U(1)$ 1-form symmetry with current $\star F$ and an electric 2-form symmetry with current $\star H_3$. For $m \neq 0$, each of these symmetries is explicitly broken to a $\mathbb{Z}_m$---the $A_1$ Wilson line of charge $m$ can end on an 't~Hooft local operator of $B_2$, and likewise the $B_2$ Wilson surface of charge $m$ can end on an 't~Hooft line of $A_1$. This reduces the symmetries to $\mathbb{Z}_m$ as discussed at the end of Section \ref{sec:bg}. Breaking these symmetries entirely further requires a complete spectrum of charged particles and charged strings, respectively: the one-to-one relationship between the Completeness Hypothesis and the absence of higher-form global symmetries appears to be intact.

The magnetic side of the story is very different, however. For any nonzero $m$, the instantons and monopoles are not only absent, but forbidden, since the symmetries they would break are gauged. Monopoles can exist if strings are added to the theory such that the current $F_2$ is broken while another linear combination involving $F_2$ is gauged, as explained around \eqref{F+str}. For $m > 1$, such strings are required to break the $\mathbb{Z}_m$ 2-form global symmetry discussed above. For $m=1$, however, there is no need to introduce fundamental strings to break any global symmetry. Hence, for $m=1$ we may construct a theory without global symmetries that lacks magnetic monopoles.

Is this a counterexample to the statement that the Completeness Hypothesis follows from the absence of higher-form global symmetries? The answer to this question depends on our definition of ``completeness.'' On the one hand, the BF theory at hand has a $U(1)$ gauge field without any magnetic charges, so naively it seems magnetically incomplete. On the other hand, such a strong notion of completeness is somewhat misguided, since this theory lacks not only monopoles, but also 't~Hooft line operators. In addition, the $U(1)$ gauge field in this theory has a mass given by $m_A=efm$, where $m$ is the BF coupling, $e$ is the electric gauge coupling of $A_1$ and $f$ is the decay constant of the axion; one might suspect that the relationship between completeness and the absence of higher-form global symmetries should apply only to massless gauge fields.

However, we should not give up so quickly on a stronger notion of completeness. To begin, we note that at energies well above $m_A$, the photon is approximately massless, so in the high-energy limit we expect to recover a massless 1-form gauge field $A_1$ and a massless 2-form gauge field $B_2$. To understand the physics at high energies, we must distinguish two possible UV completions: either we can consider $\phi$ to be a fundamental axion so that the BF theory \eqref{SBF} is valid at high energies up to the quantum gravity scale, or we can consider $\phi$ to be Nambu-Goldstone boson arising after Higgsing an abelian Higgs model.

In the abelian Higgs model, the BF theory description \eqref{SBF} breaks down at a cutoff given by the symmetry-breaking vev $v\sim f$, above which the theory enters in a Maxwell phase with a massless vector boson.\footnote{In the UV Maxwell phase, the 1-form global symmetry is spontaneously broken. This is in contrast to 0-form global symmetries, which are typically broken in the IR.} In this case, the charged strings involved in breaking the 2-form $\mathbb{Z}_m$ global symmetry are unavoidable, as they can be constructed as semiclassical solutions of the theory, with tension $T\sim f^2$. The vev of the Higgs field goes to zero at the core of the string, implying that the limit $f\rightarrow 0$ is at finite distance in field space. Such a string must be able to break in order to avoid an exact 1-form global symmetry, which in turn requires a complete spectrum of monopoles. As we saw in the scenario around \eqref{F+str} above, the current $\frac{m}{2\pi} F_2 + j_2^{\textrm{string}}$ is gauged, whereas the current $F_2$ is broken by the monopoles. Thus, avoiding higher-form global symmetries in the abelian Higgs model requires completeness of both the electric and magnetic spectra.

In contrast, in the theory with the fundamental axion, the gauge field acquires a Stueckelberg mass. There is no symmetry-restoring point at finite field distance, so $f\rightarrow 0$ is at infinite distance, and the core of fundamental strings charged under the $B$-field is singular. (This invariant distinction was suggested as the most useful way to distinguish between the Higgs and Stueckelberg mechanisms in \cite{Reece:2018zvv}.) For $m >1$, charged strings are required to break the 2-form symmetry and to satisfy the Completeness hypothesis, but for $m=1$ neither the absence of global symmetries nor the Completeness Hypothesis demand the existence of such strings. In the absence of such strings, monopoles are not allowed, either.

Nonetheless, string compactifications \emph{do} seem to produce a complete spectrum of strings and monopoles even if the gauge fields become massive due to Stueckelberg couplings. One way to understand this is by the emergence of \emph{approximate global symmetries} in the ultraviolet: at high energies in the Stueckelberg model, the BF coupling becomes negligible, and a 2-form global symmetry is restored. If, as suggested in \cite{Nomura:2019qps, COR}, we demand that approximate global symmetries (and not just exact global symmetries) must be forbidden at the Planck scale in a consistent theory of quantum gravity, we conclude that these strings must be present in the theory.\footnote{The absence of approximate global symmetries at the Planck scale in quantum gravity is also closely related to the Emergence Proposal \cite{Harlow:2015lma, Heidenreich:2018kpg, Grimm:2018ohb, Heidenreich:2017sim, Lee:2019xtm} and tower versions of the Weak Gravity Conjecture \cite{Heidenreich:2016aqi,Heidenreich:2015nta, Montero:2016tif,Andriolo:2018lvp}.} As in the abelian Higgs model, the presence of these strings leads to an exact 1-form global symmetry, whose breaking requires a complete spectrum of monopoles. Thus, whereas the strict absence of higher-form global symmetries leads to our stricter definition of completeness, the absence of approximate higher-form global symmetries implies our stronger definition, which requires monopoles even in the presence of a BF coupling.

We will discuss the proper definition of the Completeness Hypothesis, as well at its relationship to the absence of global symmetries in quantum gravity, in \cite{completenesspaper}.

\subsection{Axion Monodromy and ($-1$)-form Global Symmetries\label{sec:axionmonodromy}}

The above discussion applies to any $k=1$ Chern-Weil symmetry with current $J_p=F_p$  in any dimension $d > p$. It is interesting to consider the limiting case $d=p$, corresponding to a top-form field strength.
Much of the analysis carries over to this limiting case, and the current $J_d=F_d$ is associated with a ($-1$)-form global symmetry as discussed in Section \ref{ssec:CW}.
In this section, we will revisit the axion monodromy mechanism from this perspective and explain the gauging and breaking of ($-1$)-form global symmetries, relating this to the folk conjecture that there are no free parameters in string theory.

\begin{figure}
\begin{center}
\includegraphics[width=80mm]{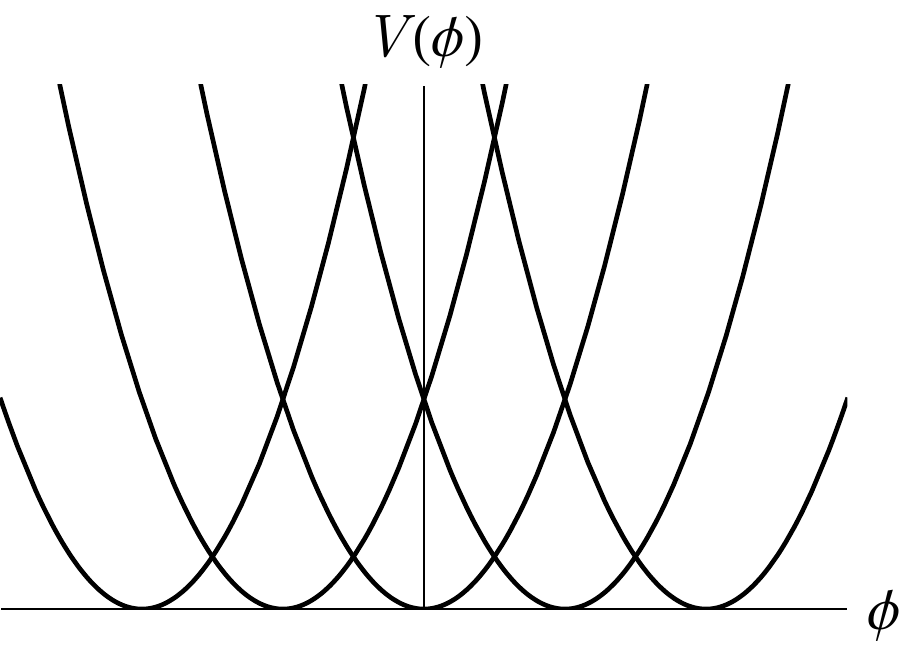}
\caption{The multi-branched potential of axion monodromy. Each branch of the potential is labeled by a distinct integer value of the discrete flux parameter $f_0$. The theory is invariant under the gauge transformation $\phi \rightarrow \phi +2\pi $, $f_0 \rightarrow f_0 + m$.}
\label{monodromypotential}
\end{center}
\end{figure}

Consider the following action in four dimensions:
\beq
\label{SKS}
S=\int \left(-\frac{1}{2t^2} F_4\wedge \star F_4-\frac{1}{2\ell^2} \rmd \phi\wedge \star \rmd \phi+\frac{m}{2\pi} \, \phi \wedge F_4\right),
\eeq
where $F_4=\rmd A_3$, $\phi$ is an axionic field, $t$ is a coupling constant with units of energy squared, $\ell$ is a coupling constant with units of inverse energy (or length), and $m \in \mathbb{Z}$. The gauge field $A_3$ has no propagating degrees of freedom in four dimensions, but it has the role of inducing a potential for the axion. The usual lore holds that one can now integrate out the 3-form gauge field via its equation of motion,
\beq
\label{f0}
\frac{1}{t^2} \rmd \, {\star F_4}= -\frac{m}{2\pi} \rmd \phi \quad \Rightarrow \quad \star F_4=t^2 \left(f_0-\frac{m}{2\pi}\phi\right),
\eeq 
where $f_0$ is an integration constant, to yield a description depending only on the axion,
\beq
\label{Saxion}
S=\int \left[-\frac{1}{2\ell^2} \rmd \phi\wedge \star \rmd \phi - t^2 \left(f_0-\frac{m}{2\pi}\phi\right)^2\right].
\eeq
Hence, the axion gets a multi-branched potential (see Figure \ref{monodromypotential}) that is invariant under the combined transformation
\beq
\phi\rightarrow \phi + 2\pi\ ,\ f_0\rightarrow f_0 + m  \ ,
\label{eq:shift}
\eeq
which is actually a gauge redundancy of the theory. This multi-branched potential has been extensively used in the context of axion monodromy to engineer large field ranges starting with a compact scalar field \cite{Silverstein:2008sg,McAllister:2008hb,Kaloper:2008fb,Kaloper:2011jz,Marchesano:2014mla}. The action \eqref{Saxion} is actually a dual description of the theory in \eqref{SKS} in which the 3-form gauge field is replaced by its ``dual ($-1$)-form gauge field'' with field strength $f_0$, and the compact scalar plus $f_0$ is equivalent to a non-compact scalar (see, e.g., \cite{Hebecker:2017wsu}). It has been argued \cite{Dvali:2005an} that any potential for an axion can be brought to the Kaloper-Sorbo-Dvali form in \eqref{SKS} if one allows a non-canonical kinetic term for $A_3$. In the presence of more axions and 3-form gauge fields, it is possible to generalize the couplings in \eqref{SKS} to describe any flux-induced scalar potential in string theory this way, including interactions as well as mass terms \cite{Bielleman:2015ina,Carta:2016ynn,Herraez:2018vae,Lanza:2019xxg}.

The branches of the potential labeled by distinct values of the parameter $f_0$ are gauge equivalent. Membranes electrically charged under $A_3$ can describe dynamical transitions among the branches, making manifest that $f_0$ is dynamical.
To avoid an exact 2-form global symmetry, these membranes must break open on strings charged under $B_2$, the 2-form gauge  field dual to the axion $\phi$. This is analogous to the situation studied in Section \ref{sec:BFgauging}, in which strings break open on monopoles, thereby violating the conservation of the current $j_2^{\rm string}$ in \eqref{F+str}. 

This theory is analogous to the BF theory in the previous section, with $\phi$ and $A_3$ playing the roles of $A_1$ and $B_2$ in the BF theory. Again, there are four global symmetries, two of which are broken and two of which are gauged by the coupling $\phi F_4$. The 3-form global symmetry with current $J_0=\frac{1}{t^2} \star F_4$ is broken, as can be seen in \eqref{f0}, while the 2-form symmetry with current $J_1=\frac{1}{2\pi} \rmd\phi$ is gauged. (In fact, it is possible to write yet another dual description of the theory involving the Stueckelberg coupling $\mathcal{L}\supset (\rmd B_2-mC_3)^2$. The 3-form gauge field then becomes massive by eating $B_2$, whose 2-form global symmetry becomes gauged.) Analogously, the 0-form continuous shift symmetry of the axion with current $J_3=\frac{1}{\ell^2} \star \rmd \phi$ is broken, while the ($-1$)-form symmetry with current $J_4=\frac{1}{2\pi}F_4$ is gauged. The conservation equation for the current $J_3=\frac{1}{\ell^2} \star \rmd \phi$ is just the equation of motion for $\phi$, so the current is not conserved in the presence of a potential, as
\beq
\frac{1}{\ell^2} \rmd(\star \rmd \phi)=\frac{m}{2\pi} F_4\ .
\eeq
 This equation also shows that the current $J_4=\frac{1}{2\pi} F_4$ is exact, indicating that the ($-1$)-form symmetry has been gauged, and the potential in \eqref{Saxion} is the analog of the Stueckelberg coupling for this ``($-1$)-form gauge field.''
Hence, axion monodromy corresponds to a concrete realization of gauging a ($-1$)-form global symmetry.

In the same way that $(-1)$-form global symmetries can be gauged, they can also be broken. As mentioned at the end of Section \ref{ssec:CW}, breaking a ($-1$)-form symmetry might appear meaningless, since the conservation of a top-form current is tautological, $\rmd F_4= 0$. However, in the context of a fundamental 3-form gauge field, we will see that there is a precise sense in which this symmetry may be broken. Consider for example the theory of a 3-form gauge field without an axion coupling $\phi \wedge F_4$. The ($-1$)-form symmetry gets broken in the presence of some additional constraint that supplements the action, e.g.,
\beq
\label{const}
f_0h=Q^{bg},
\eeq
where $f_0 = \frac{1}{t^2} \star F_4$. 
This type of constraint is very common in string compactifications, where it arises from tadpole conditions in higher dimensions. For instance, the background charge $Q^{bg}$ may arise from localized sources, while the parameter $h$ is an internal flux.

A constraint of this form is equivalent to gauging the 3-form global symmetry with current $f_0$, and in fact \eqref{const} may be viewed as a version of Gauss's Law, ensuring the cancelation of space-filling brane charge, similar to the cancelation of D9-brane charge in Type I string theory. To see the gauging more explicitly, we may incorporate the constraint in the action by means of a Lagrange multiplier $C_4$, as proposed in \cite{Lanza:2019xxg}. The action then takes the following form:
\beq
S= \int_M t^2 f_0f_0\star 1 + \int_M \left(Q^{bg}+\frac12 hf_0\right)C_4\ .
\eeq
This action is again analogous to the BF theory considered above, but now with $C_4$ and $f_0$ playing the roles of $B_2$ and $F_2$. Upon integrating out the 4-form potential $C_4$, one recovers the original action plus the constraint \eqref{const}. In \cite{Lanza:2019xxg}, it was shown that this action is dual to
\beq
S=\int_M \frac{1}{t^2} \tilde F_4\wedge \star \tilde F_4 + Q^{bg}\int_M C_4,
\eeq
where the $C_4$ couples to a space-time filling 3-brane and
 the new gauge-invariant field strengths are
$ \tilde F_4 = F_4 +h C_4$.
The theory is invariant under
\beqa
C_4\rightarrow C_4+\rmd \Lambda_3, \quad A_3\rightarrow A_3-h\Lambda_3,
\eeqa
so the 3-form global symmetry is gauged. Notice that the conservation equations for the currents read
\beqa
\label{C4}
 \rmd\, {\star \tilde F_4}=0,\quad  \rmd \tilde F_4= \rmd (hC_4),
\eeqa
referring to the 3-form and the ($-1$)-form symmetry respectively.

Taken at face value, \eqref{C4} should tell us that the $(-1)$-form symmetry with current $\tilde F_4 = t^2 \star f_0$ is broken in the presence of a tadpole constraint, just as the electric symmetry with current $\star F_2$ is broken in BF theory. However, we should be a bit more careful: as written, \eqref{C4} does not quite make sense, since it involves an equality of 5-forms in four dimensions. In order to make sense of \eqref{C4}, we are forced to consider five-dimensional manifolds. Adopting a construction familiar from the study of anomalies, given a four-manifold $M$, we may consider the cylinder $M\times [0, 1]$ (or more generally, a cobordism from $M$ to itself). Given two configurations of the theory that differ by the value of $\int_M \tilde F_4$ sitting at boundaries of the cylinder, we can ask if there exist field configurations in the interior of the cylinder that restrict to our given configurations on the boundary. If this is not possible, the symmetry is preserved; $\rmd F_4 = 0$ on the five-manifold $M \times [0, 1]$, and the charge $\int_M \tilde F_4$ is a topological invariant in the space of field configurations. Otherwise, the ($-1$)-form global symmetry is broken; there is some smooth deformation that changes the value of this charge.

We may thus understand \eqref{C4} as holding not in the spacetime $M$, but in the cylinder $M \times [0,1]$. Indeed, there are field configurations of $A_3, C_4$ on $M \times [0, 1]$ where the charge $\int_M \tilde F_4$ is different on the left and right boundaries; for example, we may take
\beq
A_3 = 0, \quad C_4 = t\ {\rm vol}_M,
\eeq
or any gauge conjugate, where $t$ is the coordinate on $[0,1]$. Thus, the $(-1)$-form symmetry with current $\tilde F_4$ is indeed broken. Once this symmetry is broken, it is no longer consistent to couple $\tilde F_4$ to an axion $\phi$, as a broken symmetry cannot be gauged. This is clear from the fact that $\tilde F_4$ is not integer valued, so a coupling $\phi \wedge \tilde F_4$ would violate the periodicity of $\phi$, just as a coupling $A_1 \wedge J_{d - 1}$ of an ordinary gauge field to a broken 0-form current, $\rmd J_{d - 1} \neq 0$, would violate the gauge symmetry of $A_1$. Thus, when this $(-1)$-form symmetry is broken, we should expect to find $\phi$ frozen to a particular value. That a tadpole constraint on $f_0$ freezes $\phi$ as well may be understood more directly: in the presence of an axion, we have the gauge redundancy \eqref{eq:shift}, and so it is only meaningful to set the gauge-invariant combination $f_0 + m \phi$ to a particular value.

This breaking of a $(-1)$-form global symmetry was interpreted in \cite{Lanza:2019xxg,Lanza:2020qmt} as an obstruction to dualizing the internal flux $f_0$ to a dynamical 3-form gauge field from the EFT perspective.\footnote{Fluxes associated to $(-1)$-form global symmetries that are gauged or broken correspond, in the terminology of  \cite{Lanza:2019xxg,Lanza:2020qmt}, to \emph{dynamical EFT fluxes} or fluxes which are \emph{fixed} from the EFT perspective, respectively.}. Clearly, we can always dualize the fluxes to 3-form gauge fields, but the obstruction refers to the impossibility of describing \emph{within the EFT} a dynamical transition that shifts the value of the flux by crossing a membrane charged under $C_3$. Since the 3-form global symmetry is gauged, these membranes should be attached to space-time filling 3-branes with a characteristic energy scale above the EFT cut-off; notice that $Q^{bg}$ is an input of the EFT. Such a scenario, by which a space-filling brane ends on a dynamical membrane, is analogous to a charged string ending on a monopole as discussed above in BF theory.\footnote{The presence of such a transition can be understood from the perspective of \cite{McNamara:2019rup}, which predicts that any two consistent theories of quantum gravity should be connected by a dynamical domain wall.}

In Section \ref{sec:String} we will describe other string theory examples in which ($-1$)-form global symmetries are broken. 
 Let us note that, even if they are broken, it still makes sense to talk about these symmetries if there is some limit in which they become \emph{approximate}. For instance, the dynamical process that breaks them might force us beyond the validity of the low energy EFT by involving energies above the cutoff, in which case a low-energy observer may see an approximate ($-1$)-form global symmetry.

To sum up, parameters in an EFT are associated to ($-1$)-form global symmetries. In the same way that $p$-form global symmetries correspond to charges defined on $(d - p - 1)$-dimensional slices of spacetime, the case of ($-1$)-form global symmetries corresponds to charges carried by the entire spacetime. The ``background gauge field'' coupling to a global ($-1$)-form symmetry are the \emph{labels} of the vacua of a family of theories, such as $\theta$-vacua or Coleman's $\alpha$-eigenstates \cite{McNamara:2020uza}. If this parameter is actually dynamical, such as in the case of axion monodromy, then the top-form current is exact and the symmetry is gauged; if there is a path in configuration space that allows us to deform the value of the charge, then the symmetry is broken, and the parameter must be set to a particular value. This occurs, for instance, in the case of a tadpole constraint. Hence, the absence of free parameters in quantum gravity is closely related to the absence of ($-1$)-form global symmetries, as also pointed out in \cite{McNamara:2020uza}. It is tempting to posit that the absence of these symmetries might be related to the uniqueness of a single quantum gravity theory \cite{McNamara:2019rup} and thus to the notion of string universality (see \cite{Kumar:2009us, Adams:2010zy, Belin:2014fna, Garcia-Etxebarria:2017crf, Kim:2019vuc, Kim:2019ths, Katz:2020ewz, Montero:2020icj, Cvetic:2020kuw} for works in this direction).

\section{$\tr F^2$ Chern-Weil Currents}\label{sec:F2}

We now turn our attention to Chern-Weil currents of the form $\tr F_2^2:= \tr (F_2 \wedge F_2)$. In contrast with the $F$ Chern-Weil currents considered in the previous subsection, these currents exist in nonabelian gauge theories as well as abelian gauge theories, since every simple Lie algebra has a nontrivial quadratic Casimir. Such a current generates a $(d-5)$-form $U(1)$ global symmetry in $d \geq 5$ dimensions. In quantum gravity, such a symmetry must be either gauged or broken, which means that the current should either be exact or else not conserved, respectively. In this section, we will consider examples of each of these.

\subsection{Gauging}\label{ssec:CWgauging}

As in the case of the $J_2=F_2$ Chern-Weil current considered in the previous section, the $J_4=\tr F_2^2$ Chern-Weil current may be gauged by coupling the current to a $(d-4)$-form $U(1)$ gauge field $C_{d-4}$:
\begin{equation}
S = \int \left(-\frac{1}{2 g_{A}^2} \tr( F_{2} \wedge \star F_{2})  -\frac{1}{2 g_{C}^2} H_{d-3} \wedge \star H_{d-3} +  \frac{1}{8\pi^2} C_{d-4} \wedge  \tr (F_2 \wedge F_2)   \right) ,
\label{eq:Wittentheory}
\end{equation}
where $H_{d-3} = \rmd C_{d-4}$.\footnote{We have written the normalization of the Chern-Weil current appropriate for an $SU(N$) group with trace taken in the fundamental representation and the Dynkin index of the fundamental normalized to $1/2$; constant factors may vary, more generally.} In the presence of this cubic Chern-Simons coupling, the equation of motion for $C_{d-4}$ is
\begin{equation}
\frac{8\pi^2}{g_C^2}\, \rmd\,{\star H_{d-3}}= \tr (F_2 \wedge F_2) = J_4.
\end{equation}
The Chern-Weil current is now exact: $J_4 = \rmd(\cdots)$, indicating that the symmetry has indeed been gauged. Note that in 4d this is a ($-1$)-form symmetry, and the gauge field $C_{d-4}$ is an axion, as in the axion monodromy scenario studied in Section \ref{sec:axionmonodromy}.

Let us now specialize to the case in which $F_2$ is the field strength for a $U(1)$ gauge symmetry.
In this case, the cubic Chern-Simons coupling also breaks the electric 1-form symmetry, as
\begin{equation}
\frac{1}{g_A^2}\, \rmd \,{\star F_2} = - \frac{1}{4\pi^2} \rmd (C_{d-4} \wedge F_2),
\end{equation}
so $\JE_{d-2} = \frac{1}{g_A^2} {\star F_2}$ is no longer conserved. One might attempt to define an improved current 
\begin{equation}
\tilde J^{\textrm{E}}_{d-2} = \JE_{d-2} + \frac{1}{4\pi^2} C_{d-4} \wedge F_2,
\end{equation}
but this $(d-2)$-form is not invariant under gauge transformations of $C_{d-4}$, so it is not a genuine operator in the theory.

Let us further specialize to the case of $d=4$. Here, the breaking of the electric 1-form symmetry can be understood via the action of a large gauge transformation $\theta \rightarrow \theta + 2 \pi $ on the line operators in the theory. Recall from Section \ref{sec:HigherForm} that pure $U(1)$ gauge theory has Wilson/'t~Hooft lines of charge $(n, m)$, where $n$ is the electric charge and $m$ is the magnetic charge. In the presence of the $\theta F_2 \wedge F_2$ coupling, 
however, a line operator of charge $(n, m)$ transforms under $\theta \rightarrow \theta + 2 \pi$ into a line of charge $(n+m, m)$. This phenomenon is known as the Witten effect \cite{Witten:1979ey} and will be discussed further in Section \ref{ssec:gaugeandbreak} below. The Witten effect implies that the electric charge of a line operator is not a gauge-invariant quantity, and correspondingly the electric 1-form symmetry is not a valid symmetry of the theory.

This is our first example of a more general phenomenon that we will study in detail below: under $\theta \rightarrow \theta + 2 \pi$, an electric charge is ``dissolved'' into an 't~Hooft line of the theory.  More generally, in $d \geq 4$, a large gauge transformation of $C_{d-4}$ dissolves an electric charge into the 't~Hooft surface of dimension $d-3$. In Section \ref{ssec:gaugeandbreak}, we will study the analog of this phenomenon for dynamical objects as opposed to the defects considered here, in which electric charge is dissolved in a magnetic monopole due to additional degrees of freedom on the monopole worldline. In Section \ref{sec:String}, we will see higher-form generalizations of this in string theory, in which lower-dimensional branes are dissolved in higher-dimensional ones.

Finally, let us briefly consider the theory of \eqref{eq:Wittentheory} in the presence of external electric sources. The equation of motion may then be written either as
$\rmd \JE_{d-2} =  j_{d-1}^{\text{Maxwell}}$
or as $
\rmd \tilde J^{\textrm{E}}_{d-2} =  j_{d-1}^{\text{Page}}$, where $ j_{d-1}^{\text{Maxwell}}$ is the Maxwell current and $j_{d-1}^{\text{Page}}$ is the Page current, as discussed in \cite{Marolf:2000cb}. These two notions of charge are distinguished in the presence of the three-field Chern-Simons term: Maxwell charge is conserved and gauge invariant but not localized, whereas Page charge is conserved and localized but not gauge invariant.

\subsection{Breaking}\label{ssec:breaking}
Breaking of quadratic Chern-Weil currents takes different forms depending on whether the underlying field strength is abelian or not. In all higher-form cases, we will only need to deal with the abelian case, which we illustrate with a $U(1)$ gauge field $d$ dimensions. Such a theory may have monopole $(d-4)$-branes, with current $j_3^{\textrm{m}}$.  These give a nonvanishing contribution to the Bianchi identity,
\begin{equation} \frac{1}{2\pi} \rmd F= j_3^{\textrm{m}}.\end{equation}
As a result, they also break the abelian Chern-Weil current:\footnote{In fact, monopoles only break the $U(1)$ Chern-Weil symmetry to a $\mathbb{Z}_2$ subgroup, as discussed in Section \ref{sec:rigid}.}
\begin{equation} \rmd (F\wedge F)= 2 F\wedge \rmd F = 4\pi F \wedge   j_3^{\textrm{m}} \neq0. \end{equation}
Just as the linear combination of currents $\frac{m}{2\pi} F_2+j_2^{\rm string}$ remains conserved in \eqref{F+str}, however, some linear combination of currents involving $F \wedge F$ may remain conserved even in the presence of monopoles. We will see examples of this in the following subsection.

The nonabelian case is more complicated. Let us consider a nonabelian semisimple group $G$ with connection $A$. In that case, the Chern-Weil current $\tr(F \wedge F)$ cannot be broken by monopoles in the same way as above. Indeed, if $G$ is simply connected, there are no topologically nontrivial monopoles whatsoever. In general, the topological classes of monopoles correspond to nontrivial bundles on the $S^2$ transverse to the monopole worldvolume, and are classified by elements in $\pi_1(G)$. Suppose such a monopole led to a non-conservation of Chern-Weil charge, say by $n$ units. Since $\pi_1(G)$ is torsion for any compact semisimple Lie group, by putting together a finite number of such monopoles, we could violate Chern-Weil charge conservation by a multiple of $n$ units with a topologically trivial field configuration, which is impossible.

If the Chern-Weil current is not broken by monopoles, how can it be broken? Consider the case of a nonabelian gauge group $G$ which is spontaneously broken to a product including semisimple factors. The usual GUT breaking 
\begin{equation} SU(5)\rightarrow \frac{SU(3)\times SU(2)\times U(1)}{\mathbb{Z}_6}\end{equation}
would be one such example, though there are many more. In the UV, there is a single Chern-Weil current, $\tr(F^{SU(5)} \wedge F^{SU(5)})$. However, after spontaneous symmetry breaking, the IR gauge group has three Chern-Weil currents,
\begin{equation} 
\tr(F^{SU(3)} \wedge F^{SU(3)}),~~~\tr(F^{SU(2)} \wedge F^{SU(2)}), ~~~F^{U(1)} \wedge F^{U(1)}.
\label{GUTcurrents}
\end{equation}
The abelian Chern-Weil current is explicitly broken by monopoles, as discussed above, but at least one of the two nonabelian Chern-Weil currents is an emergent symmetry in the IR, which should be broken once UV effects are taken into account. Indeed, an $SU(3)$ instanton can be transformed into an $SU(2)$ instanton by shrinking it to be smaller than the symmetry breaking length scale, rotating in $SU(5)$ appropriately, and re-expanding. Since the charge associated with the current $\tr(F^{SU(3)} \wedge F^{SU(3)})$ decreases in this process while the charge associated with $\tr(F^{SU(2)} \wedge F^{SU(2)})$ increases, the Chern-Weil current
\begin{equation}  \tr(F^{SU(3)} \wedge F^{SU(3)})- \tr(F^{SU(2)} \wedge F^{SU(2)})\end{equation}
is not conserved. 

This represents a special case of a more general phenomenon: IR Chern-Weil symmetries may be broken in the UV by un-Higgsing to a larger gauge group. To see why, consider the spontaneous symmetry breaking of a gauge theory with gauge group $G$ down to a subgroup, $H$.
The $G$ gauge connection splits as
\begin{equation}A^G= A^H+ A^\perp,\end{equation}
where $A^H$ is defined by the condition that it leaves the Higgs vev $\Phi$ invariant,
\begin{equation} (A^H)^a(\mathbf{t}_a)\Phi=0,\end{equation}
where the matrices $\{\mathbf{t}_a\}$ are those of whichever representation the field $\Phi$ lives in. One can define a projector
\begin{equation} \pi_{ab}\equiv \Phi^* (\mathbf{t}^*_a \mathbf{t}_b)\Phi,\end{equation}
so that $\pi_{ab} (A^H)^b=0$, which satisfies 
\begin{equation} \pi_{ab} \pi^{bc}= C_2 \vert \Phi\vert^2 \pi_{a}^c,\end{equation}
where $C_2$ is the eigenvalue of the quadratic Casimir in the Higgs representation, and indices are raised/lowered using the Killing form. As a result, one can write
\begin{equation} (A^H)_a= \left(\delta_{ab}- \frac{\pi_{ab}}{\sqrt{ C_2 \vert \Phi\vert^2}} \right) (A^G)^b\equiv \Pi_{ab}(\Phi) (A^G)^b, \label{higgsop}\end{equation}
which gives us the connection in the unbroken subgroup in terms of the connection on $G$ and the Higgs vev. From this, one may write the field strength and the Chern-Weil current of $A^H$ in terms of those of $A^G$ and Higgs field insertions. The derivative of the Chern-Weil current is in general nonvanishing, due to the terms coming from $\Pi_{ab}(\Phi)$ and its derivatives. In this way, the embedding into the nonabelian gauge group breaks the symmetry explicitly, due to the additional insertions of Higgs operators in the UV definition of the IR Chern-Weil current.

In 4d, the Chern-Weil symmetries broken by this unification process are $(-1)$-form symmetries. Any such broken symmetry cannot be consistently coupled to an axion. In the case of $SU(5)$ GUT breaking, only one global symmetry remains, so only one axion can be coupled to the three Chern-Weil currents in the IR.

\subsection{Gauging and Breaking}\label{ssec:gaugeandbreak}

In many examples, the symmetry associated with a Chern-Weil current is explicitly broken, but a linear combination of this current and another current is gauged. An example of this phenomenon arises in $SU(2)$ gauge theory Higgsed to $U(1)$ by an adjoint scalar field $\Phi$. This theory has semiclassical 't~Hooft-Polyakov monopoles, magnetically charged under the unbroken $U(1)$. These monopoles and their collective coordinates, which play a key role in our discussion below, are reviewed in the textbooks \cite{Shifman:2012zz, Weinberg:2012pjx}. 

In this theory, we can formulate a puzzle. Consider the $SU(2)$ gauge theory in the un-Higgsed phase. We may gauge the Chern-Weil current, as discussed above, by adding a $(d-4)$-form $U(1)$ gauge field $C$ with a minimal Chern-Simons coupling to $\tr(F \wedge F)$.
After Higgsing to $U(1)$, the IR theory inherits a coupling of $C$ to the $U(1)$ Chern-Weil current $F \wedge F$:
\begin{equation}
\text{UV: } \frac{1}{8\pi^2} C \wedge \tr(F \wedge F) \quad \Rightarrow \quad \text{IR: } \frac{1}{4\pi^2} C \wedge F \wedge F.
\end{equation}
Because we will keep track of constant prefactors throughout this subsection, let us be explicit about our conventions. In the UV theory, we take the trace in the fundamental representation of $SU(2)$, with the convention $F|_{\rm UV} = F^i T^i$ and $\tr(T^i T^j) = \frac{1}{2} \delta^{ij}$. In the IR theory, we take $F$ in the conventional normalization for a $U(1)$ gauge theory with a minimal charge normalized to 1. Taking $T^3$ to be the unbroken generator of $SU(2)$, the doublet representation carries charges $\pm 1/2$ under $T^3$, so our normalization is that $F|_{\rm IR} = \frac{1}{2} F^3$. This means that in the IR theory, the coupling of $C$ to the Chern-Weil current is {\em twice} the minimal allowed value.\footnote{We could, alternatively, work with $SO(3)$ gauge theory, so that the minimal charge in the UV and the IR is the same. This would  change the quantization of the Chern-Weil charge (with familiar consequences, e.g., altered periodicity of the $\theta$ term \cite{Aharony:2013hda}), and eventually lead to the same  conclusions.}

The existence of the 't~Hooft-Polyakov monopole means that the IR Chern-Weil current is not conserved:
\begin{equation}\label{breaking_by_2}
\rmd(F \wedge F) = 4\pi F \wedge j_{3}^{\textrm{m}} \neq 0.
\end{equation}
This raises a puzzle: gauging the Chern-Weil current in the UV is consistent, because it is conserved. This leads to a Chern-Simons coupling $C \wedge F \wedge F$ in the IR, which seems to gauge the IR Chern-Weil current. Yet such a gauging appears inconsistent, because the IR Chern-Weil current $F \wedge F$ is broken by the existence of monopoles. How do we resolve this apparent inconsistency?

We encountered a similar puzzle in Section \ref{sec:BFgauging} when looking at $k=1$ Chern-Weil currents. There, the resolution lay in the fact that while the Chern-Weil current $F$ itself was broken, a linear combination $(\JM_2)' = \frac{1}{2\pi} F+j_2^{\textrm{string}}$ was gauged, with $j_2^{\textrm{string}}$ the string charge current. A similar resolution applies to the puzzle at hand, so that $F \wedge F$ is broken, whereas a linear combination $J_4' := F \wedge F + \omega_4$ is gauged. In this case, however, there is not an obvious candidate for the 4-form $\omega_4$, since there is no 4-form gauge current in the bulk theory.

It turns out that a description of the gauged current $J_4'$ requires a detailed understanding of the degrees of freedom on the monopole worldvolume. The 't~Hooft-Polyakov monopole solution has a set of collective coordinates (also known as zero modes or moduli). The most obvious are the translational zero modes, which allow for the motion of the monopole worldvolume in spacetime. A less obvious zero mode is the {\em dyon collective coordinate} \cite{Jackiw:1975ep, Christ:1976cg,Jackiw:1977yn}, which allows a magnetic monopole to acquire electric charge  and become a dyon \cite{Julia:1975ff}. The dyon collective coordinate is a compact scalar field $\sigma$ localized on the monopole worldvolume. The corresponding deformation of the classical monopole solution corresponds to a rotation around the direction of $\Phi$ in field space, and has the form of a gauge transformation that does not vanish at infinity. 

For concreteness, we will review the physics of the dyon collective coordinate in the well-studied case of $d = 4$, then comment on the extension to general $d$ below. In 4d, the effective theory on the monopole worldline is just a theory of quantum mechanics, so $\sigma$ is a quantum-mechanical particle on a circle. The spectrum of this theory is labeled by the integer momentum of $\sigma$ around the circle; in this context, this integer corresponds to (one-half) the electric charge of the dyon. Furthermore, the $\theta$-term in the bulk induces a 1d $\theta$-term on the monopole worldline:
\begin{equation}
{\cal L}_{\rm M} = \frac{1}{2} \frac{m_{\rm M}}{m_W^2} \rmd_A \sigma \wedge \star \rmd_A \sigma + \frac{\theta}{2\pi} \rmd_A \sigma,
  \label{eq:monopoleEQM}
\end{equation}
where $m_{\rm M}$ and $m_W$ are the monopole and $W$ boson masses respectively and  $\rmd_A \sigma \equiv \rmd \sigma + 2A$ is the covariant derivative. The $\theta$-term shifts the spectrum of the quantum particle on a circle in precisely the correct way to implement the Witten effect \cite{Witten:1979ey} discussed in Section \ref{ssec:CWgauging}. In particular, it leads to a spectrum with magnetic and electric charges
\begin{equation}
\label{eqref}
q_\mathrm{M} = 1, \quad q_\mathrm{E} = \frac{\theta}{\pi} + n_\mathrm{E},\text{ where }n_\mathrm{E} \in 2\mathbb{Z}. 
\end{equation}
A shift of $\theta \to \theta + 2\pi$ leaves the spectrum invariant, but with a monodromy under which each state shifts to a state with two additional units of electric charge. (As a reminder, the reason that $n_\mathrm{E}$ is even is that we use a  normalization in which $SU(2)$ fundamentals, which we have not incorporated into this theory, carry one unit of $U(1)$ electric charge.)

An alternative perspective on the localized scalar field $\sigma$, valid when $d \geq 6$, is that it is a Nambu-Goldstone boson. The bulk $U(1)$ gauge symmetry may be thought of as effectively acting as a $U(1)$ {\em global} symmetry of the monopole worldvolume theory, where $A$ restricted to the worldvolume behaves as a background connection, in much the same way that bulk gauge symmetries restrict to boundary global symmetries in AdS/CFT. From the worldvolume viewpoint, the $U(1)$ global symmetry is spontaneously broken: The global part of  bulk $U(1)$ gauge transformations does not leave the monopole worldvolume vacuum invariant, and $\sigma$ is the Nambu-Goldstone boson that parametrizes these degenerate states.\footnote{In the cases with $d = 4$ or $d = 5$, the worldvolume theory does contain the massless field $\sigma$ parametrizing different {\em classical} monopole solutions, but the terminology ``Nambu-Goldstone boson'' is inappropriate for $\sigma$ due to the Coleman-Mermin-Wagner theorem: the quantum theory on the worldvolume does not have degenerate vacua, but a single vacuum sampling all values of $\sigma$ uniformly.} The Nambu-Goldstone perspective will  be useful in several later examples, where we will encounter spontaneously broken $(p-1)$-form global symmetries of worldvolume theories, in which the gauge transformation of a $p$-form gauge field $B_p$ in the bulk induces a $(p-1)$-form Nambu-Goldstone zero mode $A_{p-1}$ on the worldvolume that appears in the gauge-invariant Stueckelberg combination $\rmd_B A_{p-1} = \rmd A_{p-1} + B_p$ \cite{Gaiotto:2014kfa, Lake:2018dqm}.

Returning to the case of the 't Hooft-Polyakov monopole in a general dimension $d > 4$, there is still a compact boson $\sigma$ in the worldvolume theory and an action of the form \eqref{eq:monopoleEQM}, with $\theta$ replaced by the bulk $(d-4)$-form gauge field $C$. Though the worldvolume is no longer 1-dimensional, the underlying physics remains similar, with $\sigma$ allowing us to dissolve electric charge inside the monopole worldvolume. Now, we are equipped to resolve our puzzle. In the IR theory below the Higgsing scale, the gauge field $C$ couples to a linear combination of bulk and localized terms,
\begin{equation}
C \wedge \left[\frac{1}{4\pi^2} F \wedge F + \frac{1}{2\pi} j_{3}^{\textrm{m}} \wedge \rmd_A \sigma\right]:= C \wedge J_{4}',
\end{equation}
where one can view the magnetic current $j_{3}^{\textrm{m}}$ as the delta functions necessary to localize the coupling to $\sigma$ on the monopole worldvolume. The current $J_4'$ is conserved, since $\rmd (\rmd_A \sigma) = 2F$ on the worldvolume due to the Stueckelberg structure of the $\sigma$ Lagrangian, and in fact it is exact due to the coupling to $C$.

In the IR theory, then, the Chern-Weil current $F \wedge F$ is explicitly broken by the presence of magnetic monopoles, but the {\em sum} of this current and a current localized on the monopole worldvolume is conserved. If we view $\rmd_A \sigma$ as a 1d Chern-Weil current, we see that $C$ gauges the {\em diagonal} symmetry generated by the bulk and localized Chern-Weil currents. This resolves our puzzle. In fact, if we had not known about the dyon collective coordinate and the Witten effect before, we would have been led to discover them by thinking about how the nonabelian Chern-Weil current in the UV can be consistently gauged. We will see a number of examples below with a similar flavor, in which consistent gauging of Chern-Weil symmetries {\em requires} the existence of localized fields, which allow for the dissolution of branes within branes of a higher dimension.

From the point of view of the Swampland program, one might have dismissed Chern-Weil currents built from abelian gauge fields as uninteresting, because they are explicitly broken by the existence of magnetic monopoles. This example shows us that one should not be so hasty. We will see that there are many more examples that share the feature that localized fields lead to the conservation of a diagonal symmetry built from multiple Chern-Weil currents. Quantum gravity requires that such conserved diagonal symmetries are gauged.

\section{String/M-theory}\label{sec:String}

In this section, we will discuss how Chern-Weil symmetries are broken or gauged in several string theory setups. We will see how a variety of known string theory phenomena can actually be derived by requiring the absence of these symmetries, which suggests that they might be general features of quantum gravity rather than accidents of the string lamppost. We will also identify some universal patterns that will be useful later on when discussing the phenomenological implications.

\subsection{Kaluza-Klein and Winding Gauge Theories} 

Kaluza-Klein theory on a circle provides a simple example of an abelian gauge theory in a gravitational context. Consider a gravitational theory in $D = d+1$ dimensions, with the metric ansatz
\begin{equation}
\rmd s_D^2 = \mathrm{e}^{\lambda(x)/(d-2)} \rmd s_d^2(x) + \mathrm{e}^{-\lambda(x)} R^2 \left(\frac{\rmd y}{R} + A_1\right)^2,
\end{equation}
where $y \cong y + 2\pi R$ parametrizes the compact dimension. The Kaluza-Klein gauge field is $A_1$, with field strength $F_2 = \rmd A_1$ (locally). The scalar $\lambda(x)$ is the radion field.

In Kaluza-Klein theory, the field strength $F_2$ is not a conserved current due to the existence of the KK monopole, which is a smooth solution to the $D$-dimensional Einstein's equations \cite{sorkin:1983ns, gross:1983hb}. Unlike the 't~Hooft-Polyakov monopole discussed above, the KK monopole has no dyonic zero mode that would allow it to carry electric Kaluza-Klein charge. Recall that for the 't~Hooft-Polyakov monopole, this zero mode arose from a $U(1)$ gauge transformation that did not vanish at infinity. The KK monopole solution is translationally invariant along the compactified dimension, so the analogous operation acts trivially on the solution and does not correspond to a zero mode. As a result, in a pure gravity theory, the Chern-Weil current $F_2 \wedge F_2$ is not conserved as a simple consequence of $\rmd F_2 = j_3^{\textrm{m}}$, and there is no worldvolume degree of freedom to enliven the story.

The abelian Kaluza-Klein gauge theory can be realized within many different consistent theories of quantum gravity, and in specific examples, there is often more to say. For example, a $D$-dimensional theory that includes a gauge field $B^{(D)}_2$ coupled to string charge, when compactified on a circle, gives rise to a winding gauge field $B_1$. In such a theory, we can investigate the fate of various Chern-Weil currents.

The 2-form gauge field in the compactified theory has a modified Bianchi identity, which we will now briefly review for clarity.\footnote{This well-known example is discussed, for instance, in \S8.1 of \cite{Polchinski:1998rq}.} The winding field in $D$ dimensions (labeled with superscript $(D)$) and those in $d$ dimensions (unmarked) are related via the ansatz
\begin{equation}
B_2^{(D)} = B_2 + \frac{1}{2\pi} B_1 \wedge \left(\frac{\rmd y}{R} + A_1\right).
\end{equation}
The winding gauge field has field strength $H_2 = \rmd B_1$ (locally). However, $B_2^{(D)}$ is invariant under a gauge transformation of $A_1$, which implies that $B_2$ is not:
\begin{equation}
A_1 \mapsto A_1 + \rmd \xi, \quad B_2 \mapsto B_2 - \frac{1}{2\pi} B_1 \wedge \rmd \xi.
\end{equation}
As a result, the field strength $\rmd B_2$ is not gauge invariant. Rather, the gauge-invariant 3-form field strength is 
\begin{equation}
\widetilde{H}_3 = \rmd B_2 + \frac{1}{2\pi} H_2 \wedge A_1.
\end{equation}
The corresponding kinetic term in the compactified theory has the form $\int \widetilde{H}_3 \wedge \star \widetilde{H}_3$, and the field strength obeys a modified Bianchi identity
\begin{equation}
\rmd \widetilde{H}_3 = \frac{1}{2\pi} H_2 \wedge F_2.
   \label{eq:H3bianchi}
\end{equation}
The identity \eqref{eq:H3bianchi} indicates that the 4-form Chern-Weil current $H_2 \wedge F_2$ is exact, or equivalently, that the corresponding symmetry is gauged. This consequence of modified Bianchi identities will play an important role in our discussion below.

If the Chern-Weil current $H_2 \wedge F_2$ is gauged, what is the corresponding gauge field? The kinetic term contains terms of the form $A_1 \wedge H_2 \wedge \star H_3$. We can apply Hodge duality, $\star H_3 = {\rmd} B_{d-4}$, and integrate by parts to view this as a term $B_{d-4} \wedge (F_2 \wedge H_2)$. In other words, the field that gauges the mixed Chern-Weil current is the magnetic dual of the string gauge field $B_2$.

The observation that the Chern-Weil current $H_2 \wedge F_2$ is gauged leads to a variation on the puzzle we encountered in Section~\ref{ssec:gaugeandbreak}. We have already noted that $\rmd F_2 \neq 0$, due to the existence of KK monopoles. Similarly, we have $\rmd H_2 \neq 0$ due to the existence of objects carrying magnetic charge under $B_1$, referred to as ``H-monopoles'' \cite{Rohm:1985jv, Banks:1988rj}. Such charges are carried by NS5-branes for the 9d circle compactification of a 10d superstring theory, or NS21-branes for the 25d circle compactification of 26d bosonic string theory. The existence of independent magnetic monopoles for Kaluza-Klein and winding charge implies that $H_2 \wedge F_2$ is not even a conserved current; {\em a fortiori}, it cannot be exact. We seem to have derived a contradiction.

As in the 't~Hooft-Polyakov discussion above, the resolution to this puzzle lies in interactions that are localized on monopole worldvolumes, which we neglected in writing the modified Bianchi identity \eqref{eq:H3bianchi}. It is intuitively clear that the H-monopole should be able to carry electric Kaluza-Klein (momentum) charge: it is not wound around the compact dimension, so we should be able to boost it along that dimension. Indeed, it is known that the appropriate dyonic collective coordinates exist, so that H-monopoles can carry electric Kaluza-Klein charge, whereas KK monopoles can carry electric winding charge \cite{Sen:1997zb, Gregory:1997te}. Specializing to the 9d compactification of the Type II superstring for illustration, the full gauging via $B_5$ takes the schematic form
\begin{equation}
B_5 \wedge \left(F_2 \wedge H_2 +  j^{\textrm{e}}_3 \wedge \rmd_{B_1}\sigma^{\textrm{e}} + j^{\textrm{m}}_3 \wedge \rmd_{A_1}\sigma^{\textrm{m}}  + j_4^\textrm{NS5-w}\right),
\end{equation}
where $j_4^\textrm{NS5-w}$ denotes the current for NS5-branes wrapped on the internal dimension. In order to have no global symmetries, it must be the case that other linear combinations of the currents appearing here are broken. That is, the charges associated with each of the terms in parentheses may be dynamically converted into charges under the other terms. We will not elaborate on the details relating to the NS5 charge here, but will discuss similar phenomena in later sections.

We can gain further insight by considering circle compactifications at the self $T$-dual radius. In this case, there is an enhanced $SU(2)$ gauge symmetry (in the heterotic context) or $SU(2) \times SU(2)$ symmetry (in the bosonic string context). There are Chern-Weil currents of the form $\tr(F \wedge F)$, built out of the nonabelian gauge fields. Moving slightly away from the self-dual point in moduli space, we recover the abelian theory through Higgsing, with an adjoint Higgs (in the heterotic case) or a bi-adjoint Higgs (in the bosonic string case). Thus, we see that these examples are intimately related to the 't~Hooft-Polyakov case previously described. 

A rather different case arises for the compactification of 11d M-theory to 10d, which for small circle radius becomes Type IIA string theory, where the Kaluza-Klein photon is identified with the Ramond-Ramond gauge field $C_1$. In this context, the KK monopole is a D6-brane. Unlike the previous case we considered, there is no 1-form winding gauge field, and no scalar zero mode on the KK monopole worldvolume. However, if we consider gauge fields of different degree, there is analogous physics: the 11d 3-form $C_3$ gives rise to both a 10d 3-form with a modified Bianchi identity and a 2-form $B_2$ (the string charge) in 10d. The modified Bianchi identity ensures that the Chern-Weil current $H_3 \wedge F_2$ is gauged. Due to the change in degrees of the forms, the analogue of the compact scalar $\sigma$ above is now a 1-form gauge field $A_1$ that must exist on the D6-brane worldvolume and shift under a $B_2$ gauge transformation. Thus, the worldvolume effective action of the D6-brane can be deduced by studying properties of zero modes of the Kaluza-Klein monopole in the compactified 11d theory \cite{Imamura:1997ss}. With this analogy to the cases discussed above to guide us, we will now turn our attention to a more general discussion of the role of Chern-Weil currents in superstring theory.

\subsection{Heterotic}\label{sec:heterotic} 

 Chern-Weil symmetries arise naturally in heterotic string theory. The 10d massless heterotic bosonic fields include nonabelian gauge fields, a $B$-field, and the metric. We can construct Chern-Weil currents out of these from the characteristic classes of the gauge bundle and geometry. For instance, for the $\text{Spin}(32)/\mathbb{Z}_2$ heterotic string in 10d we have 
\begin{equation} J^F_8=\tr(F^4),\quad J^R_8=\tr(R^4),\quad J^F_4=\tr(F^2),\quad J^R_4=\tr(R^2).\end{equation}
We will take the traces in the gauge factors to be in the adjoint representation, and those involving geometry to be in the vector representation. Index theorems ensure that these charges are quantized, though not in the basis above. The precise normalization will not be important for most of our discussion. In addition, we may also define
\begin{equation} J^H_3=H_3=\rmd B_2, \quad J_6^H=\star H_3=\rmd B_6,\end{equation}
where $B_6$ is the dual potential to $B_2$. To understand which symmetries are gauged and which are broken, we must take into account the Green-Schwarz couplings in the 10d action :
\begin{equation}\int B_2 \wedge \tr(F^4),\quad \int B_6 \wedge (\tr(F^2)-\tr(R^2)).\end{equation}
The first one arises at tree level in string perturbation theory , while the second is a one-loop effect that is often expressed in terms of a modified Bianchi identity:
\begin{equation}
 \frac{1}{2\pi} \rmd H=\frac{1}{16\pi^2}\left[\frac{1}{30}\tr(F^2)-\tr(R^2)\right]+ J_{\rm NS5},\label{g00}
 \end{equation}
where we also included a current for fundamental NS5-branes. We have introduced normalization factors such that the NS5-brane charge and the periods of $H$ are integer-quantized. Thus, the above linear combination of the three 4-form currents is gauged. Since there is no other 3-form potential that could do the trick, we expect that linear combinations
\begin{equation} a\,\tr(F^2)+b\,\tr(R^2)+c\, J_{\rm NS5}\end{equation}
orthogonal to \eqref{g00}
are broken. Unlike in abelian examples, we cannot use the existence of monopoles to conclude that these currents are not closed. There is however a more indirect way of seeing why this is the case: consider an instanton of the gauge bundle, which is a finite size smooth codimension-4 gauge field configuration with nonzero $\tr(F^2)$ charge. The instanton has a moduli space which includes a rescaling modulus,  which maps a given configuration to another one of different size,
\begin{equation} A_\mu(\vec{x})\quad\rightarrow \quad \frac{1}{\rho}A_\mu(\rho\, \vec{x}).\end{equation}
As the instanton becomes smaller and smaller, the magnitude of the field strengths at its core grows bigger and bigger,  and at some point it is not a solution of heterotic supergravity in any meaningful way. Rather, this ``small instanton'' is a localized fundamental object, namely, an NS5-brane \cite{Witten:1995gx}.  Thus, when an instanton shrinks and becomes an NS5-brane, the charges associated with the currents $\tr(F^2)$ and $J_{\rm NS5}$ each change by one unit, whereas the combination 
\begin{equation}\frac{1}{480\pi^2}\tr(F^2)+J_{\rm NS5},\end{equation} 
is conserved.

 This symmetry breaking is not visible in the low-energy supergravity theory. In the full string theory, the instanton and the NS5-brane are one and the same; there is just one conserved current. The IR splitting of the 5-brane charge into $\tr(F^2)$ and $J_{\rm NS5}$ arises only after we introduce a cutoff on effective field theory, as we cannot transform an instanton into an NS5-brane without shrinking it to be smaller than the cutoff. An effective field theorist thus chooses to ``forget'' that the NS5-brane and the small instanton are actually the same object, which leads them to to see an accidental global symmetry that is not present in the full quantum gravity.

Indeed, this is perfectly analogous to the breaking of the various Chern-Weil currents in \eqref{GUTcurrents} in the IR of an $SU(5)$ GUT considered in Section \ref{ssec:breaking}. The nonconservation of these currents cannot be expressed
in terms of gauge-invariant combinations of fields in the IR gauge theory, signaling the emergence of an accidental symmetry.  Only when one goes to distance scales shorter than the cutoff of the IR gauge theory does one realize that the various types of instantons can be rotated into one another via the full $SU(5)$ gauge group, so in the UV theory only one linear combination of the IR Chern-Weil currents remains conserved.

\begin{figure}
\begin{center}
\includegraphics[width=85mm]{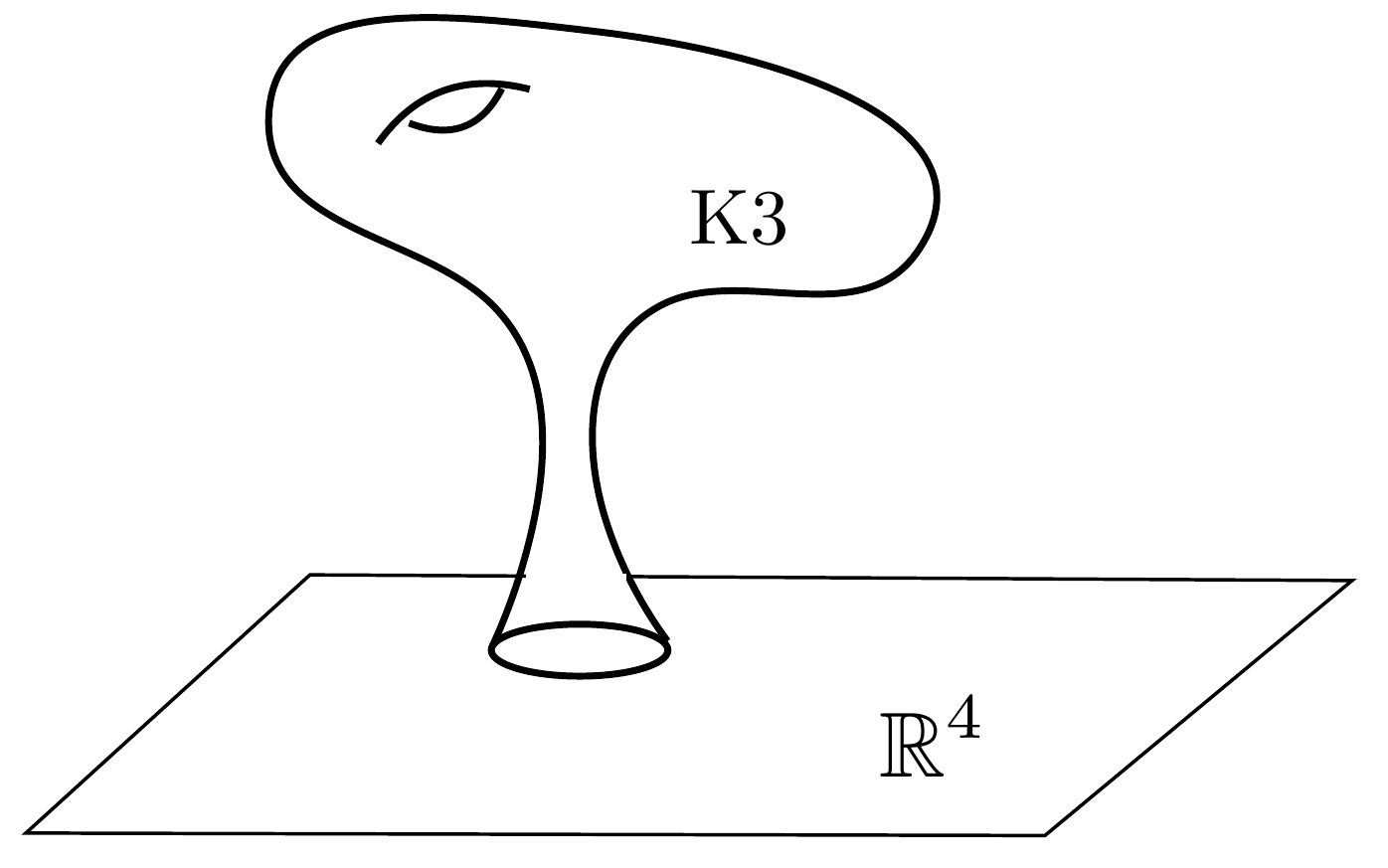}
\caption{Creating a gravitational soliton with nonzero $\tr(R^2)$ charge on $\mathbb{R}^4$ via the connected sum construction. We take a manifold with nonzero $\int \tr(R^2)$, such as K3, and glue it to $\mathbb{R}^4$ via a small tube. To a far away observer, this looks like a ``particle'' with nonzero gravitational Chern-Weil charge. This picture should be understood as a constant time snapshot, and can be generalized to other dimensions and gravitational solitons. See \cite{McNamara:2019rup} for an extended discussion.}
\label{fig:K3glue}
\end{center}
\end{figure}

The story for $\tr(R^2)$ is more complicated, as the charge under this current is carried by manifolds. Consider the current
\begin{equation}
J_4 = \frac{1}{16\pi^2}\tr(R^2)+J_{\rm NS5}
\end{equation}
in flat space. While one can put a gauge instanton of unit charge on $\mathbb{R}^n$, there is no way to introduce ``one unit of $\tr(R^2)$'' without changing the topology. The natural way to do this is via a connected sum construction, as discussed in \cite{McNamara:2019rup} (see also \cite{Hebecker:2019vyf}): one takes a compact four-manifold with nonvanishing $\tr(R^2)$, cuts out a small $S^3$, and glues the resulting manifold to $\mathbb{R}^4$ with a 4-ball $B_4$ excised, $\mathbb{R}^4 \setminus B_4$. The resulting gravitational soliton looks like flat space away from a small region where the gravitational charge is localized, just like in the case of a gauge instanton (see Figure \ref{fig:K3glue}). 

Which compact manifolds $X_4$ can we glue in this way to $\mathbb{R}^4$? Heterotic supergravity includes fermions, so we should only glue $\text{Spin}$ 4-manifolds. The spin bordism group is generated by K3, so any such manifold $X_4$ is cobordant to a connected sum of an integer number of K3's.
It is sufficient for our purposes to discuss the case of a single K3. We have 
\begin{equation}\frac{1}{16\pi^2}\int_{\textrm{K3}}\tr(R^2)=24.\end{equation}
We remind the reader that the trace on the left-hand side is taken in the vector representation. One way to see this is to note that heterotic compactifications on K3 require 24 NS5-branes to cancel the tadpole. Thus, gluing a K3 to flat space yields $-24$ units of $J_4$ charge. From our experience with similar setups, we therefore expect that this K3 manifold may be converted into 24 anti-NS5-branes by shrinking it smaller than the string scale, at which point the effective field theory breaks down and the distinction between the geometric soliton and the anti-NS5-branes is blurred. It would be interesting to investigate this further.\footnote{As circumstantial evidence, we should point out that there are related examples of conifold transitions in heterotic string theory  where some $\tr(R^2)$ is traded for NS5-branes in a controlled, supersymmetric way, although these examples involve additional ingredients.}

A similar story applies to the currents $J^{F}_8=\tr (F^4)$ and $J^R_8=\tr (R^4)$. Our remaining discussion will be more schematic, dropping constant prefactors. Green-Schwarz couplings lead to \cite{green1988superstring}
\begin{equation}\rmd H_7= J^F_8+J^R_8 +(J^F_4)^2+ (J^R_4)^2 +J_4^FJ_4^R+ j_8^{\text{F}1} := J_8,
\end{equation}
where $j_8^{\text{F}1} $ represents the contribution from fundamental strings. The current $J_8$ is exact, so the associated symmetry is gauged, whereas other currents must be broken, presumably by transitions between the objects charged under them.
Indeed, the transition of a $\tr(F^4)$ soliton to a fundamental string of unit charge was explored in \cite{Minasian:2001ib}.
 There are 8-manifolds with nonvanishing $\tr(R^4)$, so we can construct gravitational solitons as we did before. However, geometric transitions involving a change of $\tr(R^4)$ charge are not so well understood. 

Next, we turn to a discussion of Chern-Weil symmetries of the $E_8\times E_8$ heterotic string. The story is nearly identical to that of the $\text{Spin}(32)/\mathbb{Z}_2$ heterotic string, but with one important modification: while in the $\text{Spin}(32)/\mathbb{Z}_2$ theory there is a single Chern-Weil 4-form $\tr(F\wedge F)$, in the $E_8\times E_8$ theory there are two: one for each $E_8$ factor. The Bianchi identity  \eqref{g00} now involves both of these currents:
\begin{equation}
\rmd H=\tr_1(F^{(1)}\wedge F^{(1)})+\tr_2(F^{(2)} \wedge F^{(2)})-\tr(R^2)+ J_{\rm NS5}.
\end{equation}
There is a new combination,
\begin{equation}\tr_1(F^{(1)}\wedge F^{(1)})-\tr_2(F^{(2)} \wedge F^{(2)}),\label{cddd}\end{equation}
which is explicitly broken (in the same sense as above), since a small instanton of either $E_8$ can become an NS5-brane. In fact, \eqref{cddd} does not represent a conserved current in the effective theory for another reason: it is not gauge invariant. The gauge group is actually $\mathbb{Z}_2\rtimes(E_8\times E_8)$, where the $\mathbb{Z}_2$ factor acts by swapping the two $E_8$ factors. As a result, \eqref{cddd} is not gauge invariant, so it does not constitute a genuine operator of the theory.

The situation changes in compactifications where the background spontaneously breaks the symmetry between the two $E_8$ factors (which happens, for instance, in the presence of Wilson lines). As a result, the IR theory after spontaneous symmetry breaking has gauge-invariant operators $\tr_1(F^{(1)}\wedge F^{(1)})$, $ \tr_2(F^{(2)}\wedge F^{(2)})$. Nevertheless, the fact that an NS5-brane can still turn into an instanton of either gauge group means that both currents are actually broken. 

The story also changes slightly for the 8-forms $\tr(F^4)$. For a given $E_8$ factor, we have the identity 
\begin{equation} \tr(F^4)=\frac{1}{100}(\tr(F^2))^2,\end{equation}
where the traces are taken in the adjoint. This means that $J^F_8$ is, in a sense, a ``composite'' of the lower-dimensional Chern-Weil currents,
so the pointlike transition that changes the $\tr(F^4)$ charge can arise only at the intersection of NS5-branes. In Ho\v{r}ava-Witten theory, this transition corresponds to a double instanton in the end-of-the-world M9-brane; one of the instantons can condense into an M5-brane, which may then be pulled to the bulk, with a worldvolume instanton that carries the necessary M2-brane charge.

Finally, we discuss briefly the non-supersymmetric $SO(16) \times SO(16)$ string. This theory is very similar to the $E_8\times E_8$ case, with one important difference: there is no $\tr(R^4)$ term in the anomaly polynomial. As a result, the $\tr(R^4)$ charge does not correspond to fundamental string charge and is not explicitly broken by any known effects. Since there is just one massless antisymmetric tensor in the $SO(16)\times SO(16)$ theory, the current cannot be gauged independently, so it must be broken. Like the breaking of the $\tr(R^2)$ and $\tr(R^4)$ currents in the heterotic superstring, a more thorough understanding of how this occurs is beyond the scope of the present paper.

\subsection{Type II} 

In this section, we will discuss how Chern-Weil global symmetries are broken/gauged in Type II string theory. We will distinguish between abelian Chern-Weil symmetries constructed from the closed string field strengths and the nonabelian ones coming from the open string sector. 

\subsubsection{Type IIA  without branes\label{IIAwithout}}

The bosonic action of ten-dimensional massive Type IIA superstring theory in the democratic formulation is given by
\beq
\mathcal{L}=-\frac14 \sum_{p=0,2,4,6,8,10}G_p \wedge \star G_p-\frac12 H_3\wedge \star H_3,
\eeq
where the improved gauge-invariant field strengths  are 
\begin{equation}  \label{eq:gaugeinvGp}
G_p= F_p -\rmd B_2 \wedge C_{p-3}+G_0\mathrm{e}^{B_2},
\end{equation}
where $F_p = \rmd C_{p-1}$ (which is not gauge invariant). We have omitted the Einstein-Hilbert term, the dilaton kinetic term, and the dilaton dependence of the other terms, in order to focus on the Chern-Weil symmetries without cluttering our notation. The Bianchi identities and the equations of motion read
\beqa
&\rmd G-H_3 \wedge G=0\ ,\quad \rmd H_3 =0\, ,\\
&\rmd {\star G_p}+H_3 \wedge \star G_{p+2}=0.
\label{movto}
\eeqa
To avoid double counting, we impose the following duality condition:
\beq
\label{dual}
G_p=(-1)^{p/2} {\star  G_{10-p}}.
\eeq
The equation of motion for $G_p$ is then equal to the Bianchi identity of $G_{10-p}$, and the theory is self-dual.

Let us begin by discussing Chern-Weil global symmetries with $k=1$, i.e., with currents of the form $F_p$. By going to the appropriate duality frame, they reduce to the electric and magnetic global symmetries of the higher-form gauge fields, with currents constructed from the gauge-invariant field strengths $G_p$ and $H_3$. In the former case, we have $(9-p)$-form global symmetries with currents $G_{p}$, whose conservation equations read
\beqa
\rmd  G_{10}=H_3\wedge \star G_2,& \quad \rmd G_4=-H_3\wedge \star G_8, \nonumber \\
\rmd  G_8=-H_3\wedge \star G_4,& \quad \rmd  G_2=H_3\wedge \star G_{10}, \label{demcurrents} \\
\rmd   G_6=H_3\wedge \star G_6, &\quad \rmd  G_{0}=0,  \nonumber
\eeqa
where we have used \eqref{movto} and \eqref{dual}.
Note that the three currents on the left-hand side of \eqref{demcurrents} can be understood as currents for electric higher-form symmetries, whereas those on the right-hand side of \eqref{demcurrents} can be understood as currents for the corresponding magnetic symmetries (or vice versa).

The equations of motion imply that none of these currents are conserved, except for the 9-form global symmetry with current $G_0$. 
If we turn off the Romans mass by setting $G_0=0$, a 7-form global symmetry emerges instead, since then $\rmd {G_2}=0$. Hence, there is always one global symmetry that remains unbroken in the absence of branes. In order to break this symmetry, we need to add D8-branes (or D6-branes in the absence of a Romans mass), in the same way that the electric 1-form $U(1)$ symmetry of $U(1)$ gauge theory in Section \ref{sec:bg} was broken by the existence of charged particles.

The NS 2-form field $B$ has an equation of motion and Bianchi identity of the form
\beq
\label{Hcurrents}
\rmd {\star H_3}=\rmd C_{p-3}\wedge \star G_p+\partial(\mathrm{e}^B)G_0\wedge \star G\quad ,\quad \rmd H_3=0,
\eeq
so the magnetic symmetry with current $H_3$ is also left unbroken. Breaking this global symmetry requires the addition of NS5-branes.

The currents we have considered thus far are gauge invariant. The charge appearing on the non-conservation of these currents is known as the Maxwell charge \cite{Marolf:2000cb}, and it is not quantized nor conserved. On the other hand, the currents $\star F_p$ are not gauge invariant but are conserved in the absence of localized branes. The charge associated to these currents is called the Page charge; it is not gauge invariant, but it is conserved and quantized (since the field strengths are exact). 

We can next study the fate of the Chern-Weil symmetries of the form $X_p \wedge Y_q$, where $X_p$ is a $p$-form field strength and $Y_q$ is a $q$-form field strength.  The simplest examples we might consider take the form $F_4\wedge F_4$ and $F_2\wedge F_2$. We have seen that $F_p$ is not gauge invariant, however, so these currents are not gauge invariant either.

This naturally leads to us consider Chern-Weil currents of the form $G_p \wedge G_p$, since the improved field strength $G_p$ is gauge invariant. However, these currents are not conserved as one can easily see using \eqref{demcurrents},  with the exception of $G_0 \wedge G_0$ and the combination 
\beq
\label{J4}
J_4:=G_0G_4-\frac12 G_2^2
\eeq
 since
\beq
\rmd (G_0G_4)= G_0\wedge H_3\wedge G_2\ ,\quad \rmd  (G_2^2)=2H_3\wedge G_0\wedge G_2
\eeq
implying $\rmd J_4=0$.
 The other possible combination is to consider Chern-Weil symmetries of the form $G_p \wedge H_3$, which are gauge invariant and conserved, since
\beq
\rmd(G_p\wedge H_3)=H_3\wedge G_{p-2}\wedge H_3=0\ .
\eeq 
However, these currents are exact, as is clear from \eqref{demcurrents}, so the associated global symmetries are gauged. Furthermore, one can check that the current $G_2\wedge H_7$ is not conserved by using \eqref{demcurrents} and \eqref{Hcurrents}.

Finally, we can consider Chern-Weil symmetries with $k>2$. The only new conserved current arising is 
\beq
\label{J6}
J_6:= G_0^2 G_6-G_0G_2G_4+\frac13 G_2^3\ ,
\eeq
whose conservation $\rmd J_6=0$ follows from using that
\beq
\rmd(G_6)=H_3G_4\ , \quad \rmd(G_2G_4)=H_3(G_0G_4+G_2^2)\ , \quad \rmd(G_2^3)=3G_0G_2^2\ .
\eeq
We can also take products of the previously discussed conserved currents to engineer higher-form ones, like $J_4^2$ or $J_4\wedge J_6$, or products thereof with $G_0$.

One could attempt to construct a conserved 8-form current using the same logic as above, yielding $G_0G_8-G_2G_6+\frac12 G_4^2$. However, although indeed conserved, it is also exact as it appears on the right hand side of the equation of motion for $B_2$ in \eqref{Hcurrents}. Hence, the symmetry is gauged. Analogously, the possible conserved 10-form currents associated to potential $(-1)$-form symmetries are exact (and therefore, gauged) when using the equation of motion of the dilaton.\footnote{There are two possible combinations yielding 10-form conserved currents that are exact,
\beqa
\rmd (\mathrm{e}^{2\phi}\star \rmd \phi )=\frac12 H_3\wedge H_7+\frac14 (5G_0\wedge G_{10}+3G_2\wedge G_8-G_4\wedge G_6)\\
\rmd(G_2H_7+2G_0\mathrm{e}^{2\phi} \star \rmd \phi )=\frac52 G_0^2  G_{10}+\frac52 G_0G_2G_8-\frac12 G_0G_4G_6-G_2^2 G_6+\frac12 G_2G_4^2 ,
\eeqa
so the associated global symmetries are gauged.}

To summarize, the only conserved (not-gauged) currents are $G_0^k$, $G_0^kH_3$, $G_0^kJ_4$, $G_0^kJ_6$, $G_0^kJ_4^2$ and $G_0^kJ_4\wedge J_6$ with $k\geq 0$ and $J_4,J_6$ defined in \eqref{J4} and \eqref{J6} respectively. We will see in the next section that these symmetries get broken in the presence of branes.

Before closing this subsection, let us notice that, in the massless case with $G_0=0$, all the symmetries constructed from $J_4$ and $J_6$ reduce to products of $F_2$. Hence, the only conserved currents (in the absence of branes) in massless Type IIA are $H_3$ and $F_2^k$ with $k>0$.

\subsubsection{Type IIA with branes}\label{ssec:typeIICW}

As discussed in the previous subsection, we need to introduce D8-branes and NS5-branes to break the global symmetries with currents $ G_{0}$ and $H_3$ respectively. Gauge invariance of the bulk gauge fields highly constrains the worldvolume action of these branes. In fact, as shown in Appendix \ref{app:CS}, the Chern-Simons action of D-branes follows simply from the existence of worldvolume degrees of freedom that gauge the 2-form global symmetry of the NS $B$-field. The Chern-Simons action reads
\beq
\label{CS}
S_\mathrm{CS}=\int_{\mathrm{D}p}\sum_q C_q \wedge \mathrm{e}^{-\mathcal{F}_\Dp}\ .
\eeq The $B$-field then appears on  the worldvolume only in the gauge-invariant combination $\mathcal{F}_\Dp=2\pi\alpha' F_\Dp+B_2$, where $F_\Dp = \rmd A_1$ is the worldvolume gauge field strength of the D$p$-brane. We will show below that the existence of these worldvolume gauge fields can be motivated from breaking the bulk Chern-Weil symmetries in a consistent way, similar to the situation described in section \ref{ssec:gaugeandbreak}. But before that, a further observation is in order. Even if only D8-branes are required to break the bulk global symmetry with current $G_0$, the existence of these branes necessarily implies the existence of lower-dimensional D-branes as well. This can be understood as follows. Consider the above worldvolume action \eqref{CS}. A brane with non-trivial worldvolume gauge fields actually induces charge of lower-dimensional branes, as is clear from  \eqref{CS}.
By taking the zero size limit of these charge configurations, we can recover a localized lower-dimensional brane, as explained in more detail in Section \ref{subsec:openstringcurrents}.

Suppose that we did not know about the existence of the worldvolume gauge fields $A_1$ and the Chern-Simons couplings \eqref{CS}. Then, the presence of branes would raise a similar puzzle to the one discussed in Section \ref{ssec:gaugeandbreak}. Consider for example the current $G_4\wedge H_3$. 
In the presence of D4-branes, the Bianchi identity for $F_4$, hence also $G_4$, is modified:
\beq
\rmd F_4=  \delta_5^{\rm D4}\quad \rightarrow \quad \rmd G_4=H_3\wedge F_2 + G_0 \wedge H_3 \wedge B_2+ \delta_5^{\rm D4},
\label{eq:dF4}
\eeq
where $\delta_5^{\rm D4}$ is the delta-function current localized at the brane.
Therefore, the 2-form global symmetry with current $G_4 \wedge H_3$ is actually broken:
\begin{equation}
\rmd (G_4 \wedge H_3) = \delta_5^{\rm D4} \wedge H_3.\label{eomrr}
\end{equation}
This presents a puzzle, however, because the RR 3-form $C_3$ couples to $G_4 \wedge H_3$, suggesting that this current should be not only closed, but exact, and the corresponding 2-form symmetry should be gauged. In fact, we know it is gauged in the absence of brane sources, as explained in the previous subsection. An equivalent way to explain the puzzle is that \eqref{eomrr} is inconsistent with the equation of motion for $C_3$, since it would imply that
\beq
\label{puzzle}
\rmd(\rmd\star G_4)=\rmd(G_4\wedge H_3)\neq 0 \ !
\eeq

This puzzle is resolved by the presence of dynamical degrees of freedom on the D4-brane worldvolume, just as the analogous puzzle was resolved by electric degrees of freedom on the monopole worldvolume in Section \ref{ssec:gaugeandbreak}. Let us suppose that we do not already know about these degrees of freedom, and see how their existence and properties might be deduced. We can attempt to render \eqref{puzzle} consistent by hypothesizing a 2-form field $\mathcal{F}_\Dp$, localized on the brane, appearing in a term of the form $\mathcal{F}_\Dp\wedge \delta_5^\mathrm{D4}$ on the right-hand side of the equation of motion for $C_3$. If this field satisfies the equation $\rmd\mathcal{F}_\Dp=H_3$, this will guarantee $\rmd^2\star G_4=0$. We can then write $\mathcal{F}_\Dp$ as $\mathcal{F}_\Dp=B_2+2\pi \alpha' F_\Dp$, where $F_\Dp$ is a closed 2-form that must shift under a gauge transformation of $B_2$ to ensure gauge invariance of the brane action. If we assume that the worldvolume theory is weakly coupled (i.e., Lagrangian), we can then argue for the presence of a Nambu-Goldstone zero mode on the brane corresponding to a 1-form gauge field with field strength $F_\Dp$, by appealing to the discussion below \eqref{eqref}. In this case, the new worldvolume degrees of freedom on the brane correspond to gauge fields.\footnote{The existence of massless vectors also follows from supersymmetry of the D-branes and the existence of translational Nambu-Goldstone modes. To conclude the existence of massless vectors we need to assume either a large amount of worldvolume supersymmetry or a Lagrangian description in the brane worldvolume.} As explained in more generality in Appendix~\ref{app:CS}, we can extend these arguments to derive the Chern-Simons action of the D4-brane, a special case of \eqref{CS}:
\beq
S_\mathrm{CS}=\int_{\mathrm{D}4}\sum_q C_q \wedge \mathrm{e}^{-\mathcal{F}_\Dp}=\int_{\mathrm{D}4}(C_5-C_3\wedge \mathcal{F}_\Dp+C_1\wedge  \mathcal{F}_\Dp\wedge  \mathcal{F}_\Dp)\ .
\eeq
This indeed modifies the equation of motion for $C_3$ as expected:\footnote{One might ask how the story changes if one replaces the D4-brane above by an O4-plane, which does not have a worldvolume gauge field. The answer is that the orientifold projection also ensures that the right hand side of \eqref{eomrr} is trivial at the level of differential forms that we discuss in this paper, so that the current is gauged with no further complications.}
\beq
\rmd \, {\star G_4}=G_4\wedge H_3+ \mathcal{F}_\Dp\wedge \delta_5^{\rm D4}:= J_7.
\eeq
We see that while the 7-form $G_4\wedge H_3$ is not closed in the presence of D4-branes, a combination of this current plus another one localized on the D4-brane yields a new current $J_7$ which is exact. Hence, the associated 2-form global symmetry with current $J_7$ is gauged. The same occurs for other symmetries with currents $G_p\wedge H_3$ when introducing D-branes.

More generally, the Chern-Simons action of the branes always involves the right couplings to either break the Chern-Weil symmetries or gauge them by rendering their associated currents exact. For instance, the gauge-invariant 8-form 
\beq
\label{J8}
J_8=F_2 \wedge \star G_4 + F_4 \wedge G_4 + (C_3 - C_1\wedge  \mathcal{F}_\Dp) \wedge \delta_5^{\rm D4}
\eeq
is exact due to the equation of motion for $B_2$:
\beq
\rmd \,{\star H} = J_8,
\eeq
so the associated 1-form symmetry is gauged by $B_2$.
Hence, the presence of D4-branes breaks the Chern-Weil current $F_2 \wedge \star G_4 + F_4 \wedge G_4$, but the sum of this current and last term in \eqref{J8} is exact, so the corresponding symmetry is gauged.

Finally, we can investigate the fate of the Chern-Weil symmetries $J_4$, $J_6$ defined in \eqref{J4} and \eqref{J6}, and products thereof, in the presence of branes. First of all, let us consider massless Type IIA so that $G_0=0$. As explained at the end of the previous subsection, the only conserved Chern-Weil symmetries with $k>1$ are powers of $F_2^k$. In the presence of D6-branes, all these symmetries get broken as
\beq
\rmd (F_2^k)=kF_2^{k-1}\wedge  \delta_3^{\rm D6}\neq 0\ .
\eeq
In the presence of branes, in principle we could also construct new currents of the form $F_2^k \wedge   \delta_3^{\rm D6}$ but they are all exact by the previous equation, implying that the associated symmetries are gauged. However, the are other new currents localized at the branes and involving the worldvolume fields whose non-conservation occurs only when considering coincident or interesting branes. For example, $\rmd[(G_4-\mathcal{F}_\Dp F_2)\wedge \delta_3^{\mathrm{D6}}]\propto  \#  \delta_3^{\mathrm{D6}}\wedge  \delta_3^{\mathrm{D6}}\neq 0$. 

The massive case with $G_0\neq 0$ works in a similar way. The currents $J_4$ and $J_6$ defined in \eqref{J4} and \eqref{J6} are not closed in the presence of branes,
\beq
\rmd J_4=G_0\delta_5^{\mathrm{D4}}+ (G_2-G_0\mathcal{F}_\Dp)\wedge \delta_3^{\mathrm{D6}}+ (G_4-\mathcal{F}_\Dp G_2+\frac12 \mathcal{F}_\Dp^2 G_0)\wedge \delta_1^{\mathrm{D8}} \neq 0 \ 
\eeq
and similarly for $J_6$,
so the corresponding global symmetries are broken, as well as those constructed from taking products of $J_4$, $J_6$ and $G_0$. However, we can construct new currents localized on the branes of the form $\omega \wedge \delta$. For instance, at rank 3 we get a new conserved current given by
\beq
\label{J3}
J_3= (G_2-G_0\mathcal{F}_\Dp)\wedge \delta_1^{\mathrm{D8}}
\eeq 
For a scenario with non-intersecting branes, this yields a global symmetry. However, we expect the symmetry to get broken when allowing the branes to intersect and taking into account the new degrees of freedom appearing at the intersections. To motivate this, consider the modified Bianchi identity for $G_2$ in the presence of branes,
 \beq
\rmd G_2=H_3G_0+\delta_3^{\mathrm{D6}}+\mathcal{F}_\Dp \delta_1^{\mathrm{D8}}
\eeq
When computing the derivative of \eqref{J3}, we should get $\rmd J_3\propto \#\delta_3^{\mathrm{D6}}\wedge \delta_1^{\mathrm{D8}}+ \#\mathcal{F}_\Dp \delta_1^{\mathrm{D8}}\wedge \delta_1^{\mathrm{D8}}$, which is non-zero if the branes are intersecting. 

The same occurs when considering other currents of higher rank of the form $\omega \wedge \delta$. Because the generalization of the DBI+CS effective action for intersecting branes is not as well understood as in the abelian case, we conclude the analysis of Type IIA abelian currents here, with the following message. The absence of Chern-Weil global symmetries with $k=1$ requires the existence of charged branes, as is already known, but the implications do not stop here. The non-conservation of higher rank currents with $k>1$ requires the existence of worldvolume degrees of freedom on the branes, which in turn implies the existence of degrees of freedom at the intersections of branes when considering higher rank symmetries, and so on; until presumably recovering the full spectrum of Type IIA. It is very impressive how the simple quantum gravity criterion of no global symmetries seems to be enough to recover much of the rich structure of branes and worldvolume fields present in Type IIA string theory compactifications. This provides further evidence that the plethora of dynamical extended objects and interconnections typical of string theory is a generic feature of quantum gravity.

\subsubsection{Nonabelian open string currents}
\label{subsec:openstringcurrents}

Another way to engineer Chern-Weil currents in Type II string theory is via worldvolume gauge fields. The DBI + CS action for a stack of D$p$-branes includes terms of the form $\tr(F_2^2)$, which yield nonabelian Chern-Weil currents. Since the gauge fields live on the D$p$-branes, this current is associated to a $(p-5)$-form global symmetry.

\begin{figure}
\begin{center}
\includegraphics[width=\textwidth]{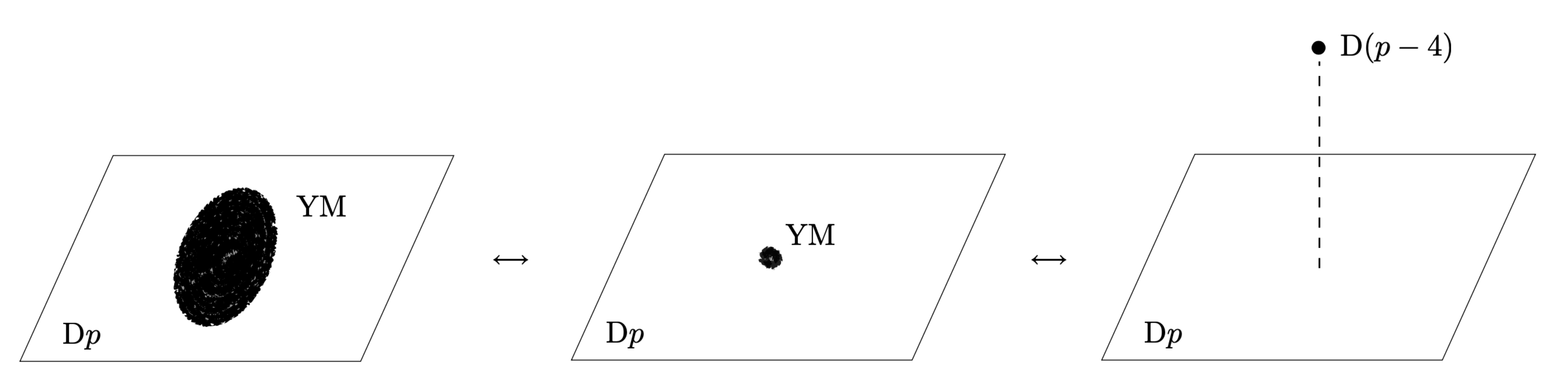}
\caption{Illustration of the process that converts between a Yang-Mills instanton localized on a stack of D$p$-branes (depicted at left) and a D$(p-4)$-brane that can move freely into the bulk (depicted at right). The instanton solution has a size modulus, and small instantons are indistinguishable from lower-dimensional D-branes. As a result, instanton charge and D$(p-4)$-brane charge are not independently conserved, avoiding a global symmetry.}
\label{fig:YMinstToDbrane}
\end{center}
\end{figure}

There is an intuitive way to understand how this Chern-Weil symmetry gets broken. Consider a gauge instanton on the worldvolume of the D$p$-brane stack, with non-zero charge $\tr(F^2)$. Taking the zero-size limit of this instanton, we end up with a localized object of the same charge. This process cannot be described in the regime of validity of supergravity, but it is well known in string theory and produces a lower-dimensional D$(p-4)$-brane \cite{Witten:1995gx, Douglas:1995bn, Douglas:1996uz, Green:1996dd}, as depicted in Figure~\ref{fig:YMinstToDbrane}. Hence, the current
\beq
\label{trFJ}
\delta_{9-p}^{\textrm{D}p} \wedge \tr(F^2)  -\delta_{13-p}^{\textrm{D}(p-4)}
\eeq
is not conserved. This is analogous to the symmetry associated to $\tr(F^2)$ in heterotic string theory discussed in Section \ref{sec:heterotic}, whose breaking is manifest by the process transforming a small instanton with $\tr(F^2)$ charge into an NS5-brane. As discussed there in detail, the distinction between the small instanton and the lower-dimensional brane is meaningless in the full string theory, implying that the global symmetry is broken. However, it remains as an approximate global symmetry in the IR, as such a process cannot be described within the low energy EFT.

A linear combination orthogonal to \eqref{trFJ} is exact, signaling the existence of a $(p-4)$-form gauge symmetry. This can be seen from the modified Bianchi identity for $F_{12-p}$. For example, we may consider an instanton with D4-charge in the worldvolume of a stack of D8-branes. Due to the Chern-Simons couplings in the brane worldvolume action, the equation of motion for $C_5$ (equivalently, the Bianchi identity for $C_3$) gets modified to
\beq
\rmd  F_4=\delta_{5}^{\rm D4}+(B^2+\tr F^2)\wedge \delta_{1}^{\rm D8},
\eeq
where we have restored the dependence on the $B$-field. Hence, an orthogonal combination to \eqref{trFJ} indeed corresponds to an exact current. This is another example of gauging a linear combination involving Chern-Weil currents while breaking the other combination.

As a final remark, notice that the limiting case of $\tr(F^2)$ living on a stack of D3-branes is an example of a $(-1)$-form global symmetry. The process of dissolving this brane into a D($-1$) instanton illustrates the breaking of the current $\delta_{6}^{\textrm{D}3} \wedge \tr(F^2)  -\delta_{10}^{\textrm{D}(-1)}$ and the gauging of the orthogonal combination.

\subsubsection{Type IIB}

We can also investigate how Chern-Weil symmetries are broken or gauged in ten dimensional Type IIB superstring  theory. It shares certain similarities with the Type IIA case, so we will keep the discussion short, focusing only on the differences.

Let us begin the analysis in the absence of branes. Excluding the Einstein-Hilbert term, the IIB pseudo-action in Einstein frame in $SL (2, \mathbb{Z})$ covariant notation is
\beq
  S \supset - \frac{1}{4
  \kappa_{10}^2} \int \left[ - \frac{1}{2} \rmd \varphi^{i j} \wedge \star d
  \varphi_{i j} + \varphi_{i j} F_3^i \wedge \star F_3^j + \frac{1}{2} 
  \tilde{F}_5 \wedge \star \tilde{F}_5 - \frac{1}{2} \varepsilon_{i j} C_4
  \wedge F_3^i \wedge F_3^j \right] .
\eeq
Here $\varepsilon_{12} = - \varepsilon_{21} = + 1$ and $\varphi^{i j} =
\varphi^{j i}$ satisfies the constraint $\det \varphi^{i j} = 1$, with inverse
$\varphi_{i j} = \varepsilon_{i k} \varepsilon_{j l} \varphi^{k l}$, whereas
$F_3^i =\rmd C_2^i$, $\tilde{F}_5 = \rmd C_4 - \frac{1}{2} \varepsilon_{i j}
C_2^i \wedge F_3^j$, and the equations of motion must be supplemented by the
constraint $\star \tilde{F}_5 = \tilde{F}_5$. In terms of more familiar fields
\beq
  \varphi^{i j} = \frac{1}{\tau_2}  \left(\begin{array}{cc}
    | \tau |^2 & \tau_1\\
    \tau_1 & 1
  \end{array}\right), \qquad \varphi_{i j} = \frac{1}{\tau_2} 
  \left(\begin{array}{cc}
    1 & - \tau_1\\
    - \tau_1 & | \tau |^2
  \end{array}\right), \qquad F_3^i = \left(\begin{array}{c}
    F_3\\
    H_3
  \end{array}\right),
\eeq
where $\tau = \tau_1 + i \tau_2 = C_0 + i \mathrm{e}^{- \Phi}$. It is also sometimes
useful to consider $\varphi^i_{\hspace{0.4em} j} \assign \varphi^{i l}
\varepsilon_{l j} = \varepsilon^{i k} \varphi_{k j}$, or explicitly:
\beq
\varphi^i_{\hspace{0.4em} j} = \frac{1}{\tau_2} \left(\begin{array}{cc}
     - \tau_1 & | \tau |^2\\
     - 1 & \tau_1
   \end{array}\right) .
\eeq

Suppose first that $SL(2, \mathbb{Z})$ is not gauged. Possible Chern-Weil
currents can be built from $F_3^i$, $\tilde{F}_5$, $\tilde{F}_7^i =
\varphi^i_{\hspace{0.4em} j} \star F_3^j$, $\varphi^{i j}$, $\rmd \varphi^{i
j}$, and $\star \rmd \varphi^{i j}$. Combinations built only from $\varphi^{i
j}$ and its exterior derivatives are exact whenever they are closed (since the
moduli space is contractible), so we can ignore these. The currents $F_3^i$
are conserved and not gauged. However,
\beq\label{sl2_singlet}
  \rmd \tilde{F}_5 = - \frac{1}{2} \varepsilon_{i j} F_3^i \wedge F_3^j = -
  F_3 \wedge H_3,
\eeq
so the only non-zero wedge product of these currents is gauged. Moreover,
\beq
 \rmd {\tilde F}_7^i = F_3^i \wedge \tilde{F}_5,
\eeq
so $\tilde{F}_7^i$ is broken, gauging the currents $F_3^i \wedge \tilde{F}_5$
in the process. Now consider the 9-form currents $\star \rmd \varphi^{i
j}$. One finds:
\beq
  \rmd \left[ \varphi^i_{\hspace{0.4em} j} \star {\rmd \varphi^{j k}} \right]
  = - F_3^{(i} \wedge \tilde{F}_7^{j)},
\eeq
hence all of these currents are broken, gauging the 10-form currents $F_3^{(i} \wedge \tilde{F}_7^{j)}$. Note that, by comparison, the
10-form current $F_3^{[i} \wedge \tilde{F}_7^{j]}$ is
broken, since (formally)
\beq
  \rmd (F_3^{[i} \wedge \tilde{F}_7^{j]}) = - F_3^{[i
 } \wedge F_3^{j]} \wedge \tilde{F}_5 \neq 0 .
\eeq
Any wedge product of conserved currents $J \wedge H$, where $J$ is a conserved current and $H$ is built from $\varphi^{ij}$ and its exterior derivatives, is exact, since $ J \wedge H = \rmd (J \wedge K)$, where $\rmd K = H$. Wedge products of non-conserved currents and $\varphi^{ij}$ and its exterior derivatives are likewise not conserved. Thus, we conclude that the only conserved, non-exact currents  before
$SL (2, \mathbb{Z})$ is gauged are $F_3$ and $H_3$. Their breaking requires the presence of D5- and NS5-branes, respectively.

Once $SL(2, \mathbb{Z})$ is gauged, the fields $F_3, H_3$ no longer define currents for global symmetries, as they themselves are no longer gauge-invariant under $SL(2, \mathbb{Z})$. In fact, the only $SL(2, \mathbb{Z})$-singlet Chern-Weil current in Type IIB is the current $F_3 \wedge H_3$, which is gauged according to \eqref{sl2_singlet}, so one might think that there are no global symmetries coming from gauge fields in Type IIB supergravity once $SL(2, \mathbb{Z})$ is gauged. This is incorrect, as there are global symmetries arising from the $SL(2, \mathbb{Z})$ gauge field itself. Given a discrete gauge group $G$, there is an associated $(d - 2)$-form global symmetry, whose charge on a $1$-manifold is given by the projection of the holonomy of the gauge field to the abelianization $G_{\rm ab}$.\footnote{These symmetries will be discussed in more detail in an upcoming paper \cite{JMgravsolitons}.} The breaking of this symmetry corresponds to adding defects of codimension two, around which the discrete gauge field has nonzero monodromy, such as Krauss-Wilczek strings in four dimensions. In the case at hand, we have $SL(2, \mathbb{Z})_{\rm ab} = \mathbb{Z}_{12}$, and so we have an $8$-form $\mathbb{Z}_{12}$ symmetry in Type IIB supergravity once $SL(2, \mathbb{Z})$ is gauged.\footnote{Properly speaking, we should consider the pin$^+$ double cover of $GL(2, \mathbb{Z})$ considered in \cite{Tachikawa:2018njr} and its abelianization, but this subtlety isn't important for the current discussion.} In order to break this symmetry, we must add 7-branes to the theory, such as D7-branes or their $SL(2, \mathbb{Z})$-conjugates.

\subsubsection{Gravitational currents}
We also comment briefly on gravitational currents in Type II. In Section \ref{sec:heterotic}, we found that these are often gauged in heterotic string theory. The situation in Type II is the converse; they are often broken. As explained in \cite{McNamara:2019rup}, in IIA the class of K3 is killed by eight O4-planes, which uplift to MO5-branes in M-theory. In this particular case it is also possible to describe the non-conservation of $\text{tr}(R^2)$ via a smooth eleven-dimensional supergravity background, since the IIA K3 compactification uplifts to $\mathrm{K3}\times S^1$ and $\Omega_5^{\text{Spin}}=0.$ 

One would also expect the same class to be broken in IIB, since it does not couple to any massless field; this is more mysterious, since the process destroying the charge must  be non-supersymmetric, as explained in \cite{McNamara:2019rup}. Note that the fact that one can have conifold transitions in Calabi-Yau manifolds \cite{Strominger:1995cz,Greene:1996cy}, which can change the value of $\tr(R^2)$, do not constitute evidence that $\int \tr(R^2)$ is broken, since $\int \tr(R^2)$ is a bordism invariant for four-manifolds, but not 6-manifolds.

In Type IIA, a certain linear combination of $\tr(R^4)$ and $\tr(R^2)^2$ is gauged by the gravitational coupling
\begin{equation}\int B_2\wedge X_8.\label{www4}\end{equation}
See \cite{Liu:2013dna} for a detailed discussion.  This means that certain two-dimensional compactifications of IIA, for instance compactification on a Bott manifold \cite{Freed:2019sco}, require the inclusion of additional F1 strings, which are pointlike on the compactification manifold. Upon circle compactification and T-duality, these become pure KK momentum, matching the fact that  the $\int A_1 \wedge X_8$ coupling that arises directly from dimensional reduction of \eqref{www4} in the Type IIA frame becomes a one-loop effect in the IIB perspective \cite{Liu:2013dna}. In the IIB frame, the wound F1 strings become pure geometry, and so the same linear combination is broken in IIB. From the IIB perspective, compactification on $\text{Bott}\times\mathbb{R}^2$ is fine, but if one replaces the $\mathbb{R}^2$ factor by a space with an $S^1$ factor, some KK momentum needs to be introduced. 

Additional currents can be gauged in the presence of D-branes or orientifold planes, due to the presence of the A-roof and L-genus term in the D-brane and orientifold action respectively  (see for instance \cite{Scrucca:1999uz}),
\begin{equation} S_{\text{D-brane}}\supset \sqrt{\frac{\hat{A}(R_T)}{\hat{A}(R_N)}} \wedge \text{ch}(F) \wedge C,\quad S_{\text{Orientifold}}\supset \int \sqrt{\frac{L(R_T/4)}{L(R_N/4)}} \wedge C,\end{equation}
where $R_T, R_N$ are the curvatures of tangent and normal bundles respectively. These couplings mean that gravitational Chern-Weil currents like $\delta_{\text{D-brane}}\wedge\tr(R^2)$ can source RR fields, and likely participate in a gauging-breaking scenario involving gravitational currents as above, which we will not explore in detail here.

\subsection{M-theory}

Next, we consider Chern-Weil symmetries and their breaking and gauging in M-theory, where the story is particularly simple and elegant. We first consider pure eleven-dimensional supergravity:
\begin{equation}
  S = \frac{1}{2 \kappa_{11}^2} \int \rmd^{11} x \sqrt{- g} R - \frac{1}{4
  \kappa_{11}^2} \int F_4 \wedge \star F_4 - \frac{1}{12 \kappa_{11}^2} \int
  A_3 \wedge F_4 \wedge F_4 .
\end{equation}
The possible Chern-Weil symmetry currents include $F_4$, $\tilde{F}_7 = \star F_4$ and
$F_4 \wedge F_4$, all of them gauge invariant. However, the $F_4$ equation of motion,
\begin{equation}
  \rmd \, {\star F_4} = - \frac{1}{2} F_4 \wedge F_4,
\end{equation}
simultaneously breaks $\tilde{F}_7$ and gauges $F_4 \wedge F_4$, so $F_4$ is
the only remaining nontrivial conserved current.

To break $F_4$, we must introduce a localized magnetic source such as an M5-brane. Even without UV input, the structure of the supergravity action requires that such a brane have a worldvolume chiral boson, as we now explain in detail. (The argument is similar to those for other worldvolume scalars previously discussed in Sections \ref{ssec:gaugeandbreak}, \ref{ssec:typeIICW}, and in Appendix~\ref{app:CS}.)
 To do so, it is convenient to work with a democratized supergravity
pseudoaction:
\begin{equation}
  S_{\text{dem}} = \frac{1}{2 \kappa_{11}^2} \int \rmd^{11} x \sqrt{- g} R -
  \frac{1}{6 \kappa_{11}^2} \int F_4 \wedge \star F_4 - \frac{1}{12
  \kappa_{11}^2} \int \tilde{F}_7 \wedge \star \tilde{F}_7,
\end{equation}
where $\tilde{F}_7 = \rmd A_6 - \frac{1}{2} A_3 \wedge F_4$ and the equations of
motion must be supplemented by the constraint $\tilde{F}_7 = \star F_4$.

Naively, the magnetic 5-brane would have the Chern-Simons charge coupling
\begin{equation}
  S_{\text{charge}} = \mu_5 \int A_6,
\end{equation}
however, $A_6 \rightarrow A_6 + \frac{1}{2} \lambda_2 \wedge F_4$ under $A_3
\rightarrow A_3 + \rmd \lambda_2$. Hence, the above action is not gauge invariant, and it is not possible to couple $A_6$ to the brane without including worldvolume degrees of freedom. Equivalently, since $\rmd F_4 = q \delta (\Sigma)$
for a charge $q$ magnetic brane with worldvolume $\Sigma$, $F_4 \wedge
F_4$ is no longer closed, requiring an additional term in the
$\tilde{F}_7$ Bianchi identity:
\begin{equation}
  \rmd \tilde{F}_7 = - \frac{1}{2} F_4 \wedge F_4 + q\mathcal{H}_3 \wedge \delta
  (\Sigma), \label{eqn:M5M2charge}
\end{equation}
for some three-form $\mathcal{H}_3$. Taking the exterior derivative, we find
$\rmd \mathcal{H}_3 = F_4$, so that $\mathcal{H}_3 = \rmd B_2 + A_3$
for some worldvolume field $B_2$.

The $F_4$ Bianchi identity can be solved using a Dirac string $\Xi$, $F_4 = \rmd A_3 + q \delta (\Xi)$, where $\partial \Xi = \Sigma$.
Using the same Dirac string, we obtain:
\begin{equation}
  \tilde{F}_7 = \rmd A_6 - \frac{1}{2} A_3 \wedge F_4 - \frac{q}{2} A_3 \wedge
  \delta (\Xi) + q B_2 \wedge \delta (\Sigma),
\end{equation}
assuming that the string does not self-intersect ($\delta (\Xi) \wedge
\delta (\Xi) = 0$). Consistency with the constraint $\tilde{F}_7 = \star F_4$ requires the brane to couple electrically as well as magnetically, via the action\footnote{The corresponding \emph{probe} action would be $S_{\text{probe}} \supset \mu_5 \int_{\Sigma}  \left[ A_6 +
  \frac{1}{2} B_2 \wedge F_4 \right] - \frac{k_5}{2} \int_{\Sigma}
  \mathcal{H}_3 \wedge \star_{\Sigma} \mathcal{H}_3$, without the $1/3$ in the first term. The difference comes about because the bulk democratic supergravity action has a hidden dependence on the worldvolume fields through the modified $F_4$ and $\tilde{F}_7$ Bianchi identities.}
\begin{equation}
  S_{\text{brane}} \supset \frac{\mu_5}{3} \int_{\Sigma}  \left[ A_6 +
  \frac{1}{2} B_2 \wedge F_4 \right] - \frac{k_5}{2} \int_{\Sigma}
  \mathcal{H}_3 \wedge \star_{\Sigma} \mathcal{H}_3,
\end{equation}
for some $\mu_5$ and $k_5$ to be determined, where the $A_6$ charge coupling is rendered gauge invariant by the transformation of the worldvolume field $B_2 \rightarrow B_2 - \lambda_2$ under the gauge transformation $A_3 \rightarrow A_3 + \rmd \lambda_2$.

Consistency of the resulting equation of motion $\rmd \star \tilde{F}_7 = -2 \kappa_{11}^2 \mu_5 \delta (\Sigma)$ with the Bianchi identity $\rmd F_4 = q \delta(\Sigma)$ and the constraint $\star \tilde{F}_7 = - F_4$ implies $\mu_5 = \frac{q}{2\kappa_{11}^2}$.
Likewise, we find 
\begin{equation}
  \rmd \, {\star F_4} = - \frac{1}{2} F_4 \wedge F_4 + \frac{q}{4} \mathcal{H}_3
  \wedge \delta (\Sigma) - 3 \kappa_{11}^2 k_5 \star_{\Sigma} \mathcal{H}_3
  \wedge \delta (\Sigma) .
\end{equation}
Consistency with~\eqref{eqn:M5M2charge} and $\star F_4 = \tilde{F}_7$ requires that
\begin{equation}
  3 \kappa_{11}^2 k_5 \star_{\Sigma} \mathcal{H}_3 = - \frac{3 q}{4}
  \mathcal{H}_3 \qquad \Longrightarrow \qquad k_5 = \frac{q}{4 \kappa_{11}^2}, \qquad \star_{\Sigma} \mathcal{H}_3 =
  -\mathcal{H}_3 ,
\end{equation}
where we assume $q>0$ for definiteness, and the constraint $\star_{\Sigma} \mathcal{H}_3 =
  -\mathcal{H}_3$ is consistent with the equation of motion $\rmd \star_{\Sigma} \mathcal{H}_3 = - F_4$. Thus, the brane carries a worldvolume chiral boson, just like a standard M5-brane. Just as in the discussion in Section~\ref{ssec:gaugeandbreak}, this boson can also be thought of as a Nambu-Goldstone mode.
  
Once such a five-brane has been introduced, there are no remaining (continuous)
global symmetries. Note that, of the brane localized currents, $\delta
(\Sigma)$ is gauged by the $F_4$ Bianchi identity, whereas $\mathcal{H}_3
\wedge \delta (\Sigma)$ is broken, gauging $F_4 \wedge \delta (\Sigma)$ in the
process.

In particular, there is no need to explicitly include M2-branes, as their
charge is carried by the chiral boson on the M5-branes, as well as being
induced by the bulk Chern-Simons term. If M2-branes are included, they modify
the $\tilde{F}_7$ Bianchi identity:
\begin{equation}
  \rmd \tilde{F}_7 = - \frac{1}{2} F_4 \wedge F_4 + q\mathcal{H}_3 \wedge \delta
  (\Sigma_5) + q' \delta (\Sigma_2) .
\end{equation}
Now $J_8 = - \frac{1}{2} F_4 \wedge F_4 + q\mathcal{H}_3 \wedge \delta
(\Sigma_5)$ is conserved but no longer gauged, so we seem to have a global
symmetry. To avoid this, M2-branes must be allowed to end on the M5-branes,
inducing worldvolume flux
\begin{equation}
  \rmd \mathcal{H}_3 = F_4- \frac{q'}{q} \delta_{\Sigma_5} (\partial \Sigma_2).
\end{equation}
Exchanging a spatial boundary for a temporal boundary, this implies that M2-branes are dissolvable in M5-branes. Thus, in some sense M2-branes are nothing
but worldvolume flux concentrated at a point, as any distinction between these
two objects would create a global symmetry. (Whether such concentrated flux is
energetically favorable is a dynamical question that cannot be answered by an
analysis of the global symmetries.)

\subsubsection{Comparison with Type IIA}

It is interesting to compare the discussion of eleven-dimensional
supergravity / M-theory above with the preceding discussion of Type IIA
supergravity / string theory, since the latter is the circle compactification
of the former. Eleven-dimensional supergravity compactified on a circle
automatically includes D0 branes (Kaluza-Klein modes) and D6-branes
(Kaluza-Klein monopoles). As discussed above, once D6-branes are included, we
automatically obtain objects carrying D4, D2, and D0 charge due to the worldvolume
gauge field on the D6. Thus, the only remaining global symmetry current is
$H_3$, broken by NS5-branes in string theory. Since Type IIA NS5-branes arise
from M5-branes transverse to the M-theory circle, this story is closely
analogous to the eleven-dimensional viewpoint. There is one global symmetry
current ($F_4$ or $H_3$, respectively), broken by including a single type of
brane (an M5-brane or NS5-brane, respectively), which is forced to carry a
worldvolume gauge field to respect bulk gauge invariance.

However, closer examination yields an interesting puzzle. In the case with a
global symmetry, hence no M5/NS5-brane, the ten-dimensional theory has a
D4-brane-like object (concentrated D6 worldvolume flux), even though it is
well known that D4-branes arise from M5-branes wrapping the M-theory circle.
Is there therefore some solitonic M5-like object in eleven-dimensional
supergravity, obviating the need to include a fundamental magnetic brane?

Let us give a simple argument that this \emph{is not} the case, and then
explain how this can be consistent with the ten-dimensional description. A
solitonic (horizon-free) M5-like object would have long-range fields
resembling those of an M5-brane. Let $\Omega_{11}$ be the region of spacetime
near the solitonic core. By extending $\Omega_{11}$ far enough away from the
solitonic core, we conclude that $\partial \Omega \cong \mathbb{R}^{5, 1}
\times S^4$ to reproduce the long-range behavior of an M5-brane. Assuming a
static, uniform solution, $\mathbb{R}^{5, 1}$ factors out entirely, so that
$\Omega_{11} =\mathbb{R}^{5, 1} \times \Omega_5$ and $S^4 = \partial
\Omega_5$. The M5 charge is measured by the integral
\begin{equation}
  Q_{\mathrm{M5}} = \oint_{S^4} F_4 = \int_{\Omega_5} \rmd F_4 = 0,
  \label{eqn:M5charge}
\end{equation}
using Stokes' theorem. Therefore, the pure supergravity equation $\rmd F_4 =
0$ precludes M5 charge.

Now let us turn to D4-brane charge dissolved in a D6-brane. We could measure
the presence of D4 charge using the ``Maxwell charge'' \cite{Marolf:2000cb}
\begin{equation}
  Q_{\mathrm{D4}}^{\text{Max}} = \oint_{S^4} G_4,
\end{equation}
but $G_4$ is not closed ($\rmd G_4 = H_3 \wedge G_2$), so the result depends
on the choice of linking sphere. Instead, consider the ``Page charge''
\begin{equation}
  Q^{\text{Page}}_{\mathrm{D4}} = \oint_{S^4} [G_4 + H_3 \wedge C_1] .
  \label{eqn:PageCharge}
\end{equation}
Provided that the $S^4$ does not intersect any D4-branes, or NS5-branes with worldvolume flux, the integrand is closed, so the result does not depend on the choice of
linking sphere. Although the same integral over a generic cycle would depend on large gauge transformations of
$C_1$, this dependence disappears when $H_3$ is topologically trivial (exact) on the cycle. This is always true on $S^4$ because $H^3(S^4) = 0$, so~\eqref{eqn:PageCharge} is well defined.

The relation between the 10d and 11d fields is
\begin{equation}
  F_4^{(11)} = G_4^{(10)} + H_3^{(10)} \wedge (\rmd y + C_1^{(10)}) .
\end{equation}
Thus, naively (\ref{eqn:PageCharge}) lifts to (\ref{eqn:M5charge}) where the
integration cycle is chosen by picking an arbitrary point on the M-theory
circle for each point on the 10d $S^4$. How, then, is it possible to have
$Q_{\mathrm{D4}} \neq 0$ when (as we argued) $Q_{\mathrm{M5}} = 0$?

\begin{figure}
\begin{center}
\includegraphics[width=0.85\textwidth]{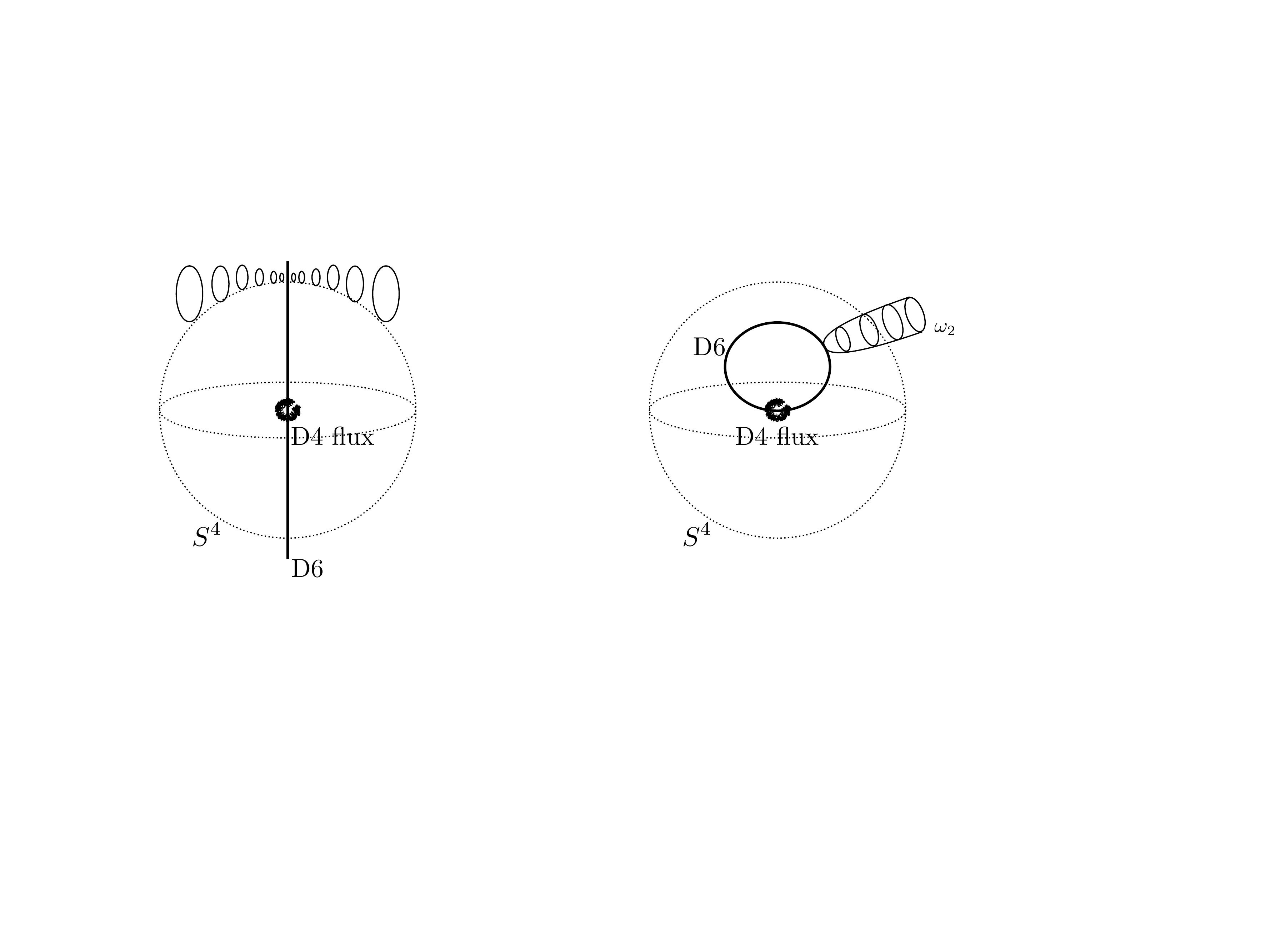}
\caption{The left panel depicts a D6-brane with dissolved D4-brane charge, represented as a blurry core.  The D4 charge is measured by the flux on a linking $S^4$, represented by the dotted sphere. This cycle does not uplift to an $S^4$ in eleven dimensions because the circle fibration over $S^4$ has no global section, as it is the double suspension of the Hopf fibration. In the right panel, we now have a spherical D6-brane. The $S^4$ no longer intersects the D6 worldvolume, so the fibration is trivial over $S^4$ and the cycle has an 11d uplift. This $S^4$ is nontrivial in homology (since it intersects the 2-cycle $\omega_2$ involving the M-theory circle and a radial direction, as shown in the figure). }
\label{fig:Mth}
\end{center}
\end{figure}

To answer this question, we consider two distinct cases of interest. First,
suppose that the D6-brane carrying the D4-charge is an infinite plane
$\mathbb{R}^{6, 1}$. In this case, because of the non-trivial graviphoton
gauge field, the transverse $S^4$ does not lift to $S^4 \times S^1$, but
rather to $S^5$.\footnote{The linking $S^2$ of the D6-brane lifts to an $S^3$
via the Hopf fibration, and so the $S^1$ fibration over $S^4$ is the double suspension of the Hopf fibration, with total space $S^5$. As a result, the eleven-dimensional space is topologically
$\mathbb{R}^{10, 1}$.} In particular, the $S^1$ fibration over $S^4$ does not have a
global section. This means that there is no four-cycle lift of the linking $S^4$ over which
to integrate $F_4$, and hence no direct connection to $Q_{\mathrm{M5}} = 0$. This is depicted in Figure \ref{fig:Mth}.

To eliminate the D6 charge, and with it the non-trivial graviphoton gauge
bundle, we instead consider a D6-brane with worldvolume $\mathbb{R}^{4, 1}
\times S^2$ and one unit of worldvolume flux threading the sphere. In the
limit where the $S^2$ shrinks to zero size, this configuration looks
indistinguishable from a D4-brane. Taking the transverse $S^4$ to have a
larger radius than the $S^2$, the M-theory lift is just the naive $S^4 \times
S^1$, so now (\ref{eqn:PageCharge}) does lift to (\ref{eqn:M5charge}).
However, because of the non-trivial topology induced by the circle fibration
in eleven dimensions, this $S^4$ is nontrivial in homology, hence Stokes' theorem
does not apply and $Q_{\mathrm{M5}} \neq 0$ is possible. To see that this is so, consider the two
cycle $\omega_2$ formed by the $S^1$ fibered over a line stretching from the D6-brane off
to infinity. The intersection number of this two-cycle with $S^4$ is one,
hence $S^4$ is nontrivial in homology, as shown in Figure \ref{fig:Mth}.

In fact, the M-theory lift of dissolving a D4-brane in the D6-brane is the
process of moving an M5-brane wrapping the $S^1$ on top of the D6-brane, where
the $S^1$ shrinks to zero size and the M5-brane self-annihilates, leaving
behind $F_4$ flux, where the linking between the $S^4$ and this two-cycle
traced out by the M5-brane reflects the change in flux as the brane is brought
in from a distance.

Thus, in summary, although M5-branes do not arise in pure eleven-dimensional
supergravity, D4-brane-like objects (with the charges of an M5-brane wrapping
the M-theory circle) do. Despite all these complications, the presence or
absence of global symmetries lines up perfectly between the different dimensions, given
a sufficiently careful treatment.

\subsection{Rigid Calabi-Yau Three-folds}
\label{sec:rigid}

In most examples of $U(1)$ gauge fields in string theory we have seen, a combination
\beq
J_4' = \frac{1}{4 \pi^2} F_2 \wedge F_2 + \frac{1}{2 \pi} j_3^{\rm m} \wedge \rmd_A \sigma,
\eeq
of the Chern-Weil current associated to a bulk $U(1)$ gauge field and the degrees of freedom on the monopole is conserved, according to the discussion in Section \ref{ssec:gaugeandbreak}. Indeed, in these cases, $J_4'$ is frequently gauged by coupling to a dynamical $(d - 4)$-form gauge field $C$. In $d = 4$, this is a manifestation of the Witten effect, and our dynamical gauge field is simply an axion field $\theta$. In this way, the examples we have considered explain the presence of a dynamical axion in $d = 4$ string models by requiring the remaining $(-1)$-form symmetry $J_4'$ to be gauged.

In this section, we consider a striking exception to this pattern, with no dynamical axion among the light fields of the theory. This is the case of Type IIB string theory compactified on a rigid Calabi-Yau three-fold,\footnote{That is, one with $h^{2, 1} = 0$.} as considered in \cite{Cecotti:2018ufg} . Such a model describes a theory of $\mathcal{N} = 2$ supergravity in $d = 4$ with only one $U(1)$ gauge field (the graviphoton) and no vector multiplets. In particular, as the $\theta$-angle can only depend on scalars in vector multiplets, not those in hypermultiplets, we see that these models must have $\theta$ frozen to a specific value in the IR. Interestingly, the authors of \cite{Cecotti:2018ufg} find that all known examples of rigid Calabi-Yau three-folds have $\theta = 0$ or $\pi$ up to an $SL(2, \mathbb{Z})$ transformation.

In these examples, we claim that the Chern-Weil current $F_2 \wedge F_2$ of the graviphoton is simply broken by monopoles. Indeed, though the $\theta$-angle appears to be a free parameter in the IR, theories with $\theta \neq 0$ or $\pi$ up to $SL(2, \mathbb{Z})$ apparently do not admit a UV completion, and as such it would be inconsistent to couple them to a dynamical axion, just as it would be inconsistent to couple a theory with an explicitly broken global symmetry to a dynamical gauge field. The fact that we find both possibilities $\theta = 0, \pi$ is consistent with the equation
\beq
\rmd \left(\frac{F_2}{2 \pi} \wedge \frac{F_2}{2 \pi} \right) = 2 \left(\frac{F_2}{2 \pi} \wedge j_3^{\rm m}\right),
\eeq
obtained by dividing \eqref{breaking_by_2} by $(2 \pi)^2$, which tells us that monopoles only violate Chern-Weil charge conservation by two, and thus only break the $U(1)$ $(-1)$-form symmetry to a $\mathbb{Z}_2$ subgroup.\

While these theories provide examples of four-dimensional compactifications of string theory with no axion in the deep IR, we should note that there will still be a heavy axion, given by a KK mode of the ten dimensional metric. Thus, the statement that the Chern-Weil symmetry $F_2 \wedge F_2$ is simply broken by monopoles holds only below the KK scale, while we expect a more typical story involving both an axion and a conserved linear combination of currents to hold above the KK scale. On the other hand, our four dimensional description of physics also breaks down once we start exciting many KK modes, so these models provide an example of a theory with no axion whose mass is below the EFT cutoff.

\subsection{Taking Stock}

Having surveyed a number of examples in string theory and M-theory, we can now step back and offer a broader perspective on what we have learned. Consistent with the lore about quantum gravity, we have not found any examples of true Chern-Weil global symmetries in a gravitational setting. We have found that Chern-Weil symmetries can simply be broken by the existence of magnetically charged objects, as  in Section~\ref{sec:rigid}. However, we find that more often, the Chern-Weil symmetries are gauged, via Chern-Simons terms. To be more precise, the gauging is usually of a linear combination of the Chern-Weil current and other currents, with other linear combinations being broken by various dynamical processes that can convert Chern-Weil charge into other charges. Whereas the presence of Chern-Simons terms in string theory might have been thought of as a requirement of spacetime supersymmetry, or as a derived consequence of the string worldsheet theory, we see that these terms also play an important conceptual role unrelated to such details, as already pointed out in \cite{Montero:2017yja}: they serve to banish a global symmetry from the theory, by gauging it.

In quantum gravity, we expect that abelian gauge theories will always have $\rmd F \neq 0$, due to the existence of magnetically  charged objects. As we first saw in our discussion of the 't~Hooft-Polyakov monopole in Section~\ref{ssec:gaugeandbreak}, this means that a Chern-Simons coupling of the form $C \wedge F \wedge F$ is consistent only in the presence of worldvolume degrees of freedom on the magnetically charged object, which allow us to dissolve electric charge inside that object. In string theory examples (e.g., with Ramond-Ramond gauge fields), electrically and magnetically charged objects are frequently different types of branes. Consistency of the bulk Chern-Simons terms requires that one kind of brane charge can be dissolved inside another kind of brane. If a charge can be carried either by branes moving freely or in a dissolved form inside another brane, there must be dynamical processes that can convert between the free charge and the dissolved charge. Otherwise, we would have a global symmetry, due to our ability to count these two kinds of charges separately (with only one linear combination being gauged). Thus, we see that the well-known string theoretic phenomenon of branes dissolving in other branes can be viewed as a mechanism by which the theory avoids unwanted global symmetries. This also has the consequence that a brane can {\em end} on another brane, with the junction of the branes serving as the location where free charge is converted into dissolved charge. Thus, much of the intricate structure of branes and their interactions within string theory can be viewed as providing a mechanism for removing potential global symmetries from the theory. This suggests that this structure could survive beyond the known string lampposts, as a general feature of theories of quantum gravity.

Another pattern we have observed in Type II string theory and M-theory is that, while many Chern-Weil symmetries are broken by Chern-Simons couplings, in each example we find at least one current that is unbroken at the level of the bulk supergravity action, corresponding to the required branes of the highest dimension. These are the D6-branes and NS5-branes in massless Type IIA, the D8-branes and NS5-branes in massive Type IIA, the D5-branes and NS5-branes in Type IIB with $SL(2, \mathbb{Z})$ ungauged, the 7-branes in Type IIB with $SL(2, \mathbb{Z})$ gauged, and the M5-brane in M-theory. Once these branes are added, branes of lower dimension may then be inferred as dissolved objects by the logic discussed in the previous paragraph. In this way, the highest-dimensional branes may be viewed as generating the full spectrum of dynamical branes. It would be interesting to provide a more complete description of this structure in general.

Finally, we note that in abelian gauge theories the Completeness Hypothesis, i.e., the existence of both electrically and magnetically charged objects spanning the full lattice of charges, can be viewed as a consequence of the absence of global symmetries associated with the currents $F$ and $\star F$. Our results suggest that the absence of global symmetries associated with higher Chern-Weil currents leads to a stronger conclusion than the Completeness Hypothesis, telling us not only about the existence of charged objects but also about relationships between them (e.g., the fact that one object can end on another). There is a close affinity between these remarks and the concept of higher-group symmetries (see, e.g., \cite{Cordova:2018cvg, Hidaka:2020iaz, Hidaka:2020izy, Brennan:2020ehu}).

\section{AdS/CFT}\label{sec:adscft}
We have seen how Chern-Weil symmetries can be gauged, broken, or a combination of both. We will now discuss examples of these structures in quantum gravity in AdS, where the dual CFT provides a novel perspective on some of these phenomena, often involving anomalies of global symmetries. More precisely, we will find that often, but not always, the Chern-Simons terms that are associated with gauging of Chern-Weil symmetries correspond to 't Hooft anomalies of the dual field theory. We will start with quantum gravity in AdS$_7$ and then briefly comment on the more familiar AdS$_5$/CFT$_4$ example.

\subsection{AdS$_7$/CFT$_6$}\label{subs:ads7}

All known six-dimensional conformal field theories are supersymmetric. These superconformal field theories (6D SCFTs) may have either $\mathcal{N}=(2,0)$ supersymmetry or $\mathcal{N}=(1,0)$ supersymmetry. We will be interested in the latter.  

6D SCFTs have been studied by many groups, dating back to the 1990s (see, e.g., \cite{Ganor:1996mu, Seiberg:1996qx, Seiberg:1996vs, Witten:1996qb}).
More recently, a classification program of 6D SCFTs has been undertaken \cite{Heckman:2013pva, Heckman:2015bfa, Bhardwaj:2015xxa, Heckman:2018pqx}, involving a combination of field theory constraints and F-theory constraints (see \cite{Heckman:2018jxk} for a review). This classification encompasses all known 6D SCFTs,\footnote{Up to subtleties involving frozen singularities in F-theory, which are now reasonably well understood \cite{Tachikawa:2015wka, Bhardwaj:2018jgp, Bhardwaj:2019hhd}.} though it relies on the assumption that each SCFT admits a tensor branch: after giving vevs to scalar fields in tensor multiplets (each of which features a 2-form gauge field $B_{\mu\nu}$ under which the dynamical strings are charged), one may perform an RG flow whose endpoint is a free field theory in the IR. The classification program then proceeds by classifying these free field theories, all of which take a rather simple form as a generalized quiver gauge theory. 

Of particular interest for us are those 6D SCFTs whose tensor branches can be described in terms of an $SU(N_i)$ quiver gauge theory. These theories may be constructed in Type IIA string theory via D6-D8-NS5-brane webs \cite{Hanany:1997gh} and may be realized as the IR fixed points obtained after giving vevs to hypermultiplets in the worldvolume theory of M5-branes probing $\mathbb{C}^2/\mathbb{Z}_n$ singularities. The IIA brane setup is
\begin{equation}
\begin{array}{c|c|c|c|c|c|c|c|c|c|c}
\text{Brane}&x_0&x_1&x_2&x_3&x_4&x_5&x_6&x_7&x_8&x_9\\\hline
\text{D}8&-&-&-&-&-&-&\times&-&-&-\\
\text{NS}5&-&-&-&-&-&-&\times&\times&\times&\times\\
\text{D}6&-&-&-&-&-&-&-&\times&\times&\times
\end{array}
\end{equation}
These theories have quivers of the form
\begin{equation}
\includegraphics[width=60mm]{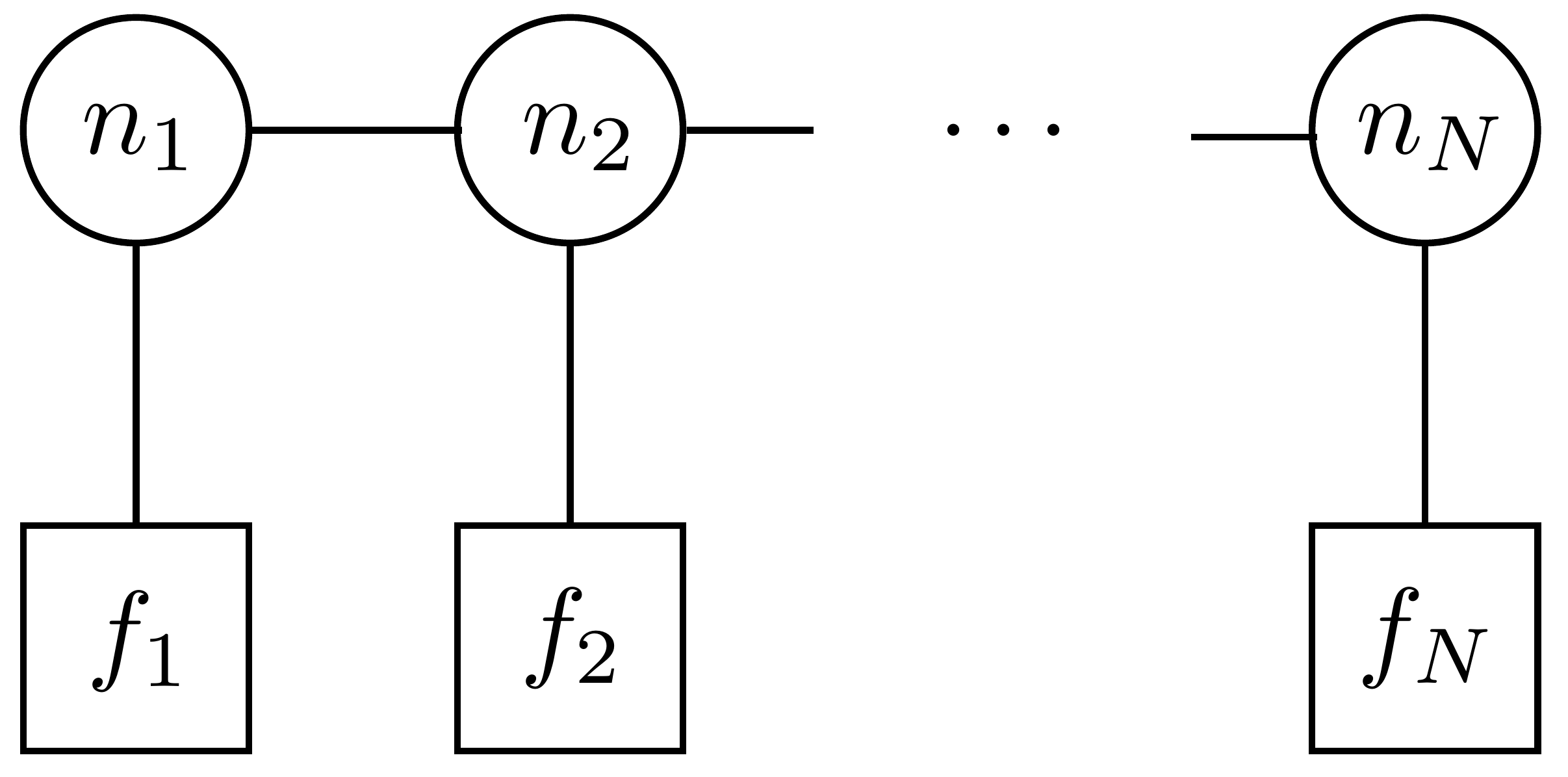}
\label{eq:Anquiver}
\end{equation}
Here, $\mathfrak{su}(n_I)$ is the $I$'th gauge algebra,
and $f_I$ indicates the number of flavors associated with the $I$'th gauge group.
There are also bifundamental hypermultiplets $(\mathbf{n}_I, \overline{\mathbf{n}}_{I+1})$ stretching between each pair of adjacent gauge groups. Anomaly cancelation, which is very stringent in six dimensions, dictates
\begin{equation}
2 n_I = n_{I-1} + n_{I+1} + f_I,
\end{equation}
with $n_0 = n_{N+1} :=0$. This condition implies that the hypermultiplet content of the theory is uniquely specified by the Dirac pairing and the ranks $n_I$ of the gauge groups, so the integers $f_I$ are in fact redundant, and often omitted from the quiver description in (\ref{eq:Anquiver}).

In the Type IIA picture, the gauge symmetries $\mf{su}(n_I)$ are associated with the worldvolume of a stack of $n_I$ D6-branes stretched between a pair of NS5-branes. The flavors $f_I$ are associated with a stack of $f_I$ D8-branes that intersects the stack of $n_I$ D6-branes. The separation between the $I$'th and $I+1$'st NS5-branes corresponds to the vev of the scalar in the $I$'th tensor multiplet. Moving to the conformal fixed point requires taking all of these vevs to vanish, and therefore coalescing all of the NS5-branes into a single stack.

Each stack of $f_I$ D8-branes gives rise to a $U(f_I)$ flavor symmetry of the 6D SCFT. However, as shown in \cite{Apruzzi:2020eqi}, a certain ``center-of-mass'' $U(1)$ suffers from an ABJ anomaly, and as a result the full global symmetry of the theory is reduced to $S[\prod_I U(f_I)] \times SU(2)_R$. At the level of the algebra, this can be written 
\begin{equation}
\bigoplus_{i=1}^m \mathfrak{u}_1^{(i)} \oplus \bigoplus_{J=1, f_J \geq 2}^N  \mathfrak{su}(f_J) \oplus \mathfrak{su}(2)_R,
\label{eq:fullglobal}
\end{equation}
where $m+1$ is equal to the number of $f_J$ satisfying $f_J \geq 1$, which is in turn equal to the number of D8-brane stacks. The anomaly polynomial $I_8$,
which is an 8-form that may be calculated for any known 6D SCFT by the prescription of \cite{Ohmori:2014kda}, 
contains terms of the form\footnote{There is also a term of the form $\int C_3\Tr F_{J}^3$, which naively would seem to produce an additional $U(1)$ by reducing $C_3= A_{C_3}\wedge \omega_2$ with $\omega_2$ the volume form of the sphere. However, this $U(1)$ actually confines and is gapped \cite{Bergman:2020bvi},  so we will not include it in the discussion below.}
\begin{equation}
I_8 \supset  \sum_{J=1, f_J \geq 3}^N a_{iJ} F_{i} \Tr F_{J}^3,
\label{eq:cubicanomaly}
\end{equation}
where the requirement $n_J \geq 3$ comes from the fact that $\Tr F^3 = 0$ for any simple Lie algebra except $\mathfrak{su}(n)$ with $n \geq 3$. Since there are $m+1$ D8-brane stacks, there can be at most $m+1$ nonabelian field strengths $F_J$ with nonvanishing $\Tr F_J^3$, whereas there are $m$ $\mf{u}_1$ factors, each of which couples to a distinct linear combination $\sum_J c_J \Tr F_{J}^3$ of nonabelian field strengths. As a result, every linear combination $\sum_J c_J \Tr F_{J}^3$ has a nonzero $F_{U(1)} (\sum_J c_J \Tr F_{J}^3)$ anomaly with the possible exception of one linear combination associated with the anomalous center-of-mass $U(1)$.

\begin{figure}
\begin{center}
\includegraphics[width=65mm]{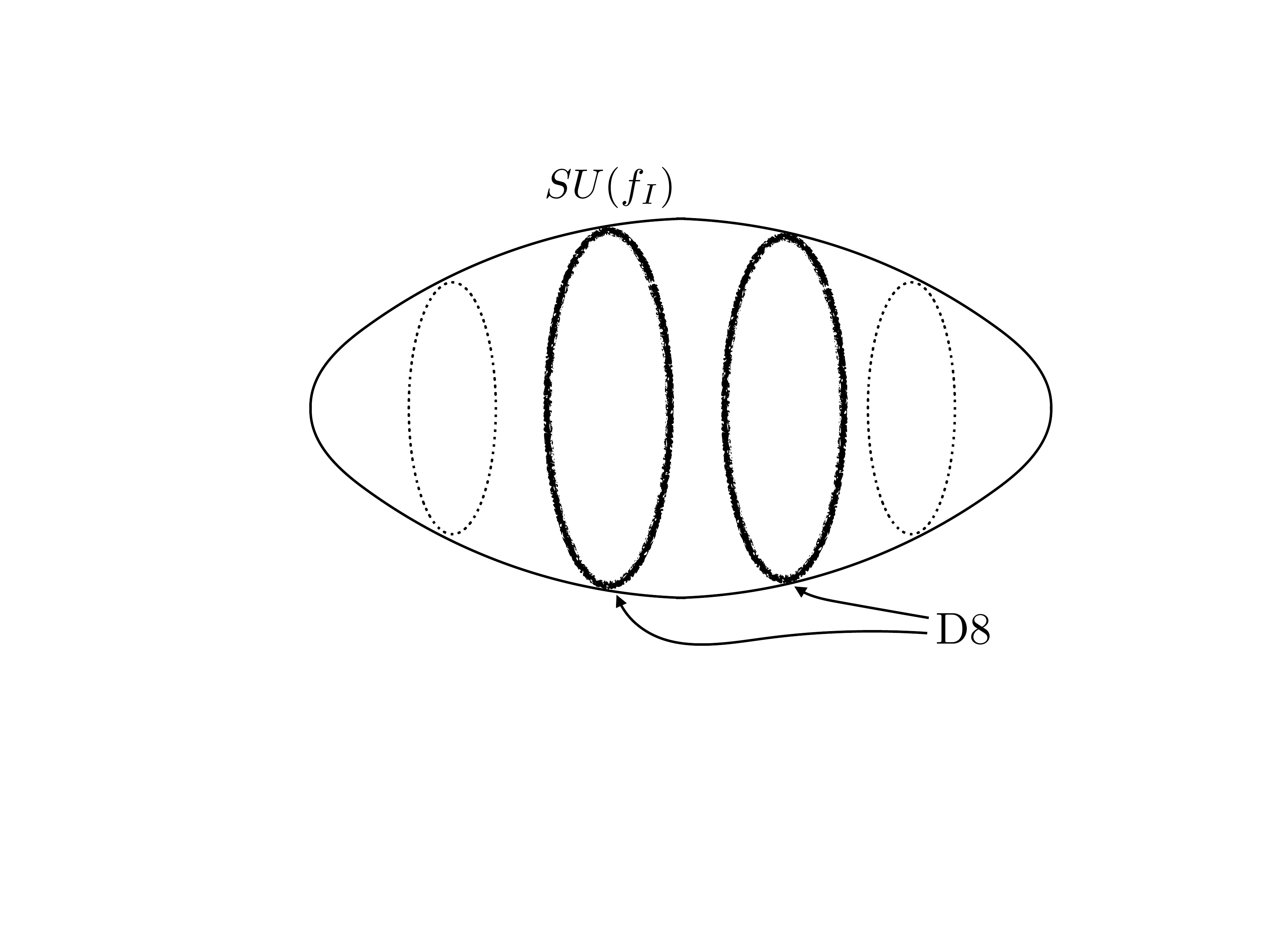}
\caption{Schematic representation of the compactification space $M_3$ (see also \cite{Cremonesi:2015bld, Apruzzi:2017nck}). It looks like an American football, with stacks of D8-branes at particular positions wrapping the angular $S^2$. A stack of $f_I$ branes provides an $SU(f_I)$ symmetry at low energies.}
\label{fig:football}
\end{center}
\end{figure}

Holographic duals of the D6-D8-NS5 $\mathcal{N}=(1,0)$ 6D SCFTs have been studied in \cite{Apruzzi:2013yva, Gaiotto:2014lca, Cremonesi:2015bld, Apruzzi:2015wna, Apruzzi:2017nck, Bergman:2020bvi}. The holographic dual theories live on the bulk spacetime AdS$_7 \times M_3$, where $M_3$ has $SU(2)$ isometry, corresponding to the $SU(2)_R$ symmetry on the boundary. $M_3$ looks like an American football, as illustrated in Figure \ref{fig:football}. The different stacks of D8-branes wrap around $S^2$'s along the symmetry axis of the football at different positions. Each $SU(f_I)$ global symmetry in the boundary SCFT corresponds to an $SU(f_I)$ gauge symmetry in the bulk. We are especially interested in the $\Tr F_I^3$ Chern-Weil currents associated with each of these symmetries. In particular, we want to know if every such current couples to an abelian gauge field in the bulk via an $A_i \wedge \Tr F_I^3$ Chern-Simons coupling. Indeed, such a coupling arises from the CS term of the $J$'th D8-brane stack \cite{Bergman:2020bvi}:
\begin{equation}
S_{\text{CS}} = T_8 \int C \wedge \Tr \left( \mathrm{e}^{2 \pi F_J + B_2} \right),\label{csacb}
\end{equation}
where $C$ is the formal sum of all RR potentials. Expanding the exponential, we find a coupling of the form
\begin{equation}
\int C_1 F_j \Tr F_J^3, 
\end{equation}
where $F_J$ represents the $SU(n_J)$ field strength and $F_j$ represents the $U(1)$ field strength that together combine into the full $U(n_J)$ gauge group of the brane. In the case at hand, the D8-brane worldvolume is AdS$_7 \times S^2$, and after integrating by parts and reducing on the $S^2$, we obtain terms in the 7d action of the form\footnote{We thank Fabio Apruzzi and Marco Fazzi for discussions on this point.}
\begin{equation}
\mathcal{L}_{\text{AdS}_7} \supset \left( \int_{S^2} F_2 \right) A_j \wedge \Tr F_J^3 ,
\label{eq:AdSterm}
\end{equation} 
where $F_2 = \rmd C_1$, and we are ignoring an overall normalization factor. As explained in \cite{Bergman:2020bvi}, both $C_1$ as well as the ``center-of-mass'' gauge $U(1)$ obtain a mass through the Stueckelberg mechanism, the latter of which is the bulk dual to the ABJ anomaly that removes the center-of-mass $U(1)$ in the boundary SCFT. What remains are precisely the $m$ massless $U(1)$ vector bosons (associated with $m+1$ D8-brane stacks) that we had in (\ref{eq:fullglobal}). The CS coupling in (\ref{eq:AdSterm}) cancels the anomaly (\ref{eq:cubicanomaly}) via anomaly inflow, so the aforementioned presence of the $F_{U(1)} (\sum_J c_J \Tr F_{J}^3)$ couplings in the anomaly polynomial ensures the presence of these CS couplings in the bulk theory. To be more precise, since every linear combination $\sum_J c_J \Tr F_{J}^3$ has a nonzero $F_{U(1)} (\sum_J c_J \Tr F_{J}^3)$ term in the anomaly polynomial, with the possible exception of the center-of-mass linear combination, every Chern-Weil current $\sum_J c_J \Tr F_{J}^3$ in the bulk theory has a nonzero Chern-Simons coupling $A_{U(1)} (\sum_J c_J \Tr F_{J}^3)$, with the possible exception of the center-of-mass linear combination, which couples to an anomalous (massive) vector gauge boson. As a result, every $\Tr F_{J}^3$ Chern-Weil current in the bulk is gauged by an abelian gauge field, at most one of which may be massive. The charge coupled to the massive $U(1)$ is only conserved at energy scales above the vector mass; the symmetry is broken at low energies.

Although we have considered here only a special class of 6D SCFTs constructible in Type IIA string theory, the conclusions are likely quite general. This is because the only $\mathfrak{su}(n), n \geq 3$ global symmetries known to arise in 6D SCFTs arise in the tensor branch description as (i) flavor symmetries of fundamentals of $\mf{su}(m), m \geq 2$ gauge algebras, (ii) flavor symmetries of fundamentals of $\mf{e}_6$ gauge algebras, (iii) flavor symmetries of spinors of $\mf{so}(10)$ gauge algebras, or (iv) leftover global symmetries after gauging part of the $E_8$ global symmetry of an E-string SCFT \cite{Ganor:1996mu}. Our above analysis carries over to symmetries of type (i) and extends straightforwardly to those of type (ii) and type (iii), since the flavor symmetry of $n$ fundamentals of $\mf{e}_6$ or $n$ spinors of $\mf{so}(10)$ is in fact $\mf{su}(n) \oplus \mf{u}(1)$, and the $\mf{su}(n)$ field strength $F$ couples to the $\mf{u}(1)$ field strength $F_{U(1)}$ via a $F_{U(1)} \,\Tr F^3$ term in the anomaly polynomial, just as we have seen in the theories studied above. Our analysis does not generalize so straightforwardly to case (iv). A better understanding of the holographic duals of theories with such global symmetries is required before an analysis of their $\Tr F^3$ Chern-Weil currents may be carried out.

We may also consider $\tr F_I^2$ Chern-Weil currents, involving two powers of the gauge fields. In this scenario, gauging would imply a coupling to a  massless three-form.  On the other hand, onee simple way to break these Chern-Weil symmetries is coupling to a massive three-form, of mass $\mu$. The equation of motion becomes
\begin{equation} \rmd \, {\star F_4}= \mu F_4 + \tr F_I^2.\end{equation}
 At scales below $\mu$, the symmetry appears to just be broken. Above $\mu$ we have an example of gauging-breaking as in Section \ref{ssec:gaugeandbreak}, where the Chern-Weil charge can transition into $F_4$ flux. 

 We will now see that all these Chern-Weil symmetries are coupled to massive 3-forms of this kind. The D8-branes in the brane construction remain as sources in the AdS$_7\times M_3$ bulk, wrapping AdS$_7\times S^2$. The Chern-Simons action \eqref{csacb} includes worldvolume couplings 
 \begin{equation} \int_{\text{D}8} \tr F_I^2\wedge C_5,\end{equation}
 which after reduction become 7d couplings to a 3-form $\int_{S^2} C_5$. This is the magnetic dual to the RR potential $C_1$ discussed earlier, so it is massive for the same reasons. 
 
However, $C_3$ only couples in this way to the linear combination
\begin{equation}\sum_I \tr F_I^2.\end{equation}
What happens to the other combinations? They too are broken, since a configuration with nonzero $\tr F_I^2$ in the worldvolume of a D8-brane can be shrunk to a pointlike D4-brane and pulled from the D8-brane worldvolume and into any other D8-brane worldvolume, 
as discussed in Section \ref{subsec:openstringcurrents}. This is possible because the different D8-brane stacks are wrapped on the same homology cycle, as depicted in Figure~\ref{fig:football}.

There is one more Chern-Weil symmetry we must discuss: We can also take $\tr F^2$ for the $\mathfrak{su}_2$ R-symmetry, which corresponds to rotations of the transverse $S^2$. The explanation in this case is slightly more complicated. As explained in \cite{Harvey:1998bx} in the context of M-theory compactifications to seven dimensions, dimensional reduction in gauged supergravity becomes subtle, and in the particular case of M-theory on a four-manifold the field strength $G_4$ of the M-theory 3-form picks up a piece proportional to the Euler class of the tangent bundle of $M_4$. This means that the triple Chern-Simons term in M theory
\begin{equation}\int C_3\wedge G_4 \wedge G_4\label{3csm}\end{equation}
 will produce, under dimensional reduction on a fluxed background, terms involving the Chern-Weil current. In \cite{Harvey:1998bx}, this was used to argue for the existence of certain topological couplings in the 7d effective action, such as
 \begin{equation}\int \tr (F\wedge F\wedge F\wedge A)+\ldots\end{equation}
 where $A$ are gauge fields for the isometries of $M_4$. The same argument leads to additional topological couplings, not considered in \cite{Harvey:1998bx} because they involve massive fields and so are not visible at the level of the supergravity action. In particular, by picking
 \begin{equation} G_4= \tr( F^2) + \omega_4\label{w333}\end{equation}
 where $\omega_4$ is proportional to the volume form of $M_4$ and represents the background $G_4$ flux, one finds a term in the effective action
 \begin{equation}\int_{\text{AdS}_7}  \tr( F^2)\wedge C_3.\end{equation}
This is a coupling to a massive three-form, since there is also a coupling of the form $\int C_3 \wedge G_4$ in seven dimensions. Thus, the symmetry is broken. 

The analysis in \cite{Harvey:1998bx} was focused on compactifications of M-theory to seven dimensions, but a variant of it involving the IIA Chern-Simons term  (which is the dimensional reduction of \eqref{3csm})
\begin{equation} \int H_3\wedge F_4 \wedge C_3.\label{dinn}\end{equation}
presumably applies in our case. One would need to argue that the IIA field strength $F_4$ picks a piece proportional to the Chern-Weil current, as in \eqref{w333}; this again follows from application of the M theory argument (itself based on fivebrane anomaly cancellation \cite{Freed:1998tg}) to a circle compactification.

\subsection{AdS$_5$/CFT$_4$}
We now consider the holographic model of AdS$_5$/CFT$_4$. The 10d IIB triple-Chern-Simons term
\begin{equation} \int F_5 \wedge F_3 \wedge B_2\end{equation}
becomes, after compactification on the $S^5$ with $N$ units of five-form flux,
\begin{equation}N \int F_3 \wedge B_2.\label{csf}\end{equation}
This is a two-field topological term, a five-dimensional analog of the BF coupling discussed in Section \ref{sec:BFgauging}. Most of the discussion from that section carries over to the case at hand, with the caveat that the theory involves two 2-form gauge fields rather than one 1-form gauge field and one 2-form gauge field, so it describes a discrete $\mathbb{Z}_N$ 1-form gauge symmetry instead of a 0-form gauge symmetry (see \cite{Hofman:2017vwr} for a thorough discussion). This topological field theory is also related to the singlet sector of the duality, where different boundary conditions are mapped to the global structure of the dual gauge group \cite{Aharony:1998qu, Witten:1998wy}. A consistent choice of boundary conditions means that only one linear combination of $B_2$ or $C_2$ gives rise to 1-form symmetry operators in the dual field theory; different choices of boundary conditions correspond to different CFT's, and changes of boundary conditions are related to gauging this 1-form symmetry (see Section 6 of \cite{Gaiotto:2014kfa}, and \cite{Aharony:2016kai} for a particularly clear discussion).

More mundane examples of Chern-Weil symmetries include the ``baryonic'' symmetries present in compactifications on Sasaki-Einstein manifolds \cite{Gubser:1998fp,Butti:2005sw}. These are dual to $\mathcal{N}=1$ SCFTs, and arise from reduction of 10d RR fields on cycles of the internal manifold. The corresponding (abelian) Chern-Weil currents are often gauged by bulk Chern-Simons terms that descend directly from 10 dimensions. More concretely, consider Type IIB string theory on a Sasaki-Einstein manifold threaded by $N$ units of five-form flux. Dimensional reduction of $C_4$ along the three-cycles of the Sasaki-Einstein manifold leads to $U(1)$ baryonic symmetries in four dimensions.  The corresponding $U(1)$ gauge fields $C_4=A\wedge \omega_3$  satisfy a modified Bianchi identity
\begin{equation} \rmd \, {\star F}= \kappa H_3 \wedge \rmd\phi,\quad C_2\equiv \phi\, \omega_2\end{equation}
where $\omega_2,\omega_3$ are closed forms and $\kappa=\int \omega_2\wedge\omega_3$. The corresponding Chern-Weil currents are explicitly broken.

On top of these, Type IIB string theory on AdS$_5\times S^5$ contains nonabelian $SU(4)$ gauge bosons, coming from the fluxed gravity reduction on the five-sphere. They are the KK vectors gauging the (double cover of the) isometry group of the five-sphere, and correspond to the R-symmetry of the dual theory. As in the AdS$_7$/CFT$_6$ example of the previous section, we can construct a Chern-Weil current
\begin{equation} J_4= \tr (F\wedge F)\end{equation}
which should either be gauged or broken. In the AdS$_7$ example, we argued that the current was coupled electrically to a massive mode of the 3-form gauge field. It is conceivable that the 5d case behaves similarly, but we have not been able to provide a proof. The 7d examples involve a twisted reduction of the M-theory 3-form which is not available in the Type II perspective, due to the self-duality of the 5-form field strength \cite{Bilal:1999ph}. Perhaps the massive coupling could be more easily described from the T-dual picture \cite{Duff:1998us}, where some of the requisite couplings can be found from reduction of the 10d Chern-Simons term \eqref{dinn}.

Unlike in the AdS$_7$ case discussed above, there is now also a nonabelian current,  associated to the traceless totally symmetric tensor $d_{abc}$ of $SU(4)$ \cite{deAzcarraga:1997ya}, \begin{equation} (J_4)_a= d_{abc} F^b\wedge F^c,\end{equation} 
where indices are lowered and raised using the Killing form. 
This current is not conserved or gauge invariant, but it does satisfy a covariant conservation equation:
\begin{equation} \rmd_A J_4=0.\label{sssss}\end{equation}
In spite of the similarity with the usual current conservation equation $\rmd J_4 = 0$, there is no reason why quantum gravity must exclude currents satisfying \eqref{sssss}. $J_4$ is neither conserved nor gauge invariant, so it does not contradict the absence of global symmetries in quantum gravity. However, in holography, this nonabelian current is in fact the current associated with $SU(4)$ color charge, due to the presence of a 5d triple Chern-Simons term,
\begin{equation}
k\int_{\text{AdS}_5} \tr(A\wedge F\wedge F),\label{susu}
\end{equation}
where $k= N^2-1$ \cite{Bilal:1999ph}. Including Chern-Simons terms, the Yang-Mills equation of motion changes to
\begin{equation}
\rmd \, {\star F}^a-ig f^a_{bc} A^b\wedge \star F^c =k\, d^a_{bc} F^b \wedge F^c,
\label{YMM}
\end{equation}
and so the current we constructed is simply the current of $SU(4)$ color charge. In this example, the nonabelian Chern-Weil current is ``broken'' in the sense that its divergence is nonvanishing. However, in contrast to the case of an abelian gauge field, no quantum gravity principle requires this ``breaking.'' It would be interesting to see if the abelian arguments could be modified so that the presence of the triple Chern-Simons term could be understood from first principles. We leave further exploration of this possibility to future work.

We finish this section with some more general comments. As we have seen, the topological terms that gauge Chern-Weil currents often correspond to 't~Hooft anomalies of the dual field theory involving currents dual to the bulk gauge fields. While this seems to be a general story whenever the bulk Chern-Weil symmetry is gauged, we lack a similar understanding when the Chern-Weil current is coupled a \emph{massive} gauge field.  For instance, consider the R-symmetry $\tr(F^2)$ current in AdS$_5$. We might expect that the coupling to a massive bulk field would be dual to a mixed anomaly involving the R-symmetry and a vector operator above the unitarity bound, which is a ``non-conserved current.'' It would be worthwhile to understand this better, and we hope to return to this point in future work.

\section{Phenomenological Outlook}\label{sec:pheno}

\subsection{Axions: Existence and Quality Problem}
\label{subsec:axionquality}

Axions play a crucial role in many scenarios for physics beyond the Standard Model. Initially proposed to solve the strong CP problem \cite{Peccei:1977hh, Peccei:1977ur, Weinberg:1977ma, Wilczek:1977pj}, such fields also provide prominent candidates for dark matter \cite{Preskill:1982cy, Dine:1982ah, Abbott:1982af} or the inflaton \cite{freese:1990rb, kim:2004rp}, among many other possible applications. In our discussion we will refer to an axion that couples to $\tr(F \wedge F)$ for QCD and solves the strong CP problem as a ``QCD axion.''\footnote{Some authors prefer to reserve the term ``axion'' for the QCD axion, and use ``axion-like particles'' or ``ALPs'' for the general case.}

Since the early days of superstring theory, string constructions of 4d gauge theories have been found to include axions \cite{Witten:1984dg, Barr:1985hk, Choi:1985je}. In supersymmetric theories, these axions form part of a complex scalar field, together with a modulus or ``saxion'' that controls the coefficient $1/g^2$ of a gauge kinetic term. One reason to expect axions to exist is the belief that all parameters of string theory are determined dynamically, and in particular continuous parameters like the theta angle are often determined the vacuum expectation values of scalar fields. However, the rigid Calabi-Yau example in Section \ref{sec:rigid} already shows that these scalar fields can in principle be very massive, with a mass around the cutoff of the EFT. 

More generally, in the non-supersymmetric context, we expect that all moduli, including axions, generically acquire a potential and become massive. It is important to understand whether string theory makes generic predictions about the existence and properties of axions, as this constitutes one of the most promising ways to confront string theory predictions with real-world data. For example, given that a nonabelian gauge group exists within a weakly-coupled 4d effective field theory, we can ask: does an axion $\theta$ exist within the effective field theory with a coupling $\theta \tr(F \wedge F)$?\footnote{The rigid Calabi-Yau example in Section \ref{sec:rigid} shows that the axion can still have a very high mass in abelian examples. However, the abelian case is also special from the Chern-Weil point of view, since here (and only here) the Chern-Weil current can be broken by the addition of monopoles. In this section we restrict our discussion to the nonabelian case.} If so, what is its decay constant, i.e., how large is the distance in field space around the $\theta$ circle? What physical effects determine its potential $V(\theta)$? What is the size and functional form of that potential? For the case of QCD, does the axion solve the strong CP problem? Such questions have been asked in many concrete string constructions, and a generic expectation has arisen that string theories contain light axions, with decay constants often of order the string scale, and masses exponentially smaller than the string scale (the literature is too large to review here, but good entry points are \cite{conlon:2006tq, svrcek:2006yi}). 

The perspective on Chern-Weil symmetries that we have taken in this paper provides a useful new angle on several of these questions, especially the existence of axions and the nature of the physical effects contributing to $V(\theta)$. The answers suggested by the Chern-Weil viewpoint suggest that quantum gravity naturally contains some of the necessary preconditions for a QCD axion solution to the strong CP problem. Combining this perspective with the Weak Gravity Conjecture for axions \cite{Arkanihamed:2006dz} could provide further insight on the relationship between the axion potential and decay constant.

Let us begin by {\em assuming}\footnote{There is no need to assume that the fields propagate in higher dimensions if we are willing to base the discussion on the absence of $(-1)$-form global symmetries in four dimensions, as explained later.} that the Standard Model gauge fields propagate in $d > 4$ dimensions, either within the bulk of the extra dimensions (as in heterotic string theory) or on a higher-dimensional brane (e.g., D7-branes within Type IIB string theory). In this circumstance, the Chern-Weil current $\tr(F \wedge F)$ for QCD would generate a forbidden $(d-5)$-form generalized global symmetry, unless it is gauged or broken. Unlike the case of a Chern-Weil current for an abelian gauge group, which can be broken by the mere existence of magnetic monopoles that lead to $\rmd F \neq 0$, it is more difficult to eliminate the Chern-Weil current for a nonabelian gauge group. Indeed, in every example that we have discussed, there is a $(d-4)$-form gauge field that couples to a linear combination of the Chern-Weil current and other currents,
\begin{equation}
C_{d-4} \wedge \left[ \frac{1}{8\pi^2}  \tr(F \wedge F) + \cdots \right].
   \label{eq:CFFgeneral}
\end{equation} 
The menu of options for what appears in the `$\cdots$' is quite limited: we can have currents associated with ``small instantons,'' like the D$(p-4)$-branes discussed in Section \ref{subsec:openstringcurrents}, which are continuously connected to YM instantons; currents for other gauge groups that are ``unifiable'' with QCD (see Section \ref{ssec:breaking}); currents for other gauge groups that are {\em interchanged} with QCD under an outer automorphism (as in the $E_8 \times E_8$ case discussed in Section \ref{sec:heterotic}); or 4-form field strengths of 3-form gauge fields.\footnote{For instance, consider the Bianchi identity for $G_6$ in the presence of D6-branes in Type IIA:
\beq
\rmd G_6=-H_3\wedge G_4+\tr\mathcal{F}_\Dp^2\wedge \delta^{\mathrm{D6}}_3
\eeq
It implies that the combination of currents on the right-hand side is gauged by $C_3$. When compactifying to four dimensions, it reduces to $hG_4 + \tr(\mathcal{F}_\Dp^2)$ gauged by the axion $\phi=\int C_3$, where $h$ is an NS flux. } While we have no rigorous proof that these are the only possibilities, any viable option should require that the $C_{d-4}$-charge can be transferred among any of the terms appearing in the brackets, so that no unbroken, ungauged currents are left over. In particular, {\em we know of no way to fully break a Chern-Weil current for a nonabelian gauge theory, without a coupling to a gauge field as in \eqref{eq:CFFgeneral}}. Although this gauge field is very massive in some cases, in most cases it has a mass at or below the cutoff; we will elaborate on this below.  Furthermore, we have discussed $C_{d-4}$ and the associated $(d-5)$-form global symmetry generally, but the discussion applies in particular to $d=4$, with the caveats discussed in Section \ref{sec:axionmonodromy} regarding $(-1)$-form symmetries.

While the idea that Chern-Weil currents are always somehow coupled to gauge fields is not rigorous, it strongly suggests not only that axions exist, but that their properties are constrained. In particular, it sheds light on one of the greatest challenges of axion model-building, the axion quality problem \cite{Barr:1992qq, kamionkowski:1992mf, holman:1992us, Ghigna:1992iv}. This problem can be summarized as follows: QCD dynamics generates an axion potential $V_\text{QCD}(\theta)$ with a minimum at $\theta  = 0$, where CP violation is absent from the strong interactions \cite{Vafa:1984xg}. Experimental constraints require that the minimum of the full potential lies at $|\theta| \lesssim 10^{-10}$. Hence, any additional contributions to $V(\theta)$ must either have a minimum closely aligned with the QCD minimum (which is highly non-generic or fine-tuned), or must simply be much smaller than the QCD contribution. This is quite challenging (see \cite{Hook:2018dlk} for a recent pedagogical overview). In models where the axion is the phase of a complex scalar $\Phi$ transforming under an approximate global Peccei-Quinn (PQ) symmetry, one must forbid or strongly suppress many PQ-violating operators like $\Phi^n + \Phi^{\dagger n}$ that could appear in the Lagrangian. This might be done by invoking a large discrete gauge symmetry under which such terms carry a nonzero charge. The philosophy behind this approach to the strong CP problem is that $U(1)_\textrm{PQ}$ is only an approximate symmetry in some range of energies, and may be broken by a wide variety of effects.

By contrast, in theories where the axion arises from the holonomy of a higher-dimensional gauge field, the situation is much better. Effects that contribute to the potential are intrinsically nonlocal in the extra dimensions, and are generally exponentially suppressed. Very schematically, if we take the instanton contributions to be cosine potentials and write their coefficients in the form of exponentials, we must consider terms like
\begin{equation}
V(\theta) \sim -\Lambda_\textrm{UV}^4 \left[ \mathrm{e}^{-S_\mathrm{QCD}} \cos(\theta) + \mathrm{e}^{-S_\mathrm{other}} \cos(\theta + \delta)\right].
\end{equation}
If the relative phase $\delta \sim O(1)$, we need $S_\mathrm{other} \simgt S_\mathrm{QCD} + 23$ in order to obtain a minimum at $|\theta| \lesssim 10^{-10}$. We emphasize that these formulas are not to be taken too literally; QCD dynamics does not generate a simple cosine potential, and the exponentially small QCD contribution cannot be thought of as a single-instanton term. The moral is simply that if terms in the potential can be controlled so that they are {\em all} exponentially small, then we have essentially taken the log of the axion quality problem: rather than explaining a small number like $10^{-10}$, we need only a modest additive difference between exponents. This is a well-known fact (discussed, for instance, in Section 2 of \cite{svrcek:2006yi}).

The additional insight provided by the Chern-Weil perspective is that the types of physics that can generate the $\mathrm{e}^{-S_\mathrm{other}}$ terms are very restricted. For example, suppose that an axion field $\theta$ couples to both the QCD Chern-Weil current and that of a hidden confining gauge group, i.e., we have a term of the form $\theta \left[\tr(F \wedge F) + \tr(H \wedge H)\right]$. Then $\theta$ gauges one linear combination of two conserved currents. The independent linear combination must be either gauged or broken. If it is gauged, we have two axions, $\theta$ and $\theta'$, with couplings
\begin{equation}
(n_F \theta + n'_F \theta') \tr(F \wedge F) + (n_H \theta + n'_H \theta') \tr(H \wedge H).
\end{equation}
If the potential generated by the strong dynamics of $H$ is larger than that generated by the strong dynamics of QCD, we can integrate out the linear combination $n_H \theta + n'_H \theta'$, leaving behind a light periodic axion coupling to $\tr(F \wedge F)$, which will then solve the strong CP problem.\footnote{A more detailed and general explanation of integrating out $N$ heavy axions may be found in \S2.2 of \cite{Fraser:2019ojt}.} Thus, additional {\em gauged} Chern-Weil currents do not contribute to the axion quality problem. Alternatively, it may be that the other linear combination of currents is simply {\em broken}. We have seen that this can happen if the $F$ and $H$ gauge groups are embedded within a unified gauge group (or one that {\em could} be unified somewhere in the moduli space of the theory). It can also happen if there is a $\mathbb{Z}_2$ exchange symmetry between $F$ and $H$, as in the case of the $E_8 \times E_8$ heterotic string theory, where there is just one $\theta$ angle for the two factors \cite{Dine:1992ya}. In both the unification and exchange scenarios, UV physics relates the different gauge groups, and potentially provides constraints on the value of $\delta$ or the relationship between $S_\mathrm{other}$ and $S_\mathrm{QCD}$. We have also seen that $S_\mathrm{other}$ may arise from small instantons (e.g., D$(p-4)$-branes in Type II string constructions). Because these are limiting cases of small-size QCD instantons, and the QCD axion potential is dominated by infrared effects (where large-size instantons overlap and the dilute gas approximation fails), it is likely that they are subdominant. 

The case with the most severe axion quality problem, then, would appear to be theories of the schematic form
\begin{equation}
\theta \left[  \tr(F \wedge F) + F_4 \right],\label{axfail}
\end{equation}
where $F_4 = \rmd C_3$. In this case, the contribution to $V(\theta)$ from QCD instantons can be overwhelmed by the axion monodromy potential, which arises from $F_4$ at tree level \cite{Dvali:2005an, Silverstein:2008sg, Kaloper:2008fb, Kaloper:2011jz}. In order for the orthogonal combination of the two currents to be broken, in this case it must be possible for a QCD instanton to dissolve into $F_4$ flux. Another way of phrasing the problem is that $\tr(F^2)$ couples to a massive axion; if the mass coming from the fluxes is very large, this will completely spoil the solution to the strong CP problem.

When is \eqref{axfail} most problematic? In particular, can it lead to an axion with mass at or above the EFT cutoff? We did find that in some examples, notably the Chern-Weil currents in AdS$_7$ and AdS$_5$ examples in Section \ref{sec:adscft}, the $C_{d-4}$-form can get a mass of the order of the cutoff of the effective field theory (i.e., the Kaluza-Klein scale). However, these examples lack scale separation. They are AdS solutions in which the internal geometry has a size comparable to the curvature radius of the AdS factor. In these cases, the low-energy theory in AdS is not very effective, and the compactification is better thought of as a higher-dimensional background \cite{Lust:2019zwm}. The usual Swampland statements do not necessarily apply in the lower-dimensional theory, but they make sense from the higher-dimensional perspective. For instance, it is possible to have discrete gauge symmetries of arbitrarily large order \cite{Lust:2019zwm}, which in flat space would be akin to a global symmetry. It is perhaps not unreasonable to expect that the physics of Chern-Weil currents is different too.

Restricting to scale-separated examples, with a reasonable low-energy EFT,  we did find, in every example we considered, that \emph{nonabelian gauge fields are always accompanied by an axion with mass below the EFT cutoff, coupling to a linear combination of $\tr(F \wedge F)$ and other Chern-Weil currents}. If this statement turned out to be true in general, it would ameliorate the problem in \eqref{axfail} and get us closer to a quality QCD axion in string theory. At this stage we do not have a general argument for it, and it could very well be that counterexamples exist. For now we will content ourselves with briefly describing some classes of examples that support this observation, hoping to return to this interesting question in the future:
\begin{itemize}
\item In $d>5$ supersymmetric theories, the statement follows from unitarity and supersymmetry. For instance, in 6d $\mathcal{N}=1$ theories, positivity of the vector field kinetic terms requires the existence of a combination of scalars with positive inner products with the vector of coefficients $b_i$ of the $b_i\int B^i_2\tr(F^2)$ topological couplings, as described in \cite{Kim:2019vuc}. This in particular means that for any given nonabelian factor, some $b_i$ has to be nonzero. More generally, the statement is a direct consequence of supersymmetry in any 4d $\mathcal{N}=2$ theory where the gauge coupling is a modulus. This is the case in every nonabelian example we know of. 

\item A more interesting class of examples is then 4d $\mathcal{N}=1$. For instance, the model-independent axion in heterotic compactifications to four dimensions sometimes obtains a large mass via the Green-Schwarz mechanism, but usually this mass is below the EFT cutoff \cite{svrcek:2006yi}. The Green-Schwarz mechanism generates a potential for charged scalar fields, which often get a vev, leading to additional axions. In this case, a linear combination of axions remains very light. We do not know how general this picture is, but we note that \cite{Aldazabal:2018nsj} related the presence of these charged fields to Swampland constraints.

\item For 4d $\mathcal{N}=1$ theories arising from Type II compactifications on Calabi-Yau orientifolds, with gauge fields on D$p$-branes wrapping a cycle $\Sigma_{p-4}$, the holomorphic gauge kinetic functions depend linearly on chiral superfields of the form $\mathrm{e}^{-\phi} \mathrm{Vol}(\Sigma_{p-4}) + i \int_{\Sigma_{p-4}} C_{p-4}$, so that the holonomy of $C_{p-4}$ is the 4d axion field. In the IIB case, we consider $p = 7$ branes, and these fields are the K\"ahler moduli $T$; in the IIA case, we consider $p = 6$, and these fields are the complex structure moduli $U$. These moduli generally remain light; indeed, understanding the mechanism by which they obtain a mass constitutes the well-studied moduli stabilization problem in these compactifications.

\item Finally, we could discuss non-supersymmetric string theories in 10 dimensions. There, the coefficient of a $\int B_6 \tr (F^2)$ in the action is determined via the Green-Schwarz mechanism. A vanishing term would mean that the anomaly polynomial factorizes as $\tr(R^2) I_8$. There is no obvious reason why this cannot happen, and yet we find that indeed this term is nonvanishing in the three tachyon-free non-supersymetric models we know of (the $SO(16)\times SO(16)$ heterotic string \cite{AlvarezGaume:1986jb}, the $U(32)$ model \cite{Sagnotti:1996qj}, and the $Sp(16)$ Sugimoto string \cite{Sugimoto:1999tx}). The statement is false for tachyonic models. The easiest way to see this is to consider again IIA on a Calabi-Yau three-fold, and consider a stack of $n$ unstable $\widehat{D3}$-branes. These have $U(n)$ worldvolume gauge fields, but couple to no RR potential  \cite{Sen:2004nf}. Thus, they lead to a nonabelian Chern-Weil current which does not couple to any axion. 

\end{itemize}

Whatever the correct statement turns out to be, the observation that string compactifications produce light axions is hardly new \cite{svrcek:2006yi}. The Chern-Weil perspective hints at the possibility that this ubiquity of axions might somehow be related to properties of $(-1)$-form symmetries in the low-energy effective field theory. 
Viewing the QCD axion as a gauge field for a linear combination of currents that includes the QCD Chern-Weil current suggests that the terms in the axion potential arise from a small menu of possibilities. The axion quality problem is not completely absent, because other instanton contributions to the potential exist. However, these instantons must be all somehow continuously connected to the Yang-Mills instantons of QCD in the ultraviolet in order to avoid an unbroken global Chern-Weil symmetry, so we expect that there are finitely many effects to consider and that they cannot take a completely arbitrary form. Large axion monodromy masses from fluxes are more problematic; in our scale-separated examples they leave an axion below the cutoff, but they may still give rise to an axion quality problem. The moral of the story is that quantum gravity provides a candidate QCD axion and {\em ameliorates} the axion quality problem; the extent to which it {\em solves} the problem is model-dependent, and deserves further study.

\subsection{Chiral Fermions}
The main message in this paper is that Chern-Weil currents generate global symmetries that are every bit as real as any other ordinary symmetry, and hence, we expect them to be broken or gauged in quantum gravity. We have explained that gauging of a Chern-Weil current means that it becomes exact. In particular, the axionic coupling $\phi \tr(F\wedge F)$ leads to an equation of motion in which the current is exact. 

From this point of view, anything that leads to the Chern-Weil current becoming exact could count as gauging. An interesting possibility in four dimensions is a chiral anomaly. Consider a theory of $N$ massless four-dimensional Dirac fermions $\psi_i$. As is well-known, we can construct independently conserved currents at the classical level, corresponding to an $SU(N)$ vector symmetry and an axial  $U(1)$ symmetry:
\begin{equation} J^a_V= \bar{\psi}_i T^a_{ij} \gamma^\mu\psi_j,\quad J^\mu_A = \bar{\psi}_i \gamma^5\gamma^\mu\psi_i.\end{equation}
These two have a mixed 't~Hooft anomaly, in such a way that coupling the first to a dynamical $SU(N)$ gauge field leads to nonconservation of the second:
\begin{equation} \rmd J_A= \partial_\mu J^\mu_A\propto \tr(F\wedge F).\label{vsw}\end{equation}
 The nonconservation is proportional to the Chern-Weil current of the nonabelian gauge field coupled to the fermion, which is therefore exact. Thus, in four dimensions, we can gauge a Chern-Weil current simply by introducing massless, charged Dirac fermions.  
 
In Section \ref{subsec:axionquality}, we focused on axions arising as holonomies of higher-dimensional gauge fields. In the language of \cite{Reece:2018zvv}, these are {\em fundamental} axions: the point in field space where the axion decay constant $f = 0$ lies at infinite distance, and the axionic strings cannot be constructed as solitonic strings in effective field theory. (Rather, they arise as fundamental objects, like D-branes wrapped on cycles in the internal dimensions.) One may wonder whether a more traditional field theory axion, arising as a pseudo-Nambu-Goldstone boson for a broken PQ symmetry, can gauge a Chern-Weil current in the same way. The present discussion makes it clear that it can, at least in quantum field theory. For example, the classic KSVZ axion model \cite{Kim:1979if, Shifman:1979if} is a renormalizable model consisting of two Weyl fermions $Q$, $\widetilde Q$ in the $\bm{3}$ and $\bm{\bar{3}}$ representations of $SU(3)$, with a Yukawa coupling $y \Phi Q {\widetilde Q}$ to a singlet complex scalar $\Phi$ that gets a vev. This vev spontaneously breaks the (anomalous) PQ symmetry of the theory, giving rise to a large Dirac mass pairing up $Q$ and $\widetilde Q$ and producing a pseudo-Nambu-Goldstone boson $\phi$ that acquires a coupling $\phi \tr(F \wedge F)$ and plays the role of an axion. In this case, we have a global PQ symmetry current which is broken {\em only} by the anomaly, so that $\rmd J_\mathrm{PQ} \propto \tr(F \wedge F)$. In the high-energy theory, the Chern-Weil current is gauged by the existence of the PQ-charged fermions. In the low-energy theory, the PQ current matches onto the axion shift symmetry current, $J_\mathrm{PQ} \sim \star \rmd \phi$, and we can say that the axion gauges the Chern-Weil symmetry, just as we did for a fundamental axion.

Conventional wisdom holds that we do not expect \eqref{vsw} to be exactly correct: because the current is anomalous, it is not conserved and there is really no associated symmetry, and hence no reason for generic symmetry-violating terms (like a mass $m {\bar \psi} \psi$, or a Planck-suppressed operator like $({\bar \psi} \psi)^2/M_{\rm pl}^2$) to be absent from the Lagrangian, unless they can  be forbidden by a (possibly discrete) gauge symmetry. This argument is at the root of the usual understanding of the axion quality problem \cite{Barr:1992qq, kamionkowski:1992mf, holman:1992us, Ghigna:1992iv}. Thus, we will obtain additional terms on the right-hand side of \eqref{vsw}:
  \begin{equation} \rmd J_A= \partial_\mu J^\mu_A\propto \tr(F\wedge F)+ \ldots,\label{vsw2}\end{equation}
where the $\ldots$ represent contributions from operators in the EFT that break the chiral symmetry explicitly in the effective Lagrangian. Due to these additional terms, it is no longer true that $\tr(F \wedge F)$ is exact, and hence the Chern-Weil symmetry is not gauged. One might take this to suggest that quantum gravity favors models of {\em fundamental} axions, like those discussed in the previous subsection, over traditional models like the KSVZ axion accompanied by PQ-violating higher-dimension operators in the Lagrangian, because only the former succeed in gauging the Chern-Weil global symmetry.

However, our preceding discussion suggests a different perspective on the gauging of Chern-Weil currents in traditional axion models. We have already seen numerous examples of a qualitatively similar form to \eqref{vsw2}, in which a linear combination of a Chern-Weil current and other currents is gauged. Such examples are compatible with the absence of global symmetries in quantum gravity if there is a mechanism through which the charges associated with the various terms on the right-hand side can be transmuted into each other, so that all of the ungauged symmetries are broken. For instance, when gauging the current via an axion, the axion equation of motion might have the following form,
  \begin{equation} 
  \rmd{\star \rmd \phi}=\tr(F\wedge F) +\sum_n c_n \cos(n\phi+\delta_n) + \cdots.
  \end{equation}
The linear combination involving $\tr(F\wedge F)$ on the right-hand side is gauged. All orthogonal currents involving $\tr(F\wedge F)$ will be broken if $\tr(F\wedge F)$ can turn into $\cos(n\phi+\delta_n)$. This could occur, for example, if the higher harmonics are generated by the coupling of $\phi$ to the Chern-Weil current of a separate nonabelian gauge group, such that both gauge groups get unified in the UV, as discussed in the previous section. 

This raises a question: can we do the same trick for the chiral fermions, such that all charges appearing in the right hand side of \eqref{vsw2} are somehow related? In other words, are there models in which the PQ symmetry is broken only by QCD instantons, together with terms into which such instantons can be continuously deformed? This would mean that any linear combination involving $\tr(F \wedge F)$ orthogonal to the right-hand side of \eqref{vsw2} must be broken, so $\tr(F\wedge F)$ can be deformed into any of the other terms in that equation. The usual argument is that an anomalous symmetry is not a symmetry at all, so all possible terms can appear. And indeed, it seems hard to imagine how {\em all} terms could be connected to each other, so it is not clear that chiral fermions suffice to gauge a Chern-Weil current without leaving behind unbroken, exact global symmetries. However, we now see that there is indeed potentially something special about a symmetry broken only by effects associated with instantons, as it gauges a Chern-Weil symmetry. In this case, a special set of operators continuously connected to the instanton charge density might appear on the right-hand side of \eqref{vsw2}, rather than the completely arbitrary set of terms expected in the conventional viewpoint. We cannot discard this possibility.
   
Assuming chiral fermions can indeed suffice to gauge the Chern-Weil current, we can envision an example of this mechanism in the Standard model: $U(1)^{B+L}$ is an anomalous chiral global symmetry of the SM without Majorana neutrino masses. Per the above arguments, it gauges a linear combination of the $SU(3)$, $SU(2)$, and hypercharge Chern-Weil currents. Since the QCD axion would presumably be gauging another linear combination, we would be left with a single linear combination, which is then explicitly broken by monopoles. This is a different scenario from the GUT breaking in Section \ref{ssec:breaking}; there, there are two linear combinations of SM Chern-Weil currents that are explicitly broken, and at most one that is gauged. This fits with the fact that $U(1)^{B+L}$ is not a symmetry in GUTs, and one expects additional higher-dimension operator contributions to the right-hand side of \eqref{vsw}.

Whether or not string theory admits constructions with global symmetries broken only by instantons and deformations thereof, in the sense that one obtains an equation of motion of the form \eqref{vsw2} with all charges on the right-hand side continuously deformable into one another, appears to be an open question. If such a construction is possible, it would significantly undermine the traditional arguments associated with the axion quality problem in models of the axion as a pseudo-Nambu-Goldstone boson. This is an exciting possibility that deserves further investigation.

\section{Conclusions}\label{sec:conclus}

In this paper, we have introduced the notion of a Chern-Weil global symmetry, and we have studied how these symmetries may be broken or gauged in quantum gravity. This has enabled us to unify many existing concepts in quantum field theory and string theory into a larger framework and to recast them in modern language.

The Chern-Weil symmetry framework also suggests that many of the phenomena observed in string theory may be general features of quantum gravity, rather than accidents of the string lamppost, as they are intimately related to the absence of these Chern-Weil symmetries. 
For instance, the breaking of $(d-p-1)$-form global symmetries with currents $F_p$ requires the existence of charged $(d-p-2)$-branes. But this is only the tip of the iceberg: these generalized global symmetries are Chern-Weil symmetries of rank one, and if we continue exploring Chern-Weil symmetries of higher rank generated by products of such field strengths, we find a \emph{matryoshka}-like structure in which degrees of freedom on the worldvolume and intersections of the branes are also required. Phenomena like tadpole conditions and branes dissolving in other branes can be further understood as manifestations of gauged and broken Chern-Weil symmetries. It is remarkable how the simple criterion of \emph{no global symmetries in quantum gravity}, when applied to these Chern-Weil symmetries, allows us to reproduce the rich and diverse web of relationships between the different types of charges and charged objects in string theory. The Chern-Weil perspective suggests that these structures may be generic features of quantum gravity.

From a phenomenological point of view, the absence of Chern-Weil symmetries in quantum gravity can be used to motivate a number of guiding principles for physics beyond the Standard Model. For instance, the ubiquitous presence of axions coupled to $\tr(F\wedge F)$ terms in string compactifications is associated to the gauging of the Chern-Weil symmetries with current $\tr(F\wedge F)$. This may shed new light on the axion quality problem in QCD. In general, one should worry about other contributions to the axion potential that could spoil the resolution of the strong CP problem. However, if $\phi$ couples to other gauge fields of a hidden gauge sector, we expect an unbroken Chern-Weil global symmetry unless the hidden sector and the Standard Model unify in the ultraviolet, so their associated instantons may be transformed into one another. A detailed analysis of how this may affect the axion potential is a question for future work.

Our work opens several interesting avenues for future exploration. For instance, we have reformulated axion monodromy as a mechanism for gauging a ($-1$)-form global symmetry, which renders a free parameter dynamical. Alternatively, we have seen how such a ($-1$)-form symmetry can also be broken if the parameter is fixed to a particular value, as happens in the presence of a tadpole constraint. This suggests that the absence of ($-1)$-form global symmetries implies the absence of free parameters in string theory. It would be nice to make this relationship more precise by developing a sharper definition of ($-1)$-form global symmetry.

We have seen a few cases of Chern-Weil global symmetries in string theory whose symmetry-breaking mechanisms are not well understood. This presents an exciting opportunity for future research, as the absence of these Chern-Weil symmetries could perhaps be used to argue for new stringy phenomena beyond the realm of low-energy effective field theory. Furthermore, it could be illuminating to study in more detail how the presence and properties of degrees of freedom at the intersection of branes follows from the absence of higher-rank Chern-Weil symmetries in string theory. 

The absence of Chern-Weil global symmetries also leads to a new perspective on some results that can also be obtained using anomaly inflow, like the presence of degrees of freedom on the worldvolume of localized objects. It would be interesting to dig further into this connection. For instance, it has been shown that anomaly inflow arguments, together with supersymmetry, may be used to constrain, e.g., the rank of the gauge groups consistent with quantum gravity \cite{Kim:2019vuc}. It would be worthwhile to try to recover such results using Chern-Weil symmetries and to see what else these symmetries might teach us about worldvolume field theories.

We have studied a number of examples of Chern-Weil symmetries whose conserved currents are given by wedge products of field strengths of abelian gauge fields. These field strengths are themselves conserved currents of lower degree. It would be interesting to extend our analysis to currents given by wedge products of other, more general currents, which are not field strengths of abelian gauge fields.

If the last century of fundamental physics has taught us anything, it is the power of symmetries. Here, we have seen an illustration of an equal but opposite truth: we have much to learn from the absence of global symmetries, and not merely their presence.

%If we have learnt something in the history of physics is that symmetries (or better, the absence of them) can teach us many things about both known and hidden properties of the theory.

%where a parameter plays the role of the background gauge field. We also explain how these symmetries can be broken if there is a path in the space of field configurations that allows us to deform the charge, which suggests that the absence of ($-1)$-form global symmetries implies the statement of not having free parameters in string theory. 

\section*{Acknowledgements}

We thank Fabio Apruzzi, Joe Davighi, Marco Fazzi, and Cumrun Vafa for useful discussions. 
BH is supported by National Science Foundation grant PHY-1914934.
JM is supported by the National Science Foundation Graduate Research Fellowship Program under Grant No.~DGE1745303.
The research of MM and IV was supported by a grant from the Simons Foundation (602883, CV). 
MR is supported in part by the DOE Grant DE-SC0013607, the NASA Grant 80NSSC20K0506, and the Alfred P.~Sloan Foundation Grant No.~G-2019-12504.
The work of TR at the Institute for Advanced Study was supported by the Roger Dashen Membership and by
NSF grant PHY-191129. The work of TR at the University of California, Berkeley, was supported by NSF grant  PHY1820912, the Simons Foundation, and the Berkeley Center for Theoretical Physics.
We acknowledge hospitality from several institutions where portions of this work were completed: the Amherst Center for Fundamental Interactions at UMass Amherst, site of the workshop ``Theoretical Tests of the Swampland''; the Aspen Center for Physics, which is supported by National Science Foundation grant PHY-1607611; the 2019 Simons Summer Workshop, at the Simons Center for Geometry and Physics at Stony Brook University; and the KITP at UC Santa Barbara, supported in part by the National Science Foundation under Grant No.~NSF PHY-1748958. \

\appendix

\section{Derivation of the Chern-Simons action\label{app:CS}}

In this appendix, we derive the Chern-Simons D-brane action by requiring gauge invariance of the bulk potentials and the presence of worldvolume gauge fields to consistently break the Chern-Weil symmetries. 

We can rewrite the invariant bulk gauge field strengths in \eqref{eq:gaugeinvGp} as
\begin{equation}
  G = \mathrm{e}^B  (\mathd [\mathrm{e}^{- B} C] + G_0) .
\end{equation}
where $G = G_0 + G_2 + \cdots + G_{10}$ and $C = C_1 + C_3 + \cdots + C_9$ are
formal sums and $\mathrm{e}^B = 1 + B + \frac{1}{2} B^2 + \cdots + \frac{1}{5!} B^5$.
A gauge transformation takes the form
\begin{equation}
  \delta C = \mathrm{e}^B  (\mathd \lambda - G_0 \sigma), \quad \delta B = \mathd
  \sigma,
\end{equation}
where $\lambda = \lambda_0 + \lambda_2 + \cdots + \lambda_8$ is another formal
sum.

To couple a charged $p$-brane, we require a $(p + 1)$-form $\mathcal{A}_{p +
1}$ whose gauge variation is closed and has integral periods. Thus,
$\mathcal{G}_{p + 2} = \rmd\mathcal{A}_{p + 1}$ must be gauge-invariant and
closed, with integral periods. We can (non-uniquely) split $\mathcal{A}_{p +
1} =\mathcal{L}_{p + 1} +\mathcal{C}_{p + 1}$, where $\mathcal{L}_{p + 1}$ is
gauge invariant. Then $\mathcal{H}_{p + 2} = \rmd\mathcal{C}_{p + 1}$ differs
from $\mathcal{G}_{p + 2}$ by an exact amount. Thus, to couple a brane we
first specify an integral cohomology class $[\mathcal{H}_{p + 2}]$, which
determines $\mathcal{C}_{p + 1}$ up to the addition of a gauge-invariant
worldvolume Lagrangian $\mathcal{L}_{p + 1}$.

In the absence of worldvolume fields, the possible $[\mathcal{H}_{p + 2}]$ are
as classified in Section \ref{IIAwithout}: $G_0, H_3, J_4, J_6, J_8$, $J_{10}$. $G_0$ would define
the coupling of a $(- 2)$-brane, which is hard to make sense of, whereas $H_3$
defines the Chern-Simons coupling $\oint B_2$, as for a fundamental string.
The remaining currents are mysterious; they seem to allow 2-, 4-, 6-, and 8-branes, but these are not the usual D-branes of IIA string theory; perhaps they
are some kind of non-BPS brane, or perhaps they simply do not exist.

In the presence of worldvolume fields, additional currents are possible. To
obtain standard D$p$-brane actions, we introduce a worldvolume gauge field
$\mathcal{F}_\Dp$ satisfying
\begin{equation}  \label{eq:calFintro}
  \rmd \mathcal{F}_\Dp = H_3,
\end{equation}
corresponding to
\begin{equation}   \label{eq:calFdef}
  \mathcal{F}_\Dp = 2 \pi \alpha' F_\Dp + B_2,
\end{equation}
in terms of potentials, where $F_\Dp = \rmd A_1$ with $A_1$ the worldvolume Maxwell field. Since we have introduced $A_1$ in a rather formal manner, it may not be clear that it should be treated as a dynamical gauge field. However, in the classical theory we can argue that it is dynamical, because its equation of motion is satisfied. The equation of motion for the $B$-field takes the form
\begin{equation} \label{eq:dstarH3}
\rmd \star H_3 = [\text{bulk terms}] + \frac{\delta \mathcal{L}_{p+1}}{\delta B_2} \wedge \delta^\Dp_{9-p},
\end{equation}
which implies that on the brane worldvolume, 
\begin{equation} \label{eq:braneconserve}
\rmd \frac{\delta \mathcal{L}_{p+1}}{\delta B_2} = 0.
\end{equation}
This is simply the familiar statement that a massless gauge field like $B_2$ can only couple to a conserved current. Because $A_1$ only enters the Lagrangian $\mathcal{L}_{p+1}$ via the combination \eqref{eq:calFdef}, the brane-localized conservation law \eqref{eq:braneconserve} is precisely the same as the equation of motion of $A_1$. This gives a justification for treating $A_1$ as a dynamical localized gauge field. The bulk and boundary currents in \eqref{eq:dstarH3} are not independently gauge-invariant, so there is no remaining ungauged global symmetry associated with \eqref{eq:braneconserve}. 

The relationship \eqref{eq:calFintro} implies that on the worldvolume, we have:
\begin{equation}
  \rmd G = \rmd\mathcal{F}_\Dp \wedge G \qquad \Rightarrow \qquad \rmd [\mathrm{e}^{-\mathcal{F}_\Dp} G]
  = 0,
\end{equation}
and we obtain new conserved currents of every even rank, equal to the
even-rank components of $\mathrm{e}^{-\mathcal{F}_\Dp} G$.

To see what the corresponding Chern-Simons coupling looks like, note that
\begin{equation}
  G = \mathrm{e}^{\mathcal{F}_\Dp}  (\mathd [\mathrm{e}^{-\mathcal{F}_\Dp} C] + G_0 \mathrm{e}^{- 2 \pi \alpha' F_\Dp}).
\end{equation}
Thus,
\begin{equation}
  \mathrm{e}^{-\mathcal{F}_\Dp} G = \mathd [\mathrm{e}^{-\mathcal{F}_\Dp} C] + G_0 \mathrm{e}^{- 2 \pi \alpha' F_\Dp}
  = \mathd \left[ \mathrm{e}^{-\mathcal{F}_\Dp} C + G_0 A_1 \wedge \frac{\mathrm{e}^{- 2 \pi \alpha' F_\Dp} - 1}{F_\Dp} \right],
\end{equation}
where
\begin{equation}
  \frac{\mathrm{e}^{- 2 \pi \alpha' F_\Dp} - 1}{F_\Dp} = - 2 \pi \alpha' + (2 \pi \alpha')^2 F_\Dp
  - \frac{4 \pi^3 \alpha^{\prime 3}}{3} F_\Dp \wedge F_\Dp + \cdots,
\end{equation}
is simply a shorthand for its Taylor series. Therefore, the desired
Chern-Simons term for a D$p$-brane is
\begin{equation}
  S_{\tmop{CS}} = \mu_p \int_{\Sigma_{p + 1}}  \left[ \mathrm{e}^{-\mathcal{F}_\Dp} C + G_0
  A_1 \wedge \frac{\mathrm{e}^{- 2 \pi \alpha' F_\Dp} - 1}{F_\Dp} \right],
\end{equation}
where $\Sigma_{p + 1}$ is the $(p + 1)$-dimensional brane worldvolume and the
integral picks out the $(p + 1)$-form component of the integrand.

How general is the above derivation? Could there be other acceptable worldvolume theories, given the supergravity gauge transformations? The general picture here is that the bulk gauge symmetries correspond to brane global symmetries, and the bulk Bianchi identities map to a higher-group symmetry structure of the brane global symmetry background connections. From this point of view, all that one needs is a worldvolume theory which couples consistently to the background connection for the higher group symmetry. While the above picture is the only one we know for this case, it would be interesting to understand this more systematically.

\bibliography{ref}
\bibliographystyle{utphys}
\end{document}